\documentclass[11pt]{article}
\usepackage{graphicx}
\usepackage{dcolumn}
\usepackage{bm}
\usepackage{amsfonts}
\usepackage{amsthm}
\usepackage{amsmath}
\usepackage{amssymb}
\usepackage{epsfig}
\usepackage{color}
\usepackage{textcomp}
\usepackage{hyperref}
\usepackage{titlesec}
\usepackage{slashed}
\usepackage{caption}
\usepackage{subcaption}
\usepackage{titlesec}
\usepackage{titletoc}

\def\ee{\end{equation}}
\def\ba{\begin{eqnarray}}
\def\ea{\end{eqnarray}}

\def\bdm{\begin{displaymath}}
\def\edm{\end{displaymath}}

\def\bq{\begin{quote}}
\def\eq{\end{quote}}

 at 10truept
\usepackage{amsmath}
\numberwithin{equation}{section}
\def\ee{\end{equation}}
\def\ba{\begin{eqnarray}}
\def\ea{\end{eqnarray}}

\def\bq{\begin{quote}}
\def\eq{\end{quote}}

 at 10truept

\newcommand{\beq}{\begin{equation}}
\newcommand{\eeq}{\end{equation}}
\newcommand{\beqa}{\begin{eqnarray}}
\newcommand{\eeqa}{\end{eqnarray}}
\newcommand{\bea}{\begin{eqnarray}}
\newcommand{\eea}{\end{eqnarray}}
\newcommand{\p}{\partial}

 \newcommand{\ep}{\epsilon}

\newcommand{\vect}[1]{\bm{\mathrm{{#1}}}}

\def\lesssim{~\mbox{\raisebox{-.6ex}{$\stackrel{<}{\sim}$}}~}

\def\ltap{\ \raise.3ex\hbox{$<$\kern-.75em\lower1ex\hbox{$\sim$}}\ }
\def\gtap{\ \raise.3ex\hbox{$>$\kern-.75em\lower1ex\hbox{$\sim$}}\ }
\def\gl{\ \raise.5ex\hbox{$>$}\kern-.8em\lower.5ex\hbox{$<$}\ }
\def\roughly#1{\raise.3ex\hbox{$#1$\kern-.75em\lower1ex\hbox{$\sim$}}}

\newcommand{\bi}{\begin{itemize}}
\newcommand{\ei}{\end{itemize}}

\def\ltap{\ \raise.3ex\hbox{$<$\kern-.75em\lower1ex\hbox{$\sim$}}\ }
\def\gtap{\ \raise.3ex\hbox{$>$\kern-.75em\lower1ex\hbox{$\sim$}}\ }
\def\gl{\ \raise.5ex\hbox{$>$}\kern-.8em\lower.5ex\hbox{$<$}\ }
\def\roughly#1{\raise.3ex\hbox{$#1$\kern-.75em\lower1ex\hbox{$\sim$}}}

\usepackage[utf8]{inputenc}
\usepackage[T1]{fontenc}
\usepackage{amsmath}
\usepackage{amsfonts}
\usepackage{amssymb}
\usepackage[version=4]{mhchem}
\usepackage{stmaryrd}
\usepackage{graphicx}
\usepackage[export]{adjustbox}
\graphicspath{{./figures/}}
\usepackage{bbold}
\usepackage{CJKutf8}

\numberwithin{equation}{section}

\begin{document}

\thispagestyle{empty}
\begin{flushright}

\end{flushright}
\vspace*{1.0cm}
\begin{center}
{\Large \bf Three Advanced Lectures on Inflation}\\

\vspace*{1.0cm} {\large Martin S. Sloth\footnote{\tt
sloth@sdu.dk}}\\
\vspace{.5cm} {\em Universe-Origins, University of Southern Denmark, Campusvej 55, 5230 Odense M, Denmark}

\end{center}

\begin{abstract} 
Lecture notes on inflation. The lectures are three double lectures, held for the first time at the {\it Nordita Winter School 2024 - Particle Physics and Cosmology}, covering an advanced introduction to the theory of primordial inflation, as well as the linear and non-linear perturbation theory of slow-roll inflation/quasi-de Sitter spacetimes.

\end{abstract}

\titleformat{\section}
  {\normalfont\Large\bfseries\scshape}
  {}
  {0pt}
  {}

\setcounter{secnumdepth}{2}

\titlecontents{section}[0em]
  {\vskip 0.5ex\bfseries\scshape}
  {}
  {}
  {}

\titlecontents{subsection}[1.5em]
  {}
  {\contentslabel{2.3em}}
  {}
  {\titlerule*[0.5pc]{.}\contentspage}

\newpage
 \tableofcontents
 \newpage

\section{Lecture 1: Background}

The strongest motivation for a phase of primordial inflation is the causality problem of the standard Big Bang model, also sometimes called the horizon problem. To understand the causal structure of the standard Big Bang model, it is useful to consider its Penrose diagram.

Penrose diagrams are also sometimes called conformal diagrams because they show the causal structure of the infinite spacetime by a conformal mapping to a diagram of finite size, preserving the causal property that two light rays only intersect if they intersect in the actual spacetime.

In the standard Big Bang model, the universe was radiation-dominated early on, so we would like to look at the Penrose diagram of a radiation-dominated Friedmann-Lema\^ itre-Roberson-Walker (FLRW) spacetime to understand the causal structure of the standard Big Bang model.

\subsection{Penrose diagrams}

To understand how to draw a Penrose diagram, let's first consider the simplest possible one, the one of Minkowski spacetime. Everyone is familiar with the Minkowski metric $(c \equiv 1)$
\beq
d s^{2}=-d t^{2}+d {\bf x}^{2}~.
\eeq
Now evidently light rays, null geodesics with $d s^{2}=0$, travelling in the $x$-direction, propagates at $45^{\circ}$ angles in the $(t, x)$ plane. So we get the typical light cones
\begin{center}
\includegraphics[max width=10cm]{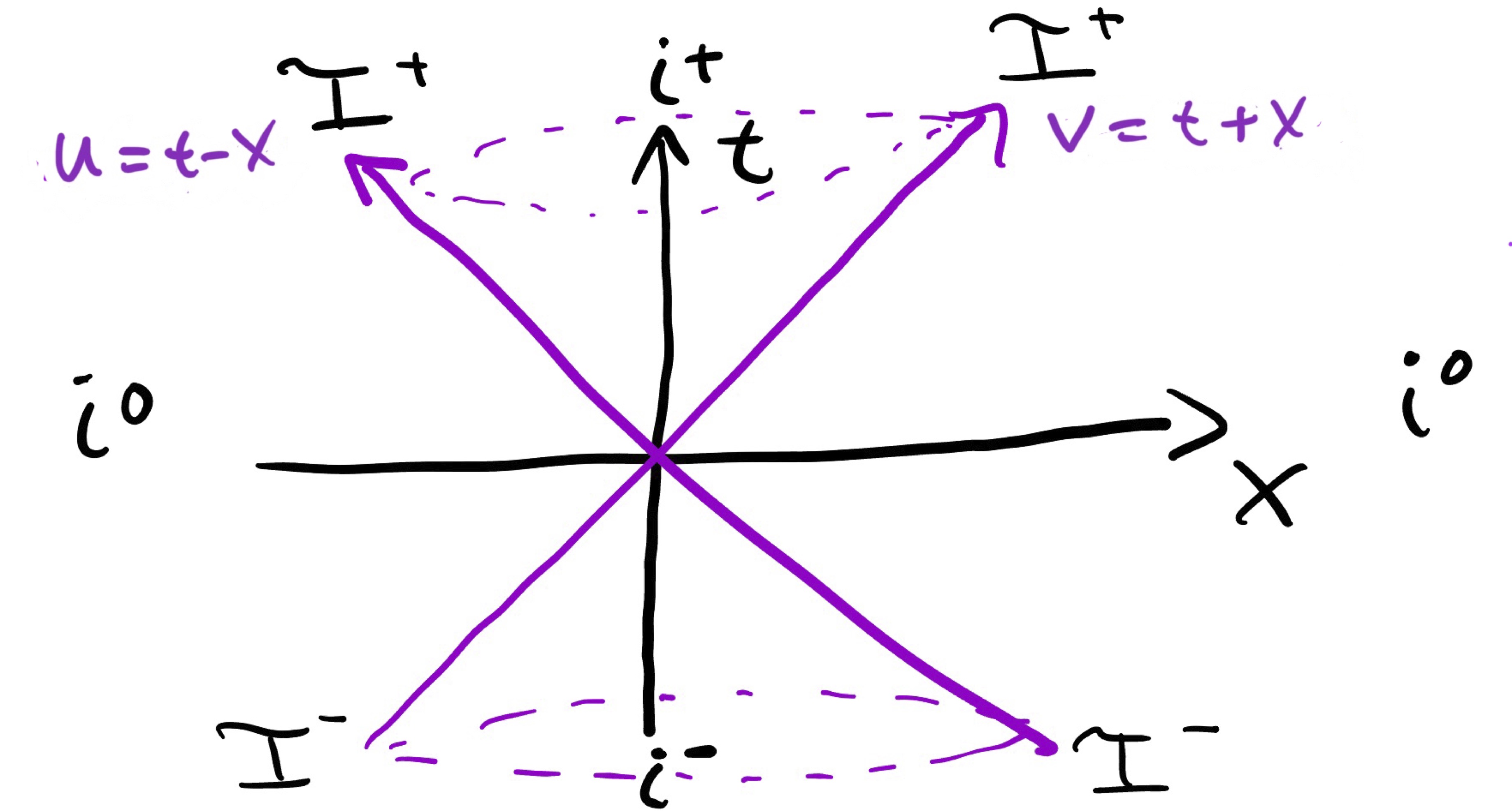}
\end{center}
Now, switching to spherical coordinates, the Minkowski metric is
\beq
d s^{2}=-d t^{2}+d r^{2}+r^{2} d \Omega^{2}\,,
\eeq
where $d \Omega^{2}=\left(d \theta^{2}+\sin ^{2} \theta d \varphi^{2}\right)$. Defining the Penrose coordinates
\beq
\tan ((T \pm R)/2)=t \pm r\,,
\eeq
with $-\pi<T-R \leq T+R<\pi$, the Minkowski metric in those coordinate becomes
\beq
d s^{2}=\frac{1}{ 4 \cos ^{2}((T+R)/2) \cos ^{2}((T-R)/2)}\left(-d T^{2}+d R^{2}+\sin ^{2}(R) d \Omega^{2}\right)\,.
\eeq
  \vspace{4pt}
    \hrule
  \vspace{4pt}
{\bf Exercise 1:} Demonstrate that this is true.
  \vspace{4pt}
    \hrule
  \vspace{4pt}

A conformal rescaling is a rescaling of the metric of the form
\beq
g_{\mu \nu}(x) \rightarrow \tilde{g}_{\mu \nu}(x)=\Omega^{2}(x) g_{\mu \nu}(x)\,.
\eeq
This is not a change of coordinates but a change of the actual geometry, which, however, preserves angles and null geodesics and thus preserves the causal structure. Making a conformal transformation, we obtain the (unphysical) metric
\beq
d \tilde{s}^{2}=-d T^{2}+d R^{2}+\sin ^{2}(R) d \Omega^{2}\,,
\eeq
which is a subsection of the Einstein static universe (spatially closed universe with $a(t)=1$ ) restricted to
\beq
-\pi<T-R \leq T+R<\pi\,.
\eeq
Notice that $T \in]-\pi, \pi[$ and $R \geq 0$. Including also the boundaries, which are not part of the physical region, it is compact.  Since it is compact, we can draw this entire spacetime in a diagram, which is the Penrose diagram of Minkowski spacetime.

\begin{center}
\includegraphics[max width=9cm]{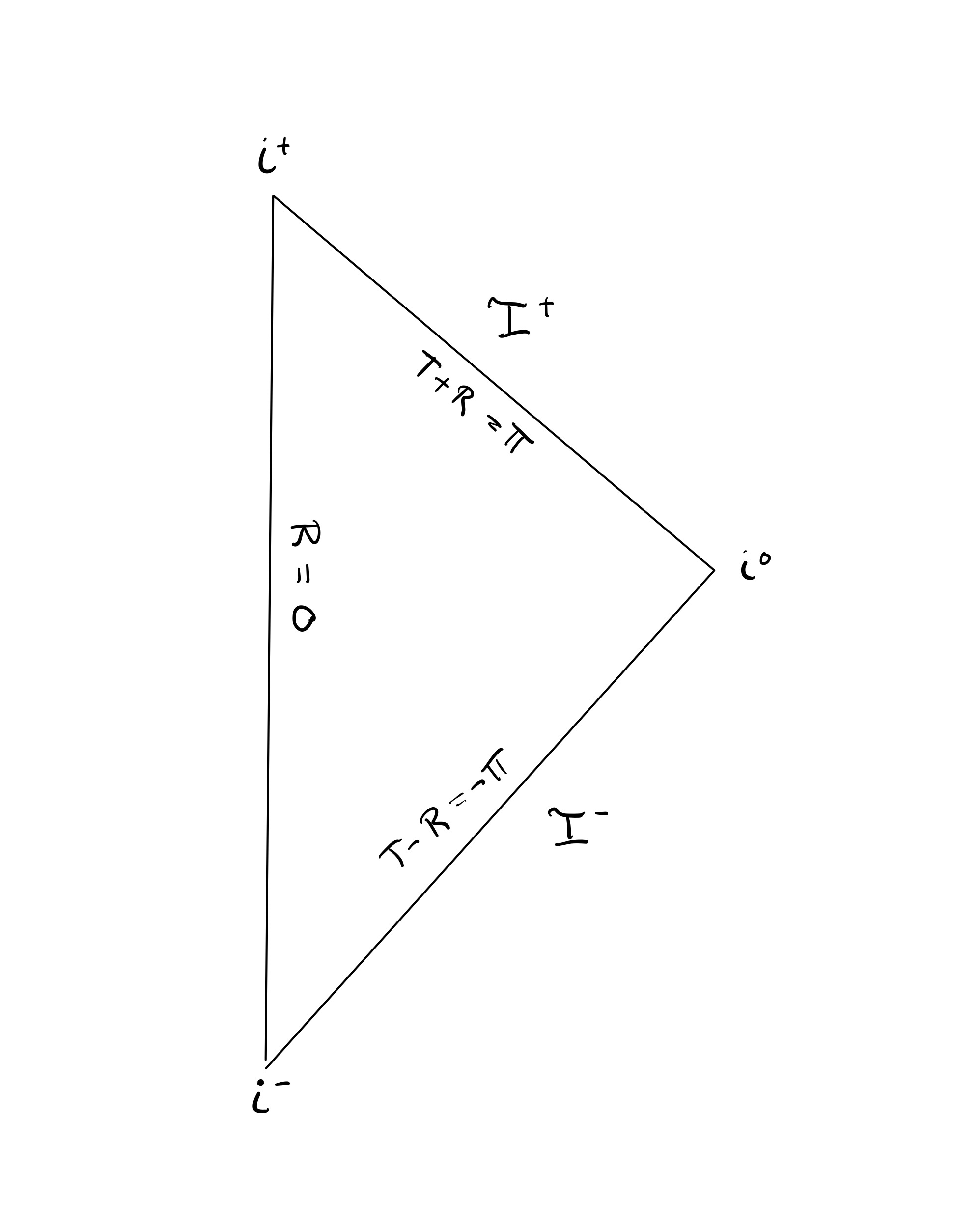}
\end{center}

Notice that the past and future null infinities $ I^{-},I^{+}$, which are manifolds at $T-R=-\pi, T+R=\pi$, the spatial infinity $i^{0}$ at $T=0, R=\pi$ and past and future time-like infinities, $i^{-}, i^{+}$ at $T=-\pi, R=0$ and $T=+\pi, R=0$ are not part of Minkowski spacetime.

Now let's turn to FLRW spacetime. Let's, for simplicity, ignore spatial curvature
\beq
d s^{2}=-d t^{2}+a^{2}(t)\left(d r^{2}+r^{2} d \Omega^{2}\right)~.
\eeq
Introducing conformal time, $\tau$, defined through the relation $a d \tau=d t$, the metric takes the conformally flat form
\beq
d s^{2}=a^{2}\left(-d \tau^{2}+d r^{2}+r^{2} d \Omega^{2}\right)~.
\eeq
We say that it is conformally flat, because after a conformal rescaling, $d s^{2} \to  d\tilde s^{2} = a^{-2} ds^2$, we obtain the Minkowski metric in conformal coordinates
\beq
d \tilde{s}^2=-d \tau^{2}+d r^{2}+r^{2} d \Omega^{2}~.
\eeq
Thus, a flat FLRW universe is conformal to Minkowski and, therefore, has the same Penrose diagram apart from one crucial point. In a radiation (or matter) dominated universe, there is a Big Bang singularity at a finite time in the past. FLRW with ordinary matter (with $p \geq 0$ ), therefore, has a Penrose diagram similar to that of Minkowski spacetime, with the difference that the lower part of the Penrose diagram is cut off by the singularity.

Penrose diagram of FLRW with $p \geq 0$
\begin{center}
\includegraphics[max width=6cm]{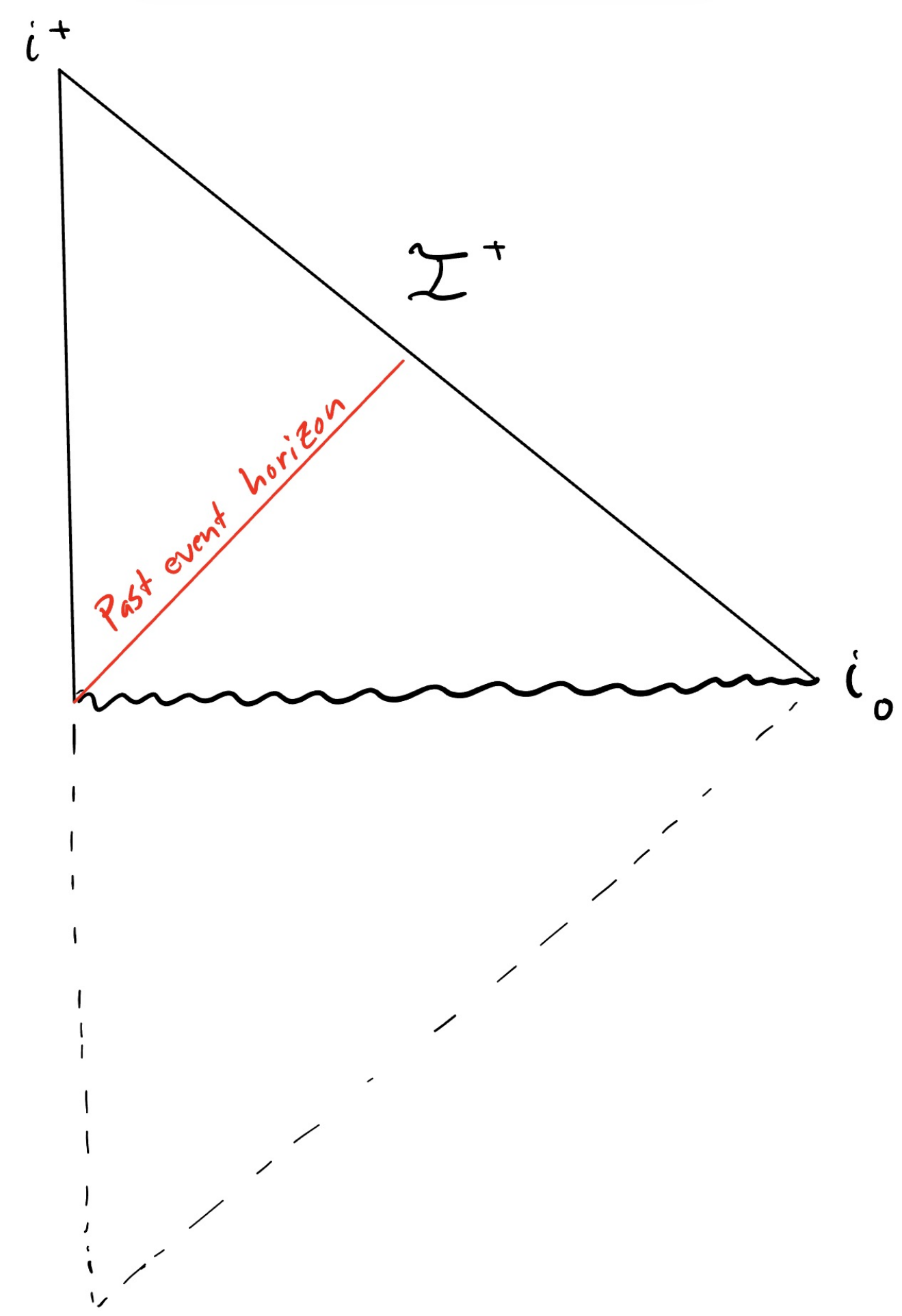}
\end{center}
We cannot influence any spacetime events until they enter the past-event horizon. Similarly, nothing can influence us unless inside our past light cone, as illustrated by the particle horizon in the figure below, which is how far a given observer can "see".
\begin{center}
\includegraphics[max width=6cm]{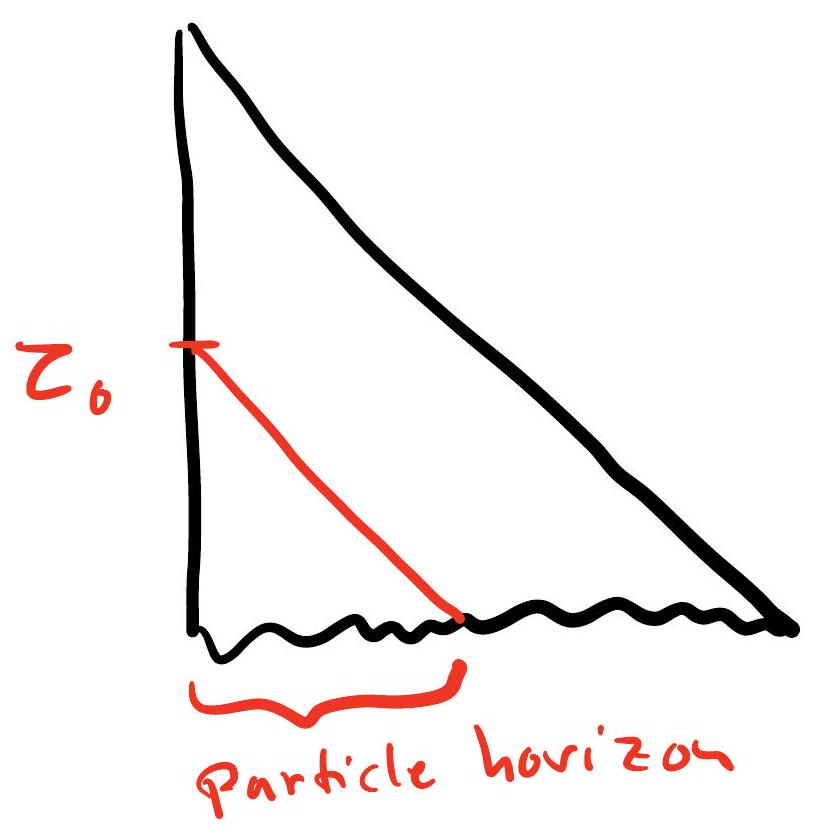}
\end{center}

While null geodesics travel at $45^{\circ}$ angles, co-moving observers are those at rest in FLRW coordinates, which have to end up at future time-like infinity but can start at any $r$. So lines of constant radial coordinates look like this (green lines)
\begin{center}
\includegraphics[max width=6cm]{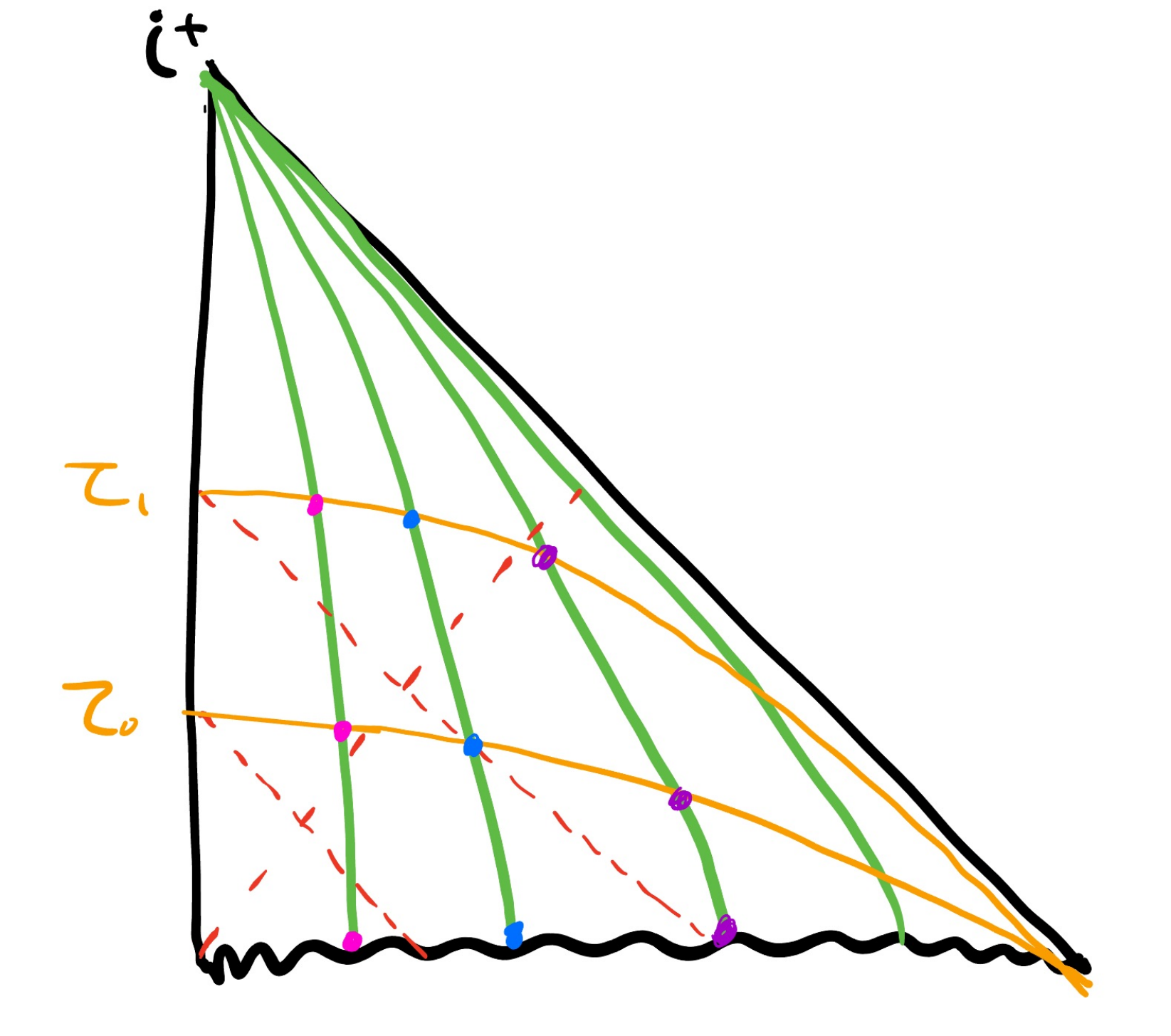}
\end{center}
Similarly, sizes of constant $\tau$ have to end on spatial infinity (orange lines).

By going back to FRW coordinates in conformal time, we can see that a comoving observer (a radial coordinate point) inside our particle horizon is also inside our past event horizon; they are the same.
\begin{center}
\includegraphics[max width=6cm]{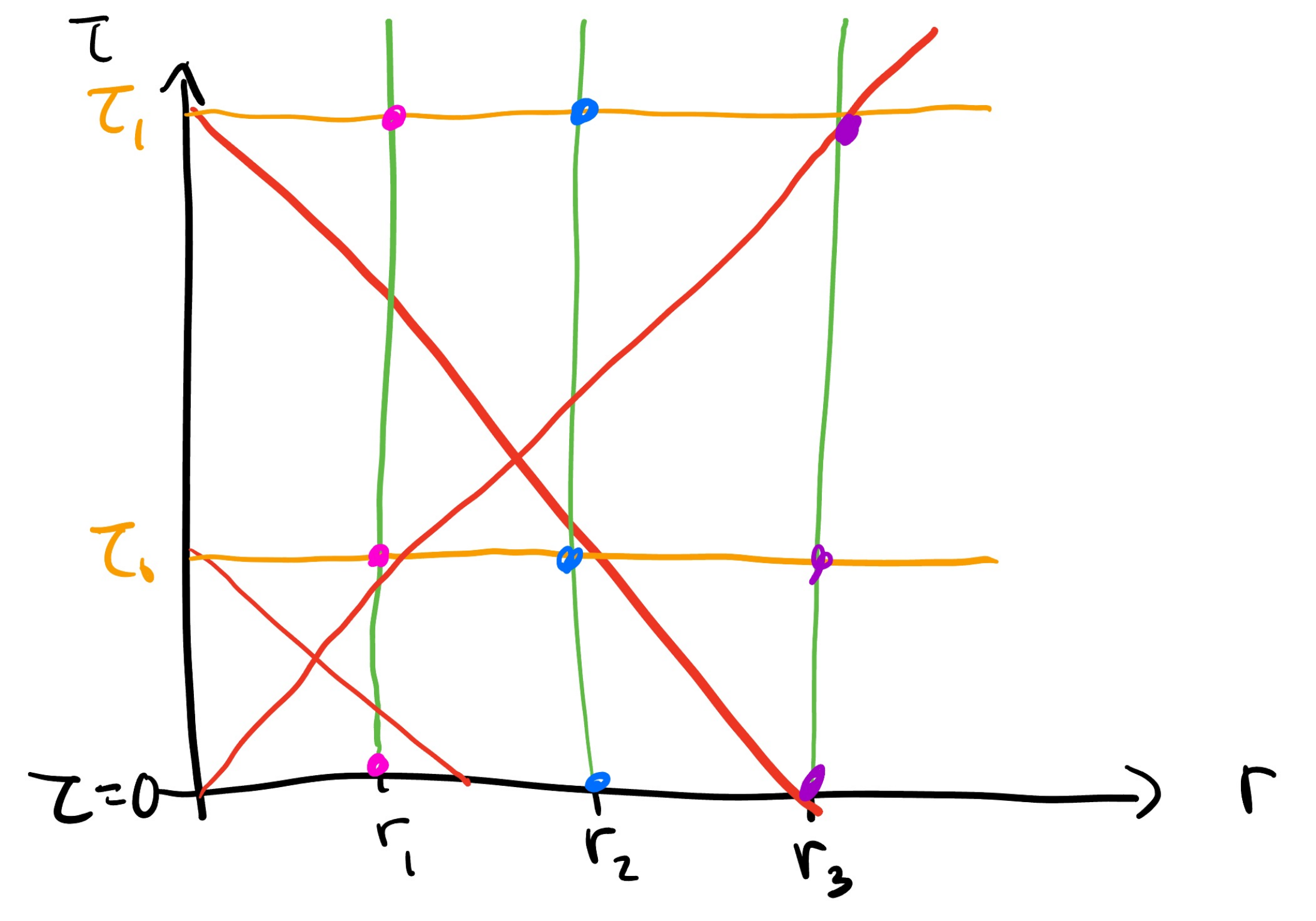}
\end{center}

We can calculate the particle horizon, $L_{\text {ph}}$, at $\tau_{0}$ as the past light cone at $\tau_{0}$ or $t_{0}$ equivalently. That is how far light can have propagated from $t=0$ to $t_{0}$ using for a null geodesic $d t=a d r$
\beq
L_{\text {ph}}=a\left(t_{0}\right) \int_{0}^{r_{0}} d r=a\left(t_{0}\right) \int_{0}^{t_{0}} \frac{1}{a(t)} d t
 =a\left(t_{0}\right) \int_{0}^{a_{0}} \frac{1}{a \dot{a}} d a \\
 =a\left(t_{0}\right) \int_{0}^{a_{0}} \frac{1}{a^{2} H} d a
\eeq
In a radiation-dominated universe $\rho_{r}=3 H^{2} M_{\rho}^{2}\propto 1/a^{4}$, so $H=H_{0}(a_{0}/a)^{2}$ and 
\beq
L_{ph}=a\left(t_{0}\right) \int_{0}^{a_{0}} \frac{1}{H_{0} a_{0}^{2}} d a=\frac{1}{H_{0}}
\eeq
So $1/H_{0}$ is setting the size of the observable universe.

\subsection{Causality problem}

The Cosmic Microwave Background (CMB) we observe consists of photons that last scattered off primordial free electrons at the surface of last scattering, when photons decoupled from matter during/shortly after the recombination of protons and electrons into neutral hydrogen atoms, about 380,000 years after the Big Bang. We observe the CMB temperature to be the same, to pretty high precision, all over the sky \cite{Fixsen:2009ug}
\beq
T_{C M B}=2.7260 \pm 0.0013 \mathrm{~K}~
\eeq

We see that different points at a slice of constant $\tau_{L S}$ (or $t_{L S}$ ) have past light cones that don't overlap, and yet at $\tau_{0} $, we observe them have the same temperature to high precision. This is an apparent paradox - how could they have agreed to have the same temperature if they never could have communicated?

This is illustrated by the green point on the last scattering surface in the figure below. It is inside our past light cone today, hence inside our observable universe today, but at the time of the last scattering, its past light cone did not overlap with our past light cone because it was outside the particle horizon. How could it then have the same temperature at that time? 
\begin{center}
\includegraphics[max width=6cm]{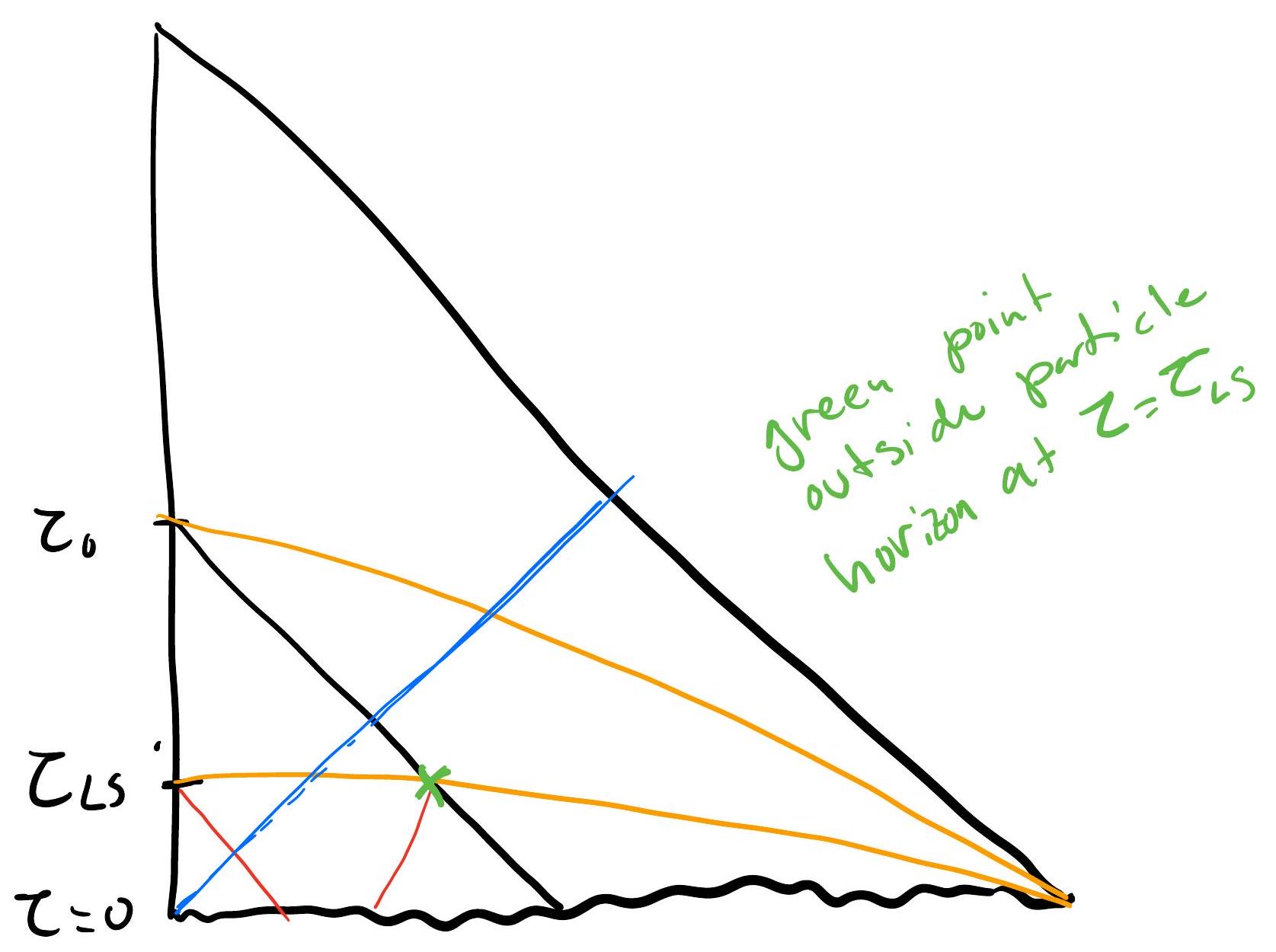}
\end{center}
This is what we call the causality problem of the standard Big Bang model!

Imagine we could move the singularity back in time or remove it; then the problem would be solved! 
\begin{center}
\includegraphics[max width=6cm]{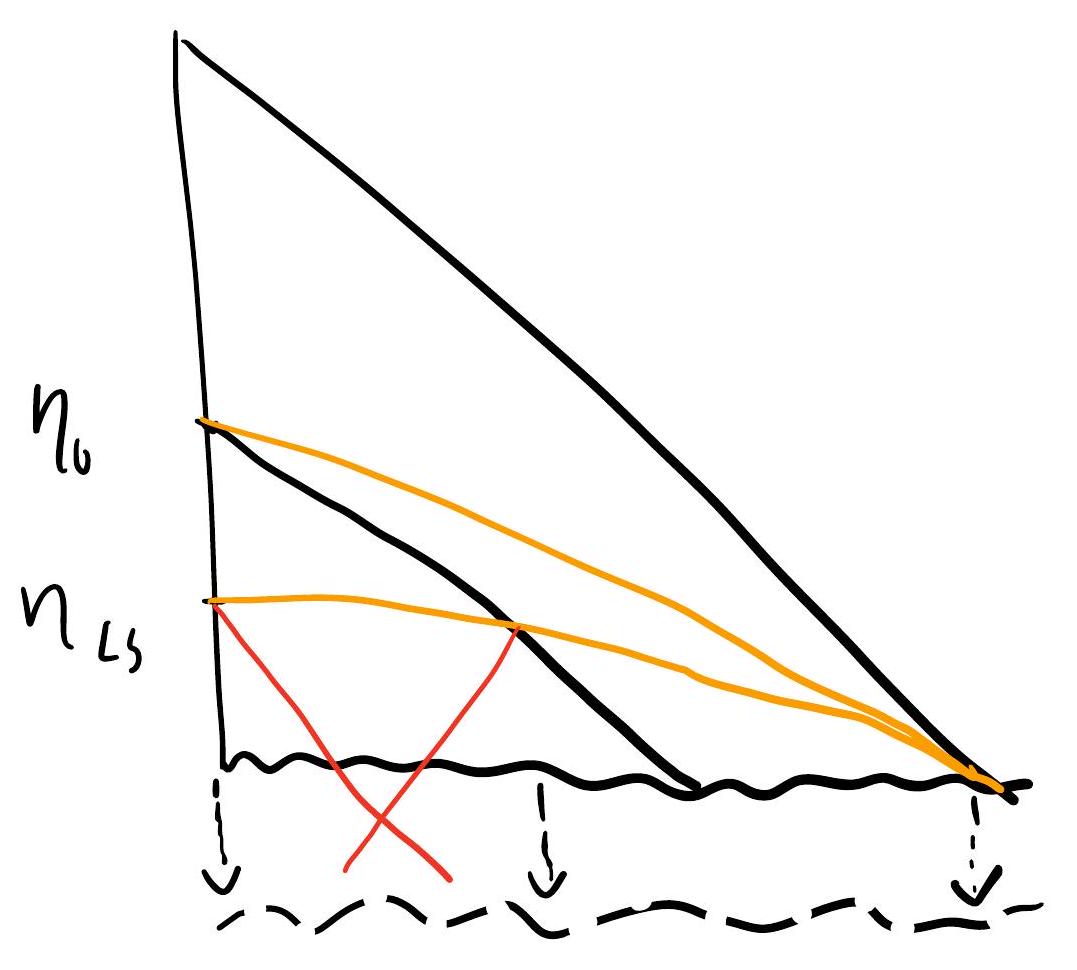}
\end{center}

In particular, completing it with a vacuum-like solution looking like Minkowski, as in the figure below, would work.
\begin{center}
\includegraphics[max width=6cm]{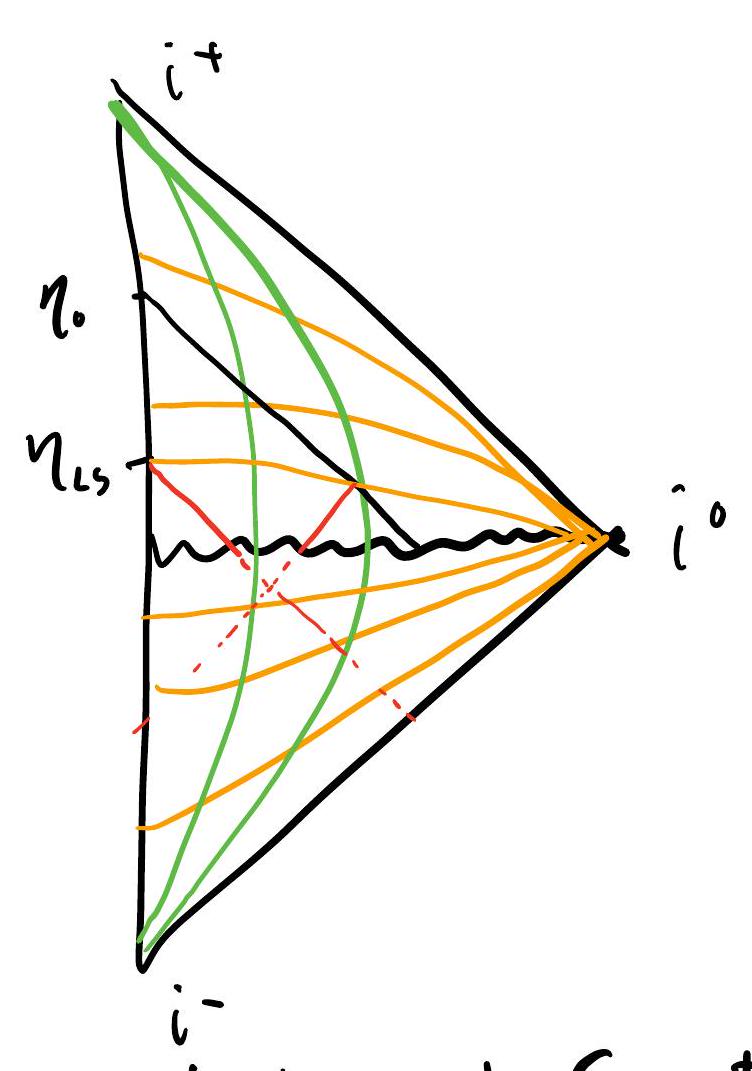}
\end{center}

However, starting before the Big Bang with a zero-density Minkowski vacuum and jumping to Planckian radiation density just at the Big Bang is not very physical. Instead, take a vacuum solution with large vacuum energy, i.e., de Sitter spacetime.

We see from the vacuum Einstein equation
\beq
R_{\mu \nu}-\frac{1}{2} g_{\mu \nu} R+g_{\mu \nu} \Lambda=8 \pi G_N T_{\mu \nu} = 0
\eeq
that we can move the $\Lambda$ term to the right-hand side, and think of it as a constant energy density with negative pressure
\beq
p_{\Lambda}=-\rho_{\Lambda}
\eeq
when comparing with the energy-momentum tensor of an ideal fluid
\beq
T^\mu_{\nu}=\textrm{diag.}(-\rho, p, p, p)
\eeq

Using the first Friedmann equation for this fluid of constant energy density
\beq
H^{2} \equiv\left(\frac{\dot{a}}{a}\right)^{2}=\frac{8 \pi G_N}{3} \rho \propto \textrm{const.} 
\eeq
implies
\beq
a \propto e^{H t} \qquad \textrm{with}\qquad H = \textrm{const.}
\eeq

The second Friedmann eq.
\beq
\frac{\ddot{a}}{a}=-4 \pi G_N\left(p+\frac{1}{3} \rho\right)
\eeq
then implies $p=-\rho$.

Thus, we have an FLRW metric with an exponentially expanding scale factor, which we know is conformal to Minkowski. Still, the energy density can be high, and it can precede the radiation-dominated phase. Also we see from $a\propto e^{H t}$ that $a \rightarrow 0$ only when $t\to -\infty$, so the singularity is indeed
moved in finitely back in $t$.

Having de Sitter FLRW from $-\infty<t<0$ and radiation-dominated FLRW from $0<t<\infty$ we then get a solution to the causality problem, and the Penrose diagram looks like this
\begin{center}
\includegraphics[max width=6cm]{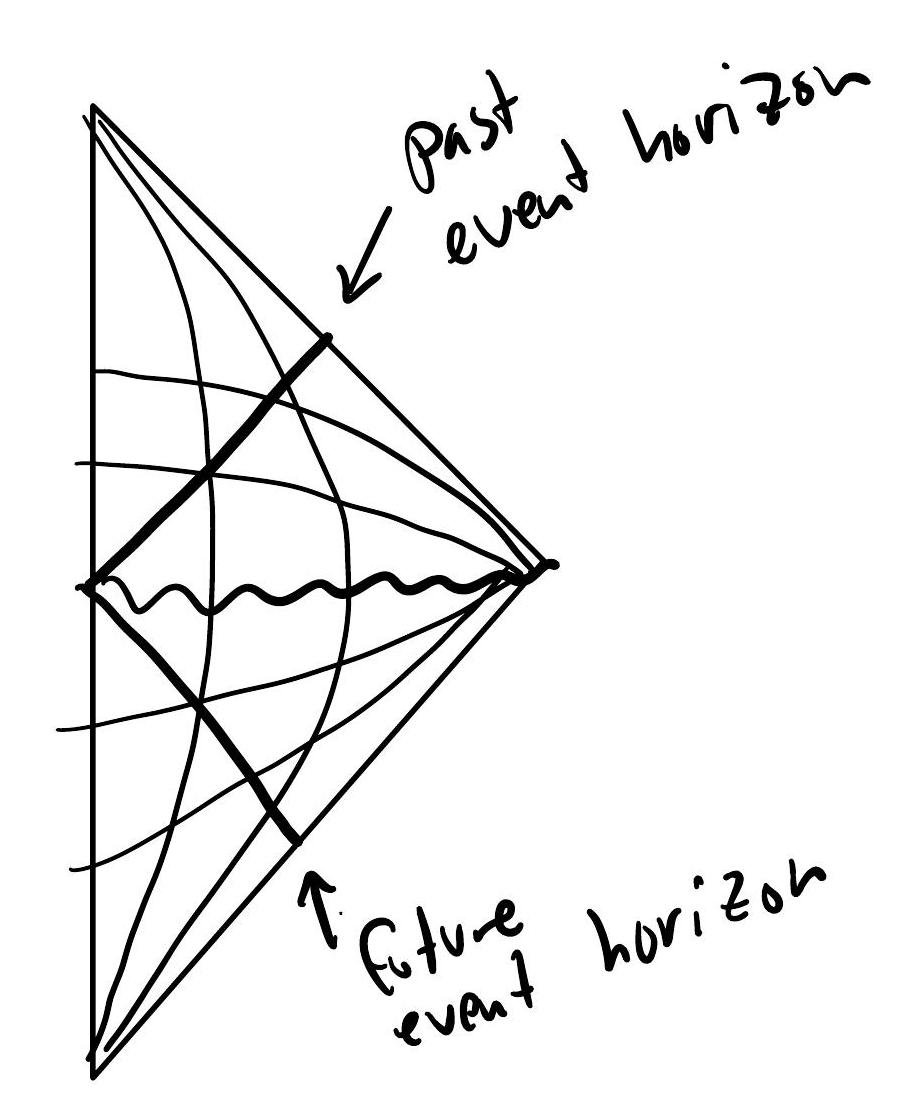}
\end{center}
So, an early de Sitter-like era solves the causality problem because, in the past, everything crossed back inside the event horizon and came in causal contact.

Another way to view it is in a diagram like this
\begin{center}
\includegraphics[max width=10cm]{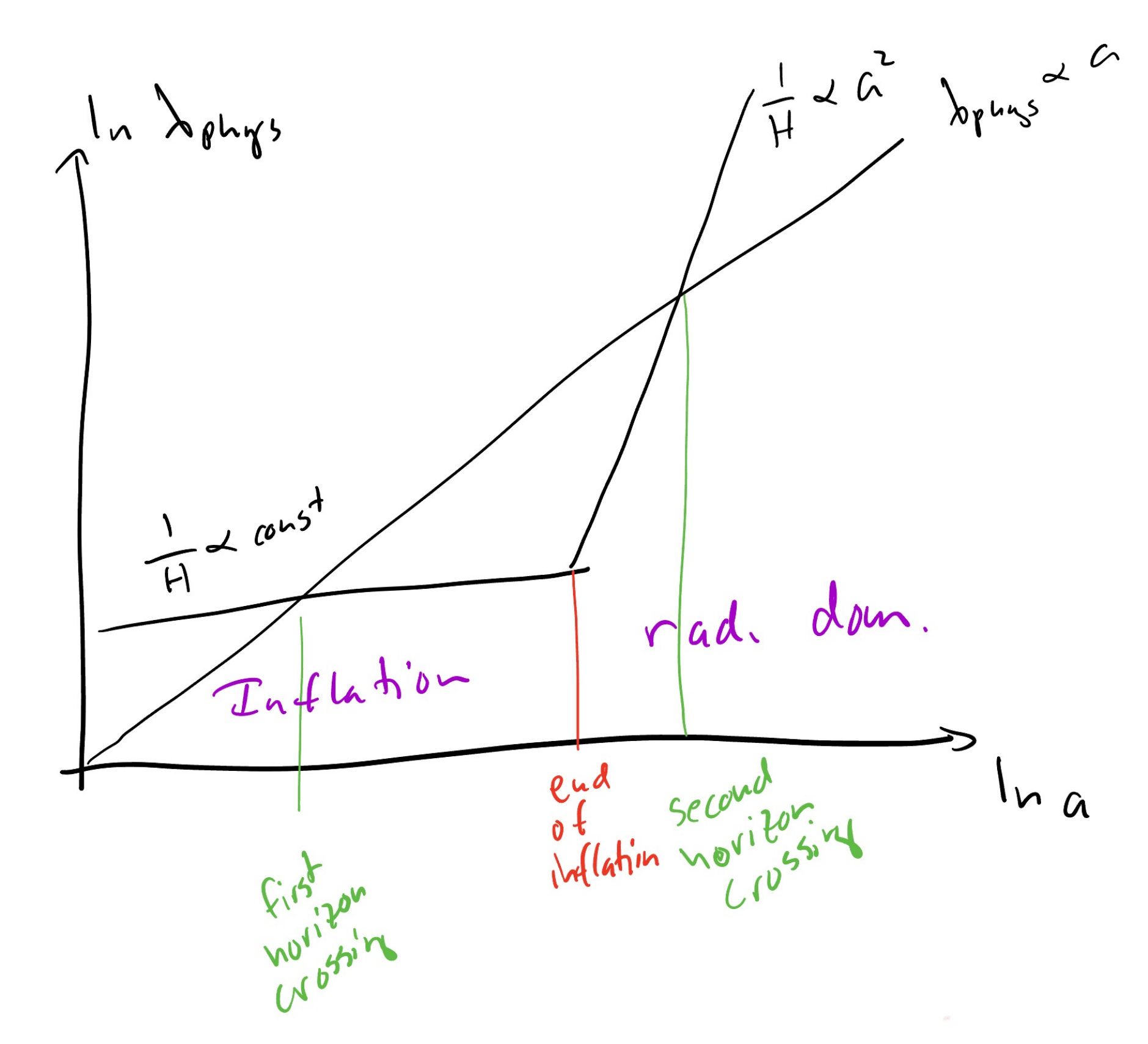}
\end{center}

\subsection{Inflation}

As illustrated in the figure above, the horizon problem requires a period initially where the physical scales $\lambda$ evolve faster than the horizon, so all of the observable universe could be in causal contact in the past. Using $\lambda \propto a, H=\dot{a} / a$, this implies
\beq \frac{d}{d t}\left(\frac{\lambda}{\left|H^{-1}\right|}\right)=\frac{d}{d t}\left(a\left|\frac{\dot{a}}{a}\right|\right)=\frac{d}{d t}|\dot{a}|>0
\eeq
which has two solutions
\beq
\dot{a}>0 \quad \text { and } \quad \ddot{a}>0
\qquad\textrm{or} \qquad \dot{a}<0 \quad\textrm{and} \quad \ddot{a}<0\,.
\eeq
Thus, we need a period of accelerated expansion (Inflation \cite{Guth:1980zm}) or a period of accelerated contraction (Pre-big bang \cite{Gasperini:1992em}, Ekpyrotic\cite{Khoury:2001wf}). Since bouncing from a contracting phase to an expanding phase requires a violation of the null energy condition in a spatially flat universe (observations require approximate spatial flatness), and the null energy condition is considered to hold in well-behaved theories, we will focus on inflation as a solution to the causality problem here. 

Assume we have a period of approximately de Sitter-like expansion with an almost constant energy density; as mentioned above, we have $a\propto e^{H t}$, i.e., exponential expansion. It is convenient to take the log when discussing the amount of expansion and measure the duration of inflation in e-folds
\beq
N=\ln \left(\frac{a\left(t_{R}\right)}{a\left(t_{i}\right)}\right)
\eeq
where $t_{R}$ is the time of reheating at the end of inflation.

To solve the causality/horizon problem, the largest observable scale today (the present horizon scale $H_{0}^{-1}$ ) must have inflated from a value $\lambda_{H_{0}}\left(t_{i}\right)$ smaller than the horizon during inflation
\beq
\lambda_{H_{0}}\left(t_{i}\right)  =H_{0}^{-1}\left(\frac{a\left(t_{R}\right)}{a\left(t_{0}\right)}\right)\left(\frac{a\left(t_{i}\right)}{a\left(t_{R}\right)}\right)\simeq H_{0}^{-1}\left(\frac{T_{0}}{T_{R}}\right) e^{-N} \lesssim H_{I}^{-1}~,
\eeq
where we used that after inflation, we have that the temperature, $T$, drops with the expansion as $T \propto 1 / a$. This leads to a bound on the required number of e-folds
\beq
 N  \gtrsim \ln \left(\frac{T_{0}}{H_{0}}\right)-\ln \left(\frac{T_{R}}{H_{I}}\right)  \sim 67-\ln \left(\frac{T_{R}}{H_{I}}\right) \gtrsim 60
\eeq

\subsection{Flatness problem}

Inflation can also explain why the universe is observed to be spatially flat to high precision, as any initial curvature will be inflated away
\beq
\Omega-1=\frac{k}{(a H)^{2}} \quad H \propto \textrm{const.} \quad \Rightarrow\quad  (\Omega-1)_R \simeq e^{-2 N} (\Omega-1)_i\,.
\eeq

All this, taken together, makes it a good assumption that there was a period of quasi-de Sitter expansion in the early universe  \cite{Guth:1980zm}.

\subsection{de Sitter interlude}

We saw that there is a set of coordinates where de Sitter looks like the lower part of Minkowski
\begin{center}
\includegraphics[max width=6cm]{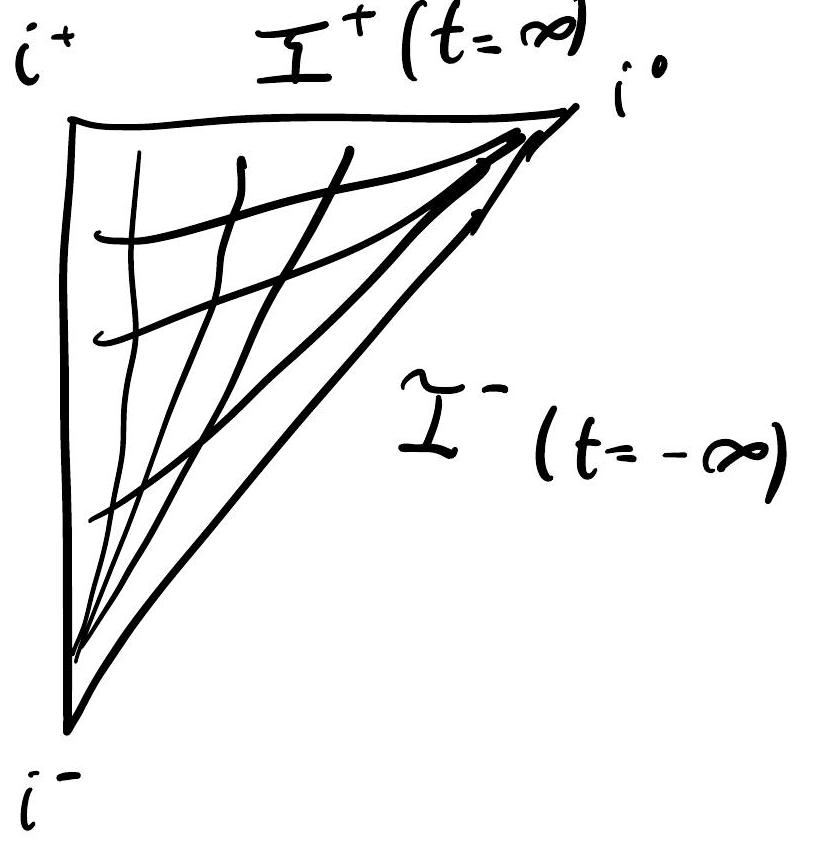}
\end{center}
where we see that, using $a=e^{H t}$ and $a d \tau\equiv d t$, we have 
\beq
\tau=\int \frac{1}{a} d t=\int e^{-H t} d t=\frac{-1}{a H} ~,
\eeq
and therefore
\beq
 \tau \rightarrow 0\quad \textrm{for}\quad t \rightarrow \infty \quad(a \rightarrow \infty) \qquad\textrm{and}\qquad
 \tau \rightarrow-\infty \quad\textrm{for}\quad t \rightarrow-\infty \quad(a \rightarrow 0)~.
\eeq

However, just like the Schwarzschild metric is not geodesically complete, as in-falling observers are crossing the horizon in finite time, these coordinates of de Sitter are not geodesically complete. They only cover half of the de Sitter spacetime.

The original singularity theorem of Hawking and Penrose assumes the strong energy condition
\beq
\rho+3 p \geq 0
\eeq
which is violated during inflation, and so does not apply to inflation. Borde, Guth, and Vilenkin, however, showed that inflation is also not geodesically complete \cite{Borde:2001nh}.

For null geodesics, one cannot use the proper time to parametrize their curve, so one needs to use an affine parameter, $\lambda$. One can show that the geodesic equation is satisfied for null geodesics in FLRW spacetimes if
\beq
d \lambda \propto a d t
\eeq
  \vspace{4pt}
    \hrule
  \vspace{4pt}
{\bf Exercise 2:} Show that this is true.
  \vspace{4pt}
    \hrule
  \vspace{4pt}  
  \vspace{4pt}
  \vspace{4pt}
    
Normalizing the affine parameter by choosing
\beq
d \lambda=\left.\frac{a(t)}{a\left(t_{f}\right)} d t \quad \Rightarrow \quad \frac{d \lambda}{d t}\right|_{t=t_{f}}=1~,
\eeq
multiplying by $H=\dot{a}/{a}$ and integrating
\beq
\int_{\lambda\left(t_{i}\right)}^{\lambda\left(t_{f}\right)} H(\lambda) d \lambda  =\int_{t_{i}}^{t_{f}} \frac{d a}{d t} \frac{1}{a} \frac{a}{a_{f}} d t \\
 =\int_{a\left(t_{i}\right)}^{a\left(t_{f}\right)} \frac{1}{a\left(t_{f}\right)} d a =1-\frac{a\left(t_{i}\right)}{a\left(t_{f}\right)} \leq 1
\eeq
Now define the averaged $H_{\textrm{av}}$
\beq
 H_{a v}=\frac{1}{\lambda\left(t_{f}\right)-\lambda\left(t_{i}\right)} \int_{\lambda\left(t_{i}\right)}^{\lambda\left(t_{f}\right)} H(\lambda) d \lambda \leq \frac{1}{\lambda\left(t_{f}\right)-\lambda\left(t_{i}\right)}
\Rightarrow  \lambda\left(t_{f}\right)- \lambda\left(t_{i}\right)\leq\frac{1}{H_{a v}}
\eeq

So any backward null-geodesic in spacetime with $H_{a v}>0$ must have a finite affine length. One can also show that this holds for time-like geodesics. Thus, de Sitter in the FLRW coordinates is only past eternal for observers at rest in co-moving coordinates, while other observers will see a universe that has only existed a finite time. This is because the FLRW coordinates of de Sitter, also called the Poincar\'e patch or flat slicing, only cover half of the de Sitter spacetime.

The simplest way to obtain de Sitter spacetime is to realize it as a hypersurface in a 5-d Minkowski spacetime describing a hyperboloid
\beq
-X_{0}^{2}+X_{1}^{2}+X_{2}^{2}+X_{3}^{2}+X_{4}^{2}=l^{2} \quad\left[\Lambda=\frac{3}{l^{2}}\right]
\eeq
in a flat five-dimensional space with metric
\beq
d s^{2}=-d X_{0}^{2}+d X_{1}^{2}+d X_{2}^{2}+d X_{3}^{2}+d X_{4}^{2}~.
\eeq
\begin{center}
\includegraphics[max width=10cm]{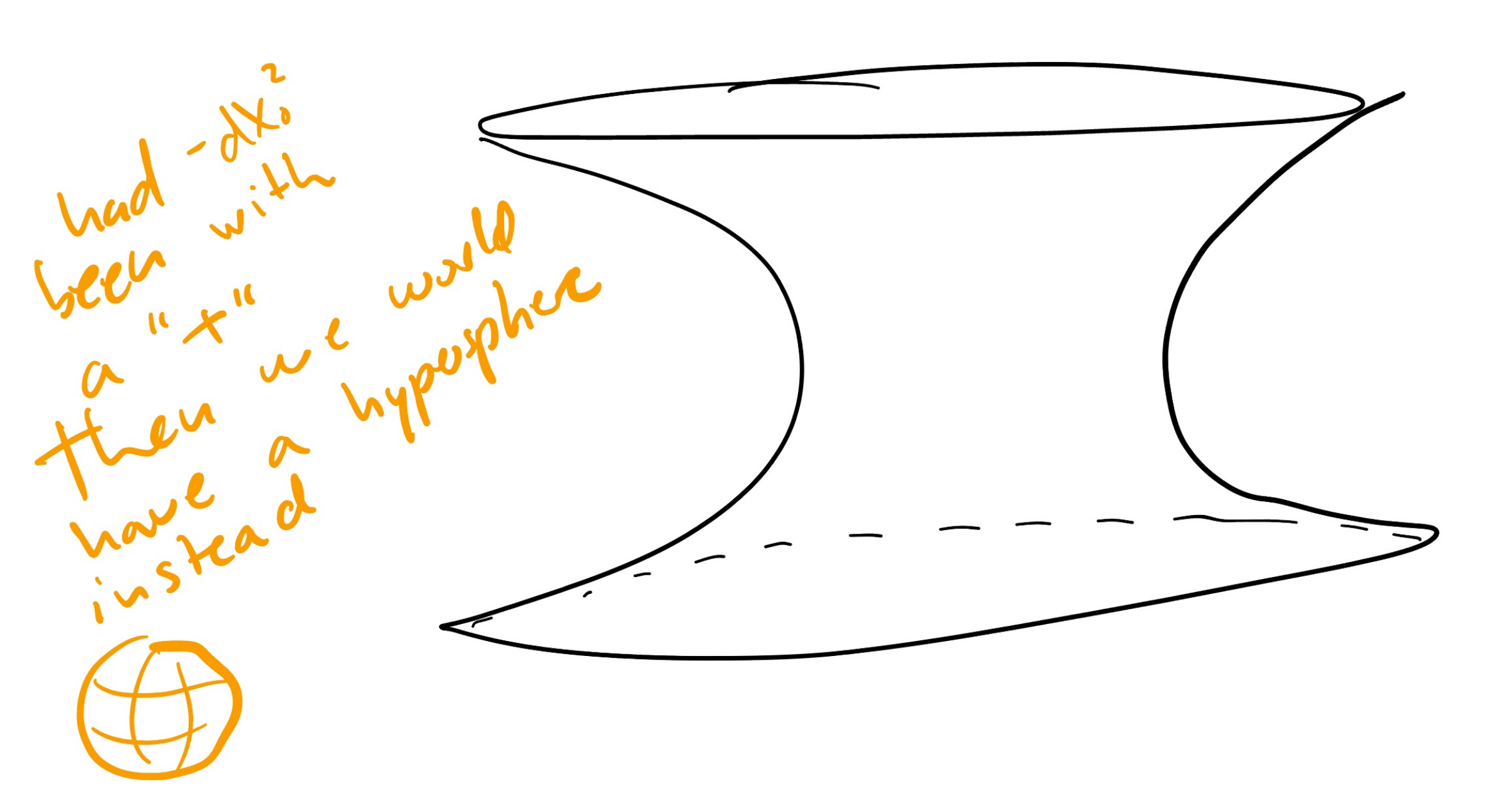}
\end{center}
Note, if we switch the sign on $ dX_{0}^{2}$, then we have Euclidean de Sitter describing a hypersphere instead.

We can now introduce coordinates on the hyperboloid by
\bea
l \sinh (\hat{t} / l)={X}_{0} \quad &&l \cosh (\hat{t} / l) \cos \chi={X}_{1}\\ \nonumber
&& l \cosh (\hat{t} / l) \sin \chi \cos \theta=X_{2} \\ \nonumber
&& l \cosh (\hat{t} / l) \sin \chi \sin \theta \cos \varphi=X_{3} \\ \nonumber
&& l \cosh (\hat{t} / l) \sin \chi \sin \theta \sin \varphi=X_{4}  \\ \nonumber
\eea
in terms of which the metric becomes
\beq
 d s^{2}=-d \hat{t}^{2}+l^{2} \cosh ^{2}(\hat{t} / l)\left[d \chi^{2}+\sin ^{2} \chi\left(d \theta^{2}+\sin ^{2} \theta d \varphi^{2}\right)\right]
\eeq

The singularities at $\chi=0, \pi$ and $\theta=0, \pi$ are just the usual coordinate singularities of spherical coordinates. Apart from that, these global coordinates cover the entire de Sitter hyperboloid
\beq
 -\infty<\hat{t}<\infty, \quad 0 \leq \chi \leq \pi, \quad  0 \leq \theta \leq \pi,\quad  0 \leq \varphi < 2 \pi~.
\eeq

\begin{center}
\includegraphics[max width=6cm]{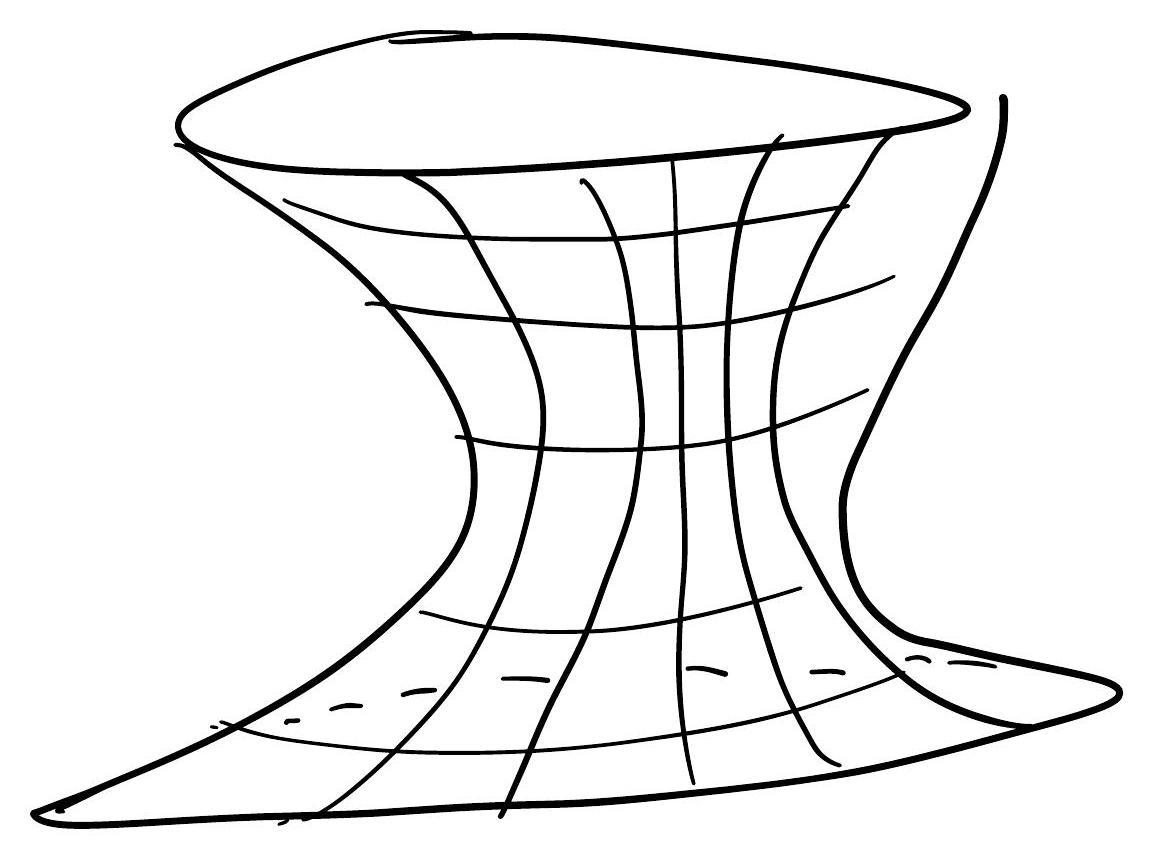}
\end{center}

To obtain the Penrose diagram and understand the causal structure of de Sitter, we can define a new time coordinate
\beq
\cosh (\hat{t}/l)=\frac{1}{\cos \hat{T}}
\eeq
so $-\pi / 2<\hat{T}<\pi / 2$ and the metric becomes
\beq
d s^{2}=\frac{l^{2}}{\cos ^{2}(\hat{T})}\left(-d \hat{T}^{2}+d\chi^2+\sin ^{2}\chi d \Omega^{2}\right)
\eeq

which is conformal to
\beq
d \tilde{s}^{2}=-d \hat{T}^{2}+d\chi^2+\sin ^{2} \chi d \Omega^{2}
\eeq

This is the same metric as the conformal or Penrose coordinates of Minkowski, except in this case, we don't have $-\pi<T-R \leq T+R<\pi$, but instead
\beq
-\frac{\pi}{2}<\hat{T}<\frac{\pi}{2}~, \qquad 0 \leq \chi \leq \pi
\eeq

\begin{center}
\includegraphics[max width=16cm]{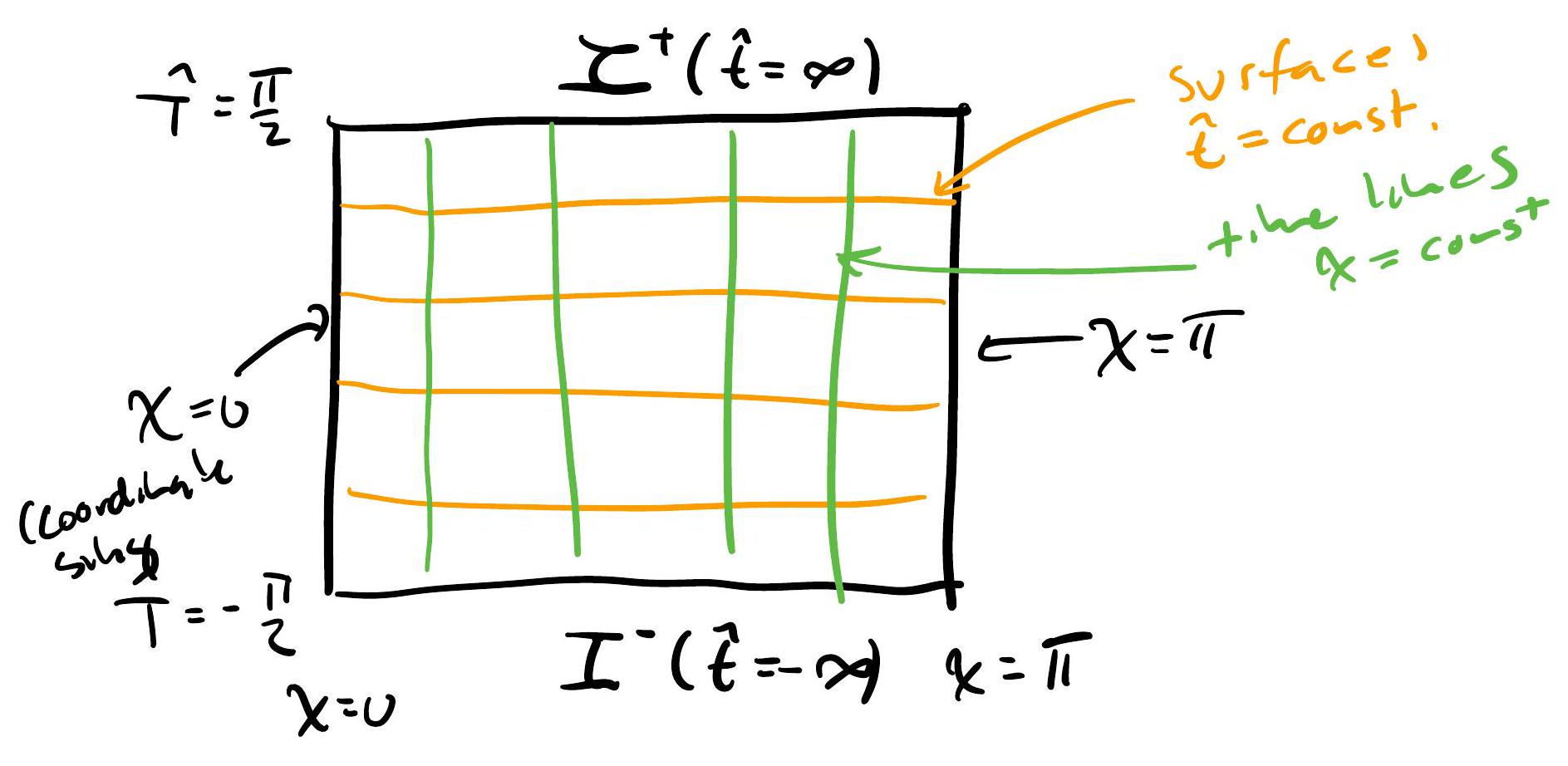}
\end{center}

Now clearly, the FLRW or Planar or Poincar\'e coordinates from before, obtained from the definitions
\beq
t=l \log \frac{X_{0}+X_{1}}{l}, \quad x=\frac{l X_{2}}{X_{0}+X_{1}}, \quad y=\frac{l X_{3}}{X_{0}+X_{1}},\quad  z=\frac{l {X}_{4}}{X_{0}+X_{1}}\,,
\eeq
which implies
\beq 
d s^{2}=-d t^{2}+\exp (2 t / l) d {\bf x}^{2} =-d t^{2}+e^{2 H t} d {\bf x}^{2}\,, \quad H=\frac{1}{l}\,,
\eeq
only covers half of the global de Sitter spacetime (the patch with $X_0+X_1 >0$):
\begin{center}
\includegraphics[max width=\textwidth]{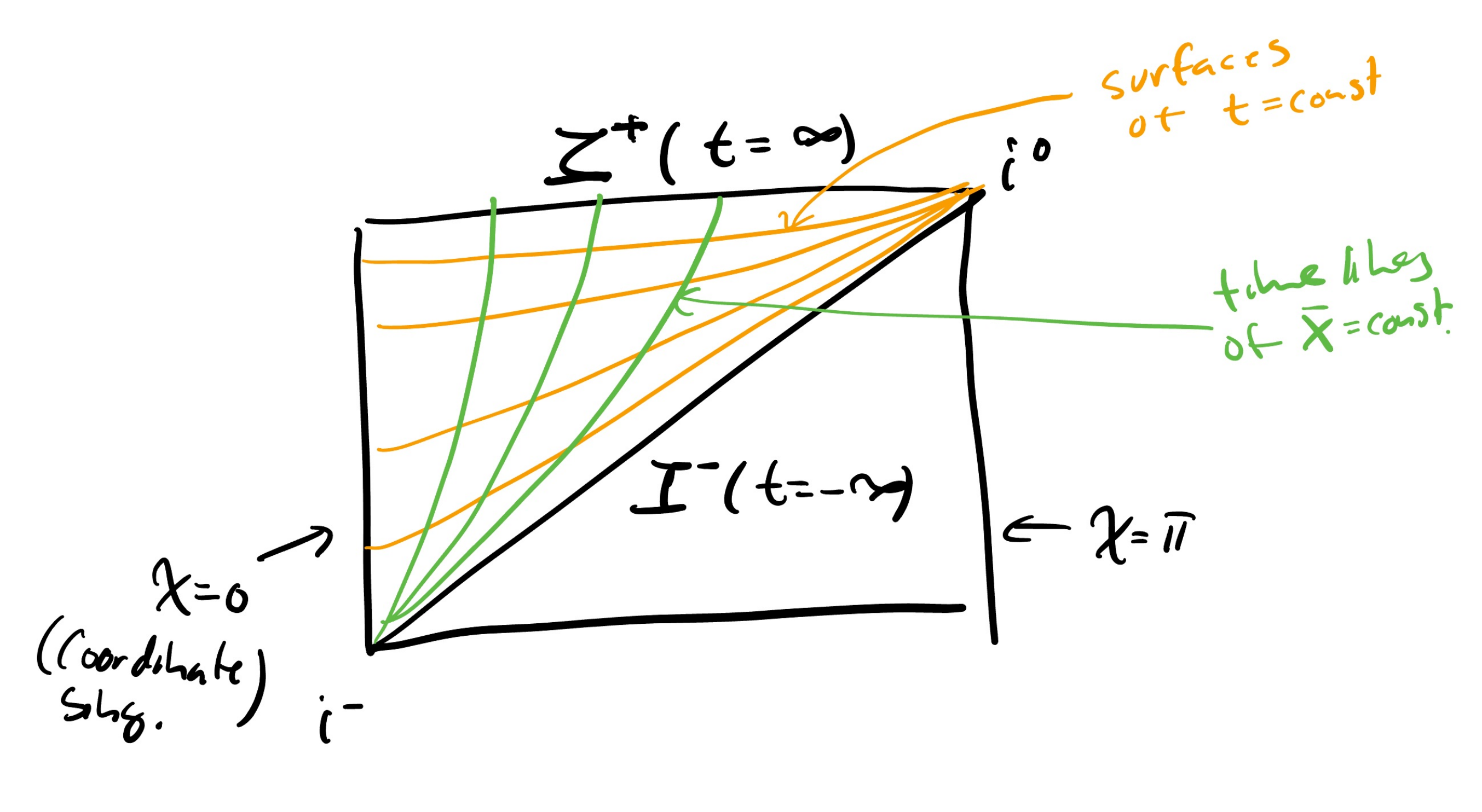}
\end{center}

Now it is clear why de Sitter in FLRW coordinates is geodesically incomplete. Note that sometimes people discuss observer-dependent issues like particle production and disagree because they compare results in the global coordinates versus FLRW coordinates. However, a cosmologically co-moving observer is at rest in FLRW coordinates.

Nevertheless, in FLRW coordinates, we do not avoid cosmological observers experiencing an initial singularity (geodesic incompleteness), and an initial condition is required at the boundary.

Of course, when pasting de Sitter as the lower to FLRW as the upper part to get the inflationary spacetime, we cut de Sitter at some finite $t=0$ and do not include the $t \rightarrow \infty$ part of de Sitter. So, inflation cannot be exactly de Sitter, but must include some ways of ending inflation at $t=0$.

\subsection{Models of Inflation (Microphysics)}

To end inflation, we need some dynamics playing the role of a clock telling us when to go from de Sitter-like inflation to radiation-dominated FLRW.

\subsubsection{Old Inflation}

Pure de Sitter is in global thermal equilibrium and, therefore, quite dead. Nothing happens, and the temperature is constant
\beq
T_{d s}=\frac{H}{2 \pi}\,.
\eeq

However, imagine that there is also some additional radiation with a changing temperature. Note that the radiation may still be in local thermal equilibrium such that the local temperature determines the state of the radiation field at each point of spacetime as if it were in a thermal bath. Then, inspired by our understanding of particle physics, and for example the electroweak phase transition, it is natural to think of some scalar field, with a potential that develops a new "true" vacuum below some critical temperature $T_{c}$.
\begin{center}
\includegraphics[max width=8cm]{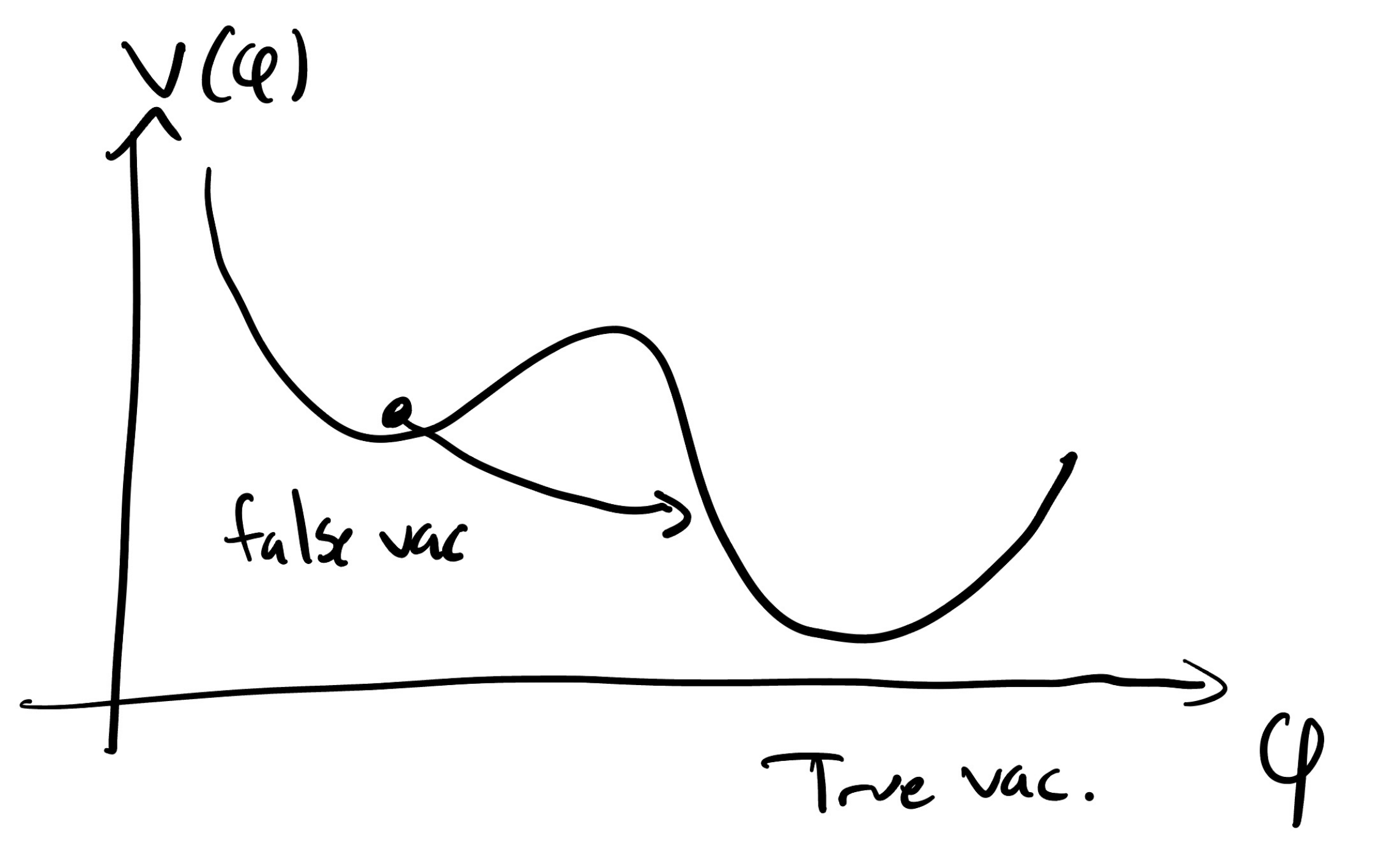}
\end{center}

Initially stuck in the false minimum, the field's potential energy is like a cosmological constant, and if the radiation density is small enough, the universe will be dominated by an approximate cosmological constant. After the temperature dips below $T_{c}$ and the second minimum appears, the field will tunnel, and the false vacuum energy is converted into radiation in a first-order phase transition.
\begin{center}
\includegraphics[max width=8.5cm]{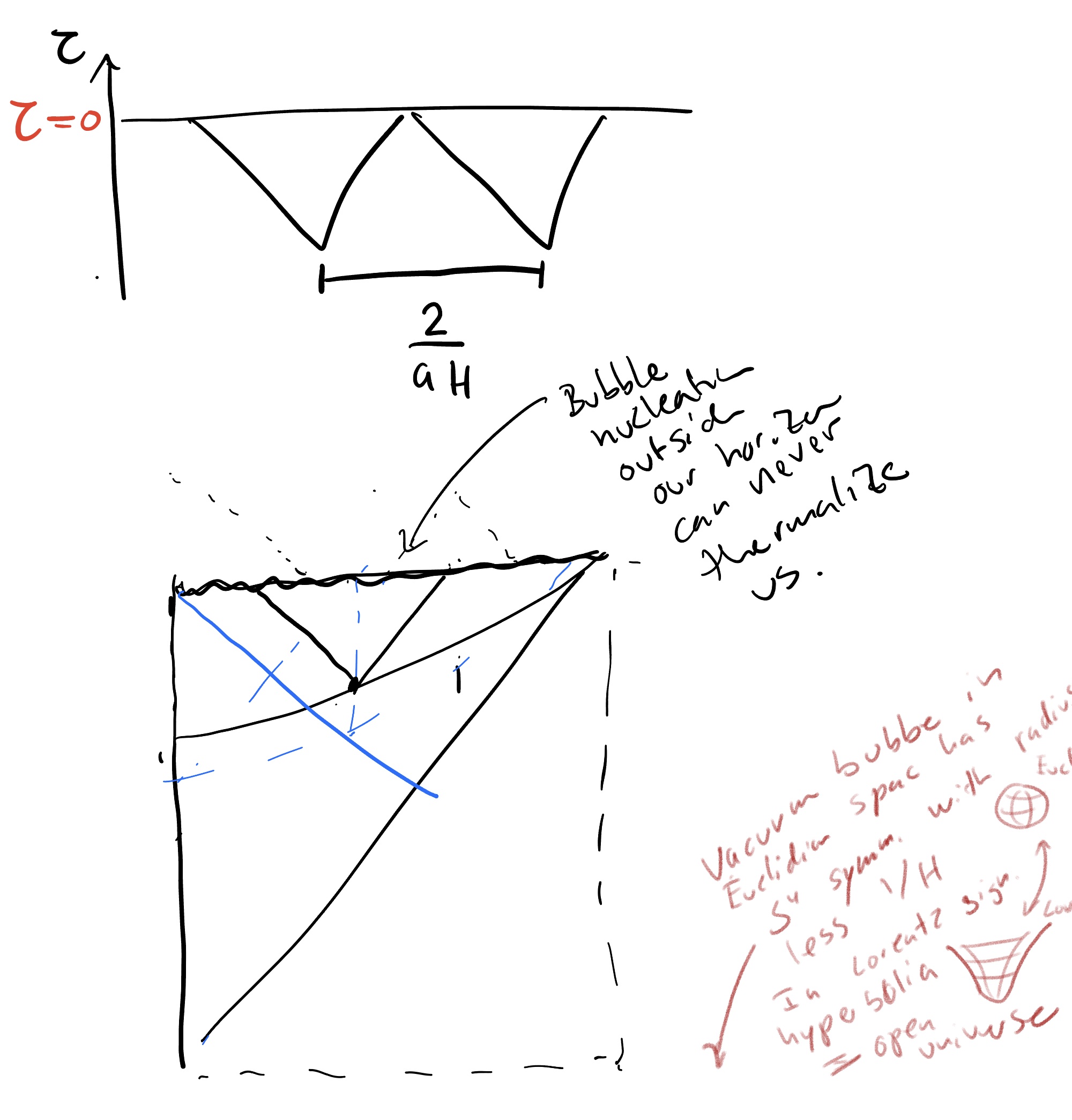}
\end{center}

The phase transition is completed after bubbles of true vacuum nucleate and expand at the speed of light when the bubbles meet and percolate. For that to happen, one needs more than one bubble per event horizon since a bubble can never grow larger than this. However, in that case, the nucleation rate, $\Gamma$, is so high that one never achieves enough inflation. In the other case of one bubble per horizon volume or less, the phase transition never completes, as illustrated in the figure above.

Also, we can not live inside a single bubble, as that would be an open universe with curvature radius = initial size of the bubble, which must be less than $1 / H$ for causality reasons, leading to a curvature-dominated universe. Altogether, this is called the graceful exit problem!

\subsubsection{Slow-roll Inflation}

The idea is to have a slow rollover phase transition instead.

\begin{center}
\includegraphics[max width=8cm]{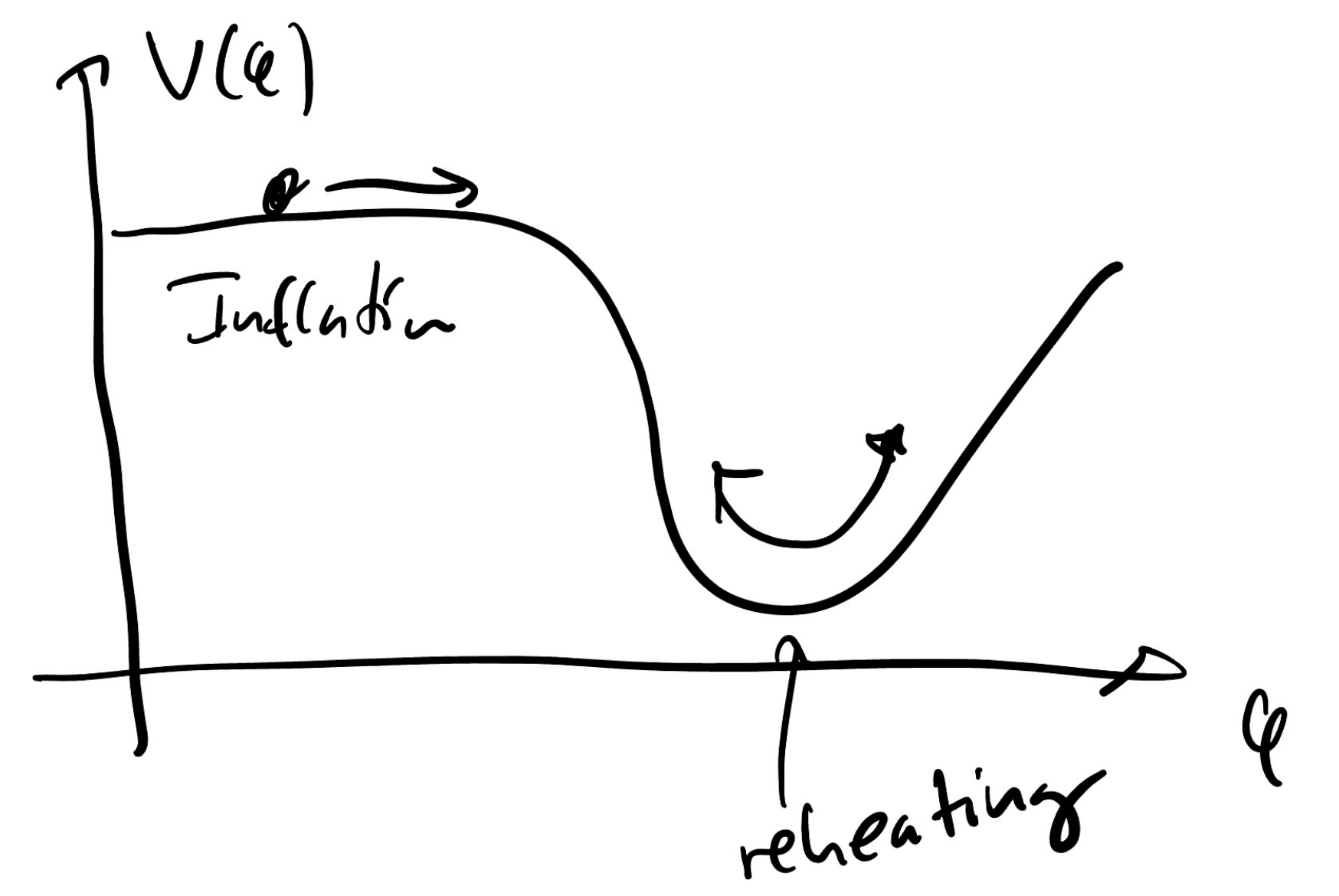}
\end{center}

Assume that in addition to gravity, we have a scalar field called the inflaton, such that the total action is 
\beq
S=S_{\text {grav }}+S_{\phi}
\eeq
with
\beq
S_{\text {grav }}  =\frac{1}{16 \pi G_N} \int d^{4} x \sqrt{-g} R
\eeq
and
\beq
S_{\phi}  =\int d^{4} x \sqrt{-g} \mathcal{L}_{\phi}
 =-\int d^{4} x \sqrt{-g}\left[\frac{1}{2} \partial_{\mu} \phi \partial^{\mu} \phi+V(\phi)\right]
\eeq

The field equation of motion is
\beq
\frac{\delta S}{\delta \phi}=0 \qquad \Rightarrow \qquad \ddot{\phi}+3 H \dot{\phi}-\frac{1}{a^{2}}\nabla^2\phi+V_\phi(\phi)=0
\eeq
where $V_\phi \equiv dV/d\phi$.

From the definition of $T_{\mu \nu}$
\beq
T_{\mu \nu}=-\frac{2}{\sqrt{-g}} \frac{\delta S_{\phi}}{\delta g^{\mu \nu}}
\eeq
and using Jacobi's formula $\delta g =g^{\mu\nu}\delta g_{\mu\nu}$ to get
\bea \delta \sqrt{-g}&=&-\frac{1}{2} \frac{1}{\sqrt{-g}} \delta g=\frac{1}{2} \sqrt{-g} g^{\mu\nu} \delta g_{\mu\nu} \\ \nonumber
 \delta g^{\mu \nu}&=&-g^{\mu \alpha} g^{\nu \beta} \delta g_{\alpha \beta} \\
 \Rightarrow \quad T^{\mu \nu}&=&\partial^{\mu} \phi \partial^{\nu} \phi+g^{\mu \nu} \mathcal{L}_{\phi}
\eea

Lowering indices and assuming the ideal fluid form
\beq
T_{\mu \nu}= \text{diag.}(\rho, a^2p, a^2p, a^2p)\,,
\eeq
one obtains the energy and pressure of the scalar field
\bea
\rho_{\varphi} & =&\frac{1}{2} \dot{\phi}^{2}+\frac{1}{2} \frac{1}{a^{2}}\left(\partial_{i} \phi\right)^{2}+V(\phi) \\
p_{\phi} & =&\frac{1}{2} \dot{\phi}^{2}-\frac{1}{6} \frac{1}{a^{2}}\left(\partial_{i} \phi\right)^{2}-V(\phi)
\eea

Since the gradient terms quickly redshift away compared with the almost constant potential energy during inflation, we can assume that the scalar field is homogeneous and has $\partial_{i}\phi\sim 0$. From the first Friedmann equation, we then find
\beq
H^{2}=\frac{8 \pi G_{N}}{3} \rho=\frac{8 \pi G_{N}}{3}\left[\frac{1}{2} \dot{\phi}^{2}+V(\phi)\right]
\eeq
and for
\beq
V(\varphi) \gg \dot{\varphi}^{2}
\eeq
we have quasi-de Sitter expansion
\beq
p \simeq -\rho
\eeq

The slow-roll approximation is, therefore, to assume $\dot{\phi}$ small and also $\ddot{\phi}$ small to keep $\dot{\phi}$ small long enough to have enough inflation to solve the causality problem. So assuming $|\ddot{\phi}|\ll |3 H \dot{\phi}|$, we have 
\beq
H^{2}=\frac{8 \pi G_N}{3} V(\phi),~ \qquad 3 H \dot{\phi}=-V_\phi
\eeq
from the first Friedmann equation and the equation of motion. These are called the slow-roll equations.

To keep control of the slow-roll approximation, we introduce the slow-roll parameters
\beq
 \epsilon=4 \pi G_{N} \frac{\dot{\varphi}^{2}}{H^{2}}~, \qquad \eta=\frac{1}{8 \pi G_{N}}\left(\frac{V_{\phi\phi}}{V}\right)
\eeq

where $\eta-\epsilon=-\frac{\ddot{\phi}}{H \dot{\phi}}$ to leading order in slow-roll. The slow-roll approximation is then equivalent to requiring
\beq
\epsilon \ll 1, \quad|\eta| \ll 1
\eeq
In particular, for a canonical homogenous scalar field, $\epsilon = -\dot{H}/H^2=4\pi G_N (\dot\phi^2/H^2)$, and  inflation, $\ddot{a}>0$, is equivalent to $\epsilon<1$.

The constraint on the number of e-folds for solving the causality and flatness problem becomes a constraint on the potential
\beq
N  \equiv \ln \left(\frac{a\left(t_{f}\right)}{a\left(t_{i}\right)}\right)=\int_{t_{i}}^{t_{f}} H d t \simeq 8 \pi G_{N} \int_{\phi_{f}}^{\phi_{i}} \frac{V}{V_{\phi}} d \phi \gtrsim 60
\eeq
where we used $dt = (1/\dot\phi)d\phi$ and $\dot\phi = -V_\phi/(3H)$.

While the details of inflationary model building are a huge topic, we will mention four classes of inflation models

\begin{enumerate}
  \item Large-field models

  \item small-field models

  \item Hybrid models

  \item Curvaton models

\end{enumerate}

If we first focus on single-field models, a single-field inflation potential can be described by two energy scales, the height of the potential, $\Lambda$, and its width, $\mu$, related to the field excursion of the inflaton field $\Delta \phi$ during inflation
\beq
V(\phi)=\Lambda^{4} f\left(\frac{\phi}{\mu}\right)
\eeq

\subsubsection{Large-field models}

In large-field models, $\Delta \phi \gg M_{p}$, so one of the energy scales describing the potential is Planckian. This can be achieved if the inflaton field starts high up in a monomial-type potential
\beq\label{monomial}
V(\phi)=\Lambda^{4}\left(\frac{\phi}{\mu}\right)^{n}~, \qquad n \geq 1
\eeq
as illustrated for a quadratic potential in the figure below. Naively, it looks like the potential is steep for large field values and that the field will roll down fast. However, for large field values, the relative slope $V_\phi/V=n/\phi$ is small, and for $\phi >> M_p$ the potential is shallow in Planck units. Also, $H$ is large, and therefore, the friction term $H\dot\phi$ in the slow-roll equation is large, which also helps in preventing the field from rolling down fast until the field reaches a smaller-than-Planckian field value, starts to oscillate around the minimum, and reheats the universe.
\begin{center}
\includegraphics[max width=8cm]{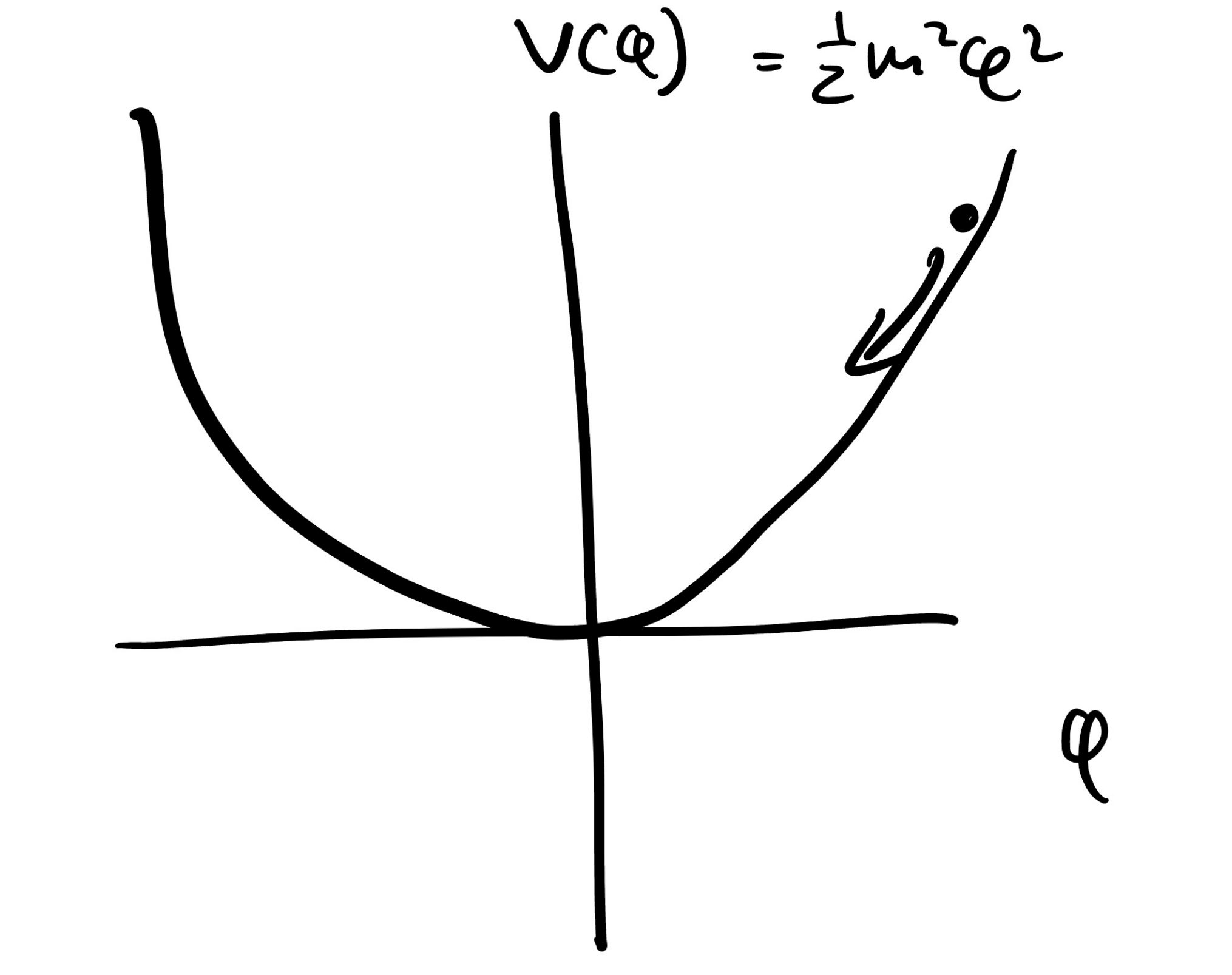}
\end{center}

For the potential in eq.(\ref{monomial}), the slow-roll parameters become
\beq
\epsilon  =\frac{n^{2} M_{p}^{2}}{2 \phi^{2}}~,\qquad \eta=\frac{n(n-1) M_{p}^{2}}{\phi^{2}}~.
\eeq
We can rewrite them in terms of the number of e-folds
\beq
N  =\int_{\phi_f}^{\phi} \frac{V}{V'} \frac{d \phi}{M_{p}^2}=\frac{1}{n M_{p}^2} \int_{\phi_f}^{\phi} \phi d \phi=\frac{\phi^{2}-\phi_f^2}{2 n M_{p}^{2}}\approx \frac{\phi^{2}}{2 n M_{p}^{2}}
\eeq
which implies
\beq
\epsilon=\frac{n}{4 N}~,\qquad \eta=\frac{n-1}{2 N}
\eeq
or $\epsilon \sim \eta \sim 1/N$. We now see that during slow-roll inflation, when $\epsilon \ll 1 $, we must have $\phi \gg M_{p}$.

As long as $\rho \simeq m^{2} \phi^{2} \ll M_{p}^{4}$ $(n=2)$ we don't need quantum gravity, and the model is safe as a Quantum Field Theory (QFT) on a background spacetime.  So this just requires $\phi \ll M_{p}^{2}/m$, which can be larger than $M_{p}$ if $m \ll M_{p}$.

However, some symmetry does need to protect the potential from getting corrected by an infinite tower of higher-dimensional operators of the form induced by graviton loops
\beq
\mathcal{L}\supset \frac{\varphi^{n}}{M_{p}^{n-4}}
\eeq
that becomes in important for $\phi \gg M_{p}$. One way is to use spontaneously broken approximate shift symmetry and let the inflaton be the associated pseudo-Nambu-Goldstone boson (pNGB).

Prime examples of large-field models of inflation are chaotic inflation \cite{Linde:1983gd} and axion monodromy inflation \cite{Silverstein:2008sg,McAllister:2008hb,Kaloper:2008fb}.

\subsubsection{small-field models}

Here, the potential is inverted, so we start at a small-field value and roll away, as in the figure below.
\begin{center}
\includegraphics[max width=8cm]{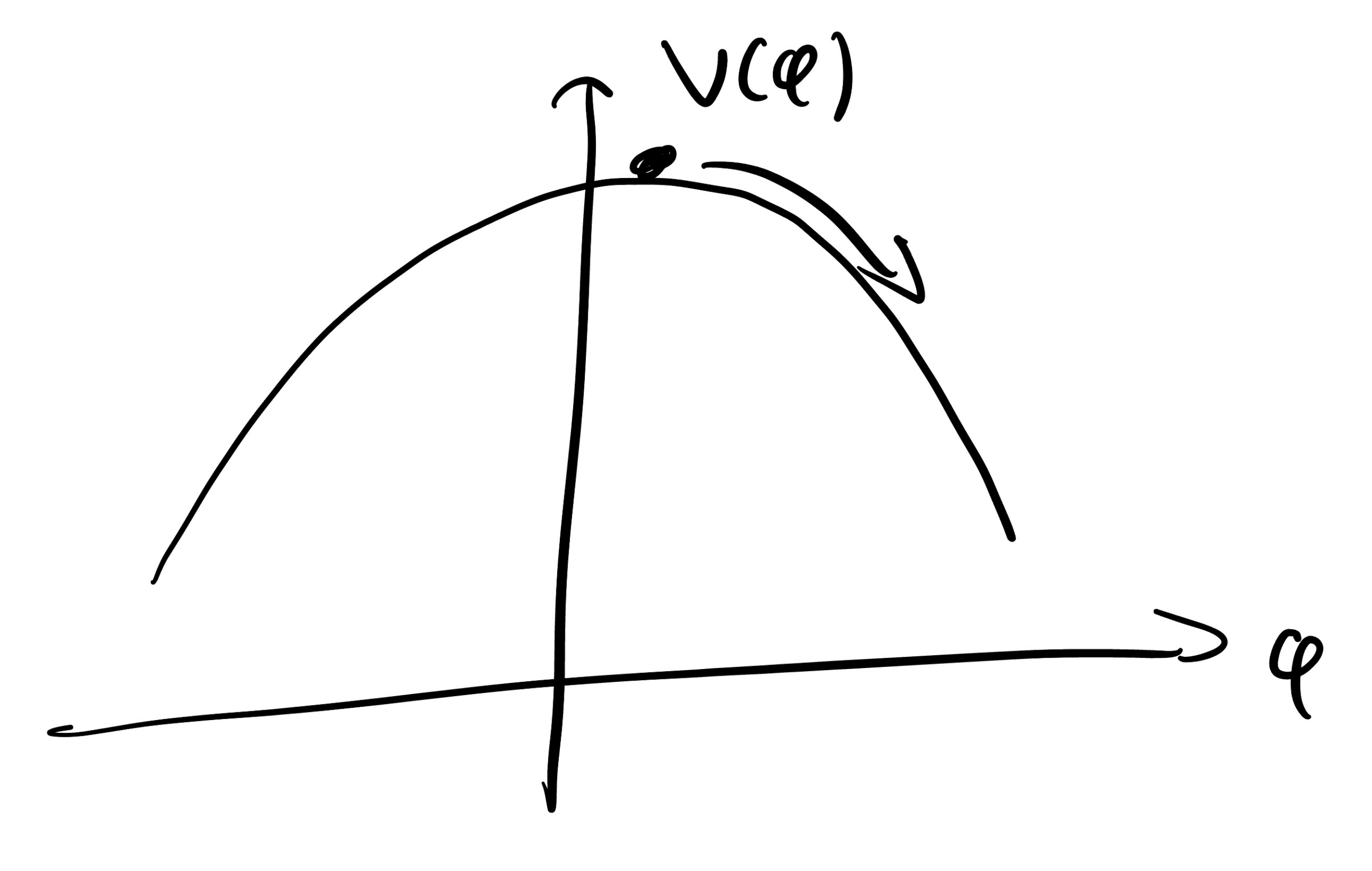}
\end{center}

One can think of this as effectively a tachyonic mass $
V_{\phi \phi}<0$ implying $\eta<0$. This type of potential is inspired by spontaneous symmetry breaking in particle physics. 
\begin{center}
\includegraphics[max width=6cm]{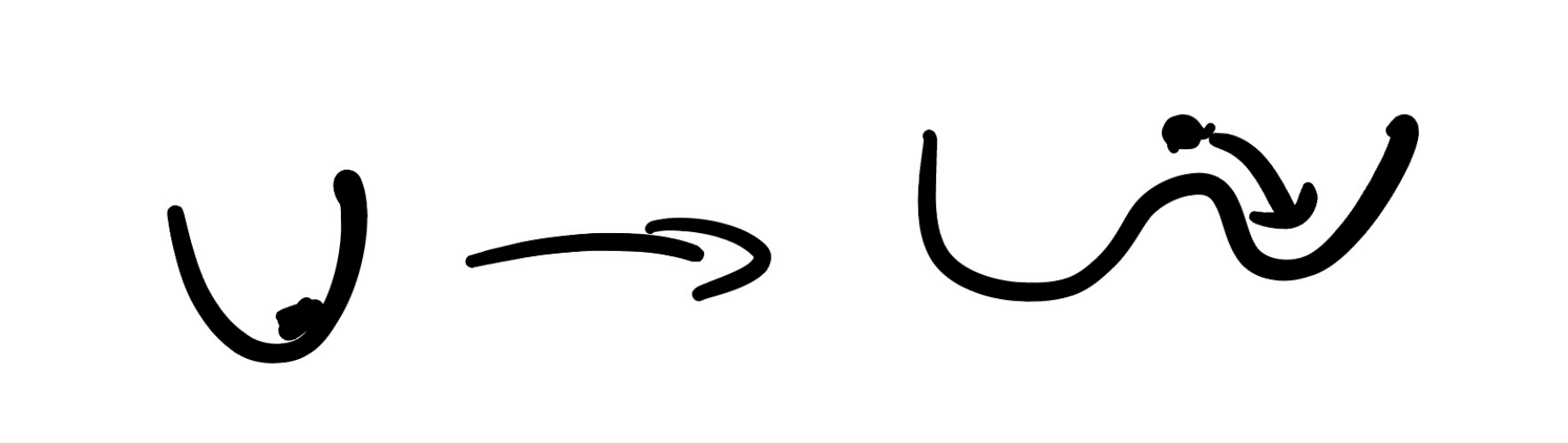}
\end{center}

If one Taylor expand the function $f\left(\frac{\phi}{\mu}\right)$ for $\phi \ll \mu$ one get a potential of the form
\beq
V(\phi)=\Lambda^{4}\left[1-\left(\frac{\phi}{\mu}\right)^{n}+\ldots\right], \quad n \geq 2~.
\eeq
Obviously, this breaks down when $\phi \sim \mu$, which is where inflation ends, $\phi_{f} \sim \mu$. Since $\phi$ is small and to leading order $V(\phi) \approx \Lambda^{4}$, one typically has $\epsilon \approx 0 $.

Examples of small-field models are new inflation \cite{Linde:1981mu,Albrecht:1982wi} and hilltop inflation \cite{Boubekeur:2005zm}. 

\subsubsection{Hybrid models}

Hybrid inflation has been much discussed in the context of supersymmetry and supergravity. It is effectively a single-field model, but where the end
of inflation is triggered by a second field, the waterfall field, with a typical potential of the form \cite{Linde:1993cn}
\beq
V(\phi, \psi)=\frac{1}{2} m^{2} \phi^{2}+\frac{1}{2} g^2\phi^{2} \psi^{2}+\frac{1}{4} \lambda\left(M^{2}-\psi^{2}\right)^{2}
\eeq
\begin{center}
\includegraphics[max width=8cm]{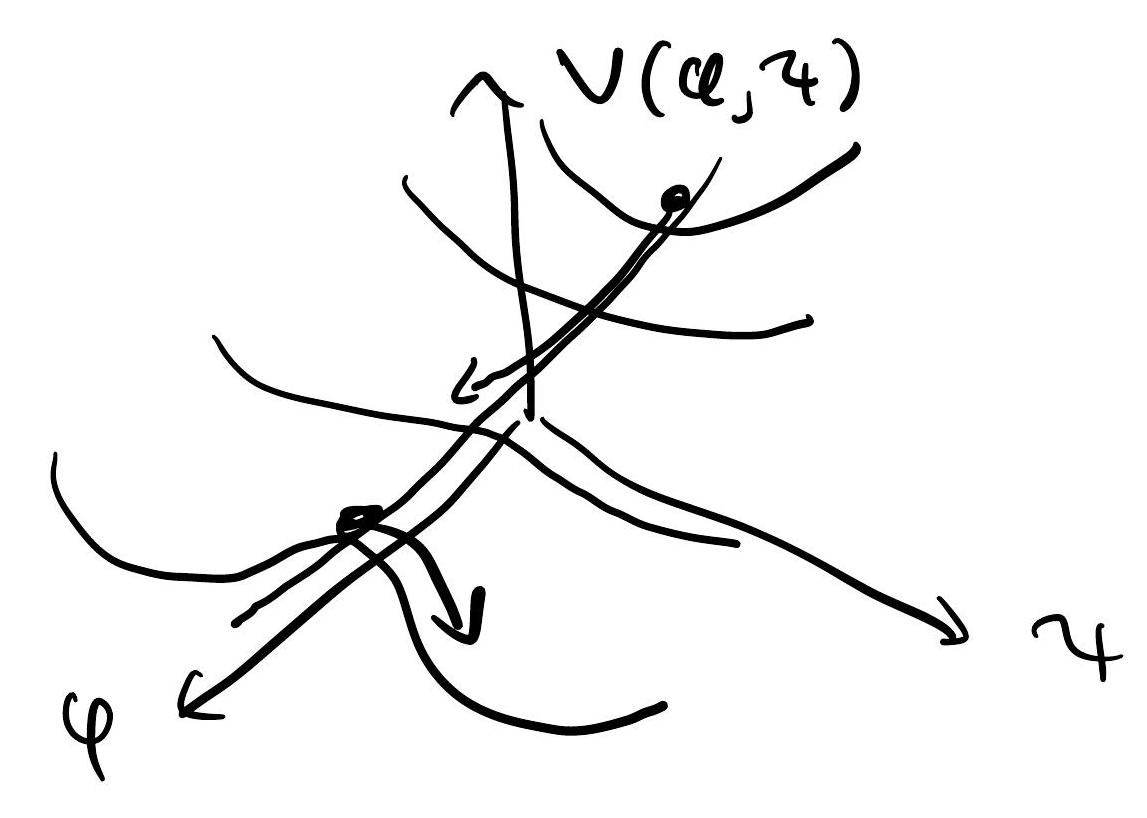}
\end{center}
One could also consider variants where the phase transition at the end is first order.

Along the inflationary valley $\psi=0$, one gets effectively
\beq
V(\phi)=\Lambda^{4}\left[1+\left(\frac{\phi}{\mu}\right)^{n}+\ldots\right], \quad n \geq 2~,
\eeq
so typically one has $\eta=2(M_{p}/{\mu})^{2}>0$ $(n=2)$, and the slow-roll conditions then implies $\mu>M_{p}$ and $\epsilon=(\phi/\mu)^{2} \eta \ll \eta$.

\subsubsection{Curvaton model}

The models above are strongly constrained by observations, since fluctuations in the inflaton field itself generate CMB perturbations in those models. Since CMB perturbations are close to scale-invariant, this implies that the inflaton potential also needs to be sufficiently flat, i.e., close to de Sitter with $\epsilon, \eta \ll 1$. This is what f.ex. rules out the Pre-big bang scenario \cite{Gasperini:1992em} and the old Ekpyrotic scenario \cite{Khoury:2001wf} in the absence of the curvaton.

The curvaton is another light field that remains frozen during inflation, and because of being almost massless during inflation, it acquires a close to flat spectrum. After inflation, the curvaton becomes dynamically important and decays into radiation, so all the CMB perturbations are created by the curvaton instead of by the inflaton. Although the inflaton model is set free in the curvaton model, we will, for simplicity, assume that the potential of the inflaton, $\phi$, and the curvaton, $\sigma$, is
\beq
V(\phi, \sigma) = \frac{1}{2}m_\phi^2 \phi^2 +\frac{1}{2}m_\sigma^2 \sigma^2 ~.
\eeq 
Often the curvaton, $\sigma$, is assumed to be an axion-like particle, as it has to be relatively light, $m_\sigma \ll H$, during inflation. 

In the curvaton scenario, the inflaton potential is therefore allowed to be much steeper. Also, pre-big bang and the new Ekpyrotic scenario require a curvaton to be compatible with observations.

In fact, the curvaton mechanism was first introduced in 2001 by Enqvist and Sloth \cite{Enqvist:2001zp}, in the context of explaining how the pre-big bang scenario could generate a nearly flat spectrum of adiabatic CMB perturbations, using one of the axions of string theory in the role of what shortly thereafter would be called the curvaton by Lyth and Wands \cite{Lyth:2001nq}. Soon after that, Moroi and Takahashi discussed a closely related realization in which the role of the curvaton is played by a cosmological modulus field \cite{Moroi:2001ct}.

The same is true for the Ekpyrotic scenario. It also requires the curvaton mechanism to be compatible with observations. The new Ekpyrotic scenario \cite{Buchbinder:2007ad}, introduced later, was therefore also called a "ghost condensate theory with the curvaton heart" by Andrei Linde \cite{linde2007inflationary}.

\newpage

\section{Lecture 2: Linear perturbation theory}

As hinted at in the previous section discussing the curvaton, perturbations are important for understanding and constraining models of inflation.

In flat space, the only propagating modes of gravity are the two polarizations of the graviton. The equation of motion of the graviton in flat space is obtained in linearized gravity by writing
\beq
g_{\mu \nu}=\eta_{\mu \nu}+h_{\mu \nu}, \quad, \qquad \left|h_{\mu \nu}\right| \ll 1\,,
\eeq
and only working to linear order in $h_{\mu\nu}$. The perturbation $h_{\mu\nu}$ is a symmetric $4\times 4$ tensor and therefore has 10 components. However, not all of these components describe propagating degrees of freedom. In the ADM decomposition, the lapse and shift, corresponding at the linearized level to $h_{00}$ and $h_{0i}$, are non-dynamical: they enter the action without independent time derivatives and act as Lagrange multipliers. Their equations of motion impose four constraints, namely one Hamiltonian constraint and three momentum constraints. The contracted Bianchi identities,
\beq
\nabla^\mu G_{\mu\nu}=0,
\eeq
ensure that these constraints are preserved under time evolution, reflecting the underlying diffeomorphism invariance of the theory. Thus, after accounting for the four constraint equations, one is left with $10-4=6$ dynamical components before gauge fixing.

Now consider an infinitesimal coordinate transformation
\beq
x^{\mu} \rightarrow x^{\prime \mu}=x^{\mu}+\xi^{\mu}(x)
\eeq
such that derivatives of $\xi$ are no larger than $h_{\mu\nu}$. Using that the metric transforms like a tensor
\beq
g_{\mu \nu}(x) \rightarrow g_{\mu \nu}^{\prime}\left(x^{\prime}\right)=\frac{\partial x^{\alpha}}{\partial x^{\prime \mu}} \frac{\partial x^{\beta}}{\partial x^{\prime \nu}} g_{\alpha \beta}(x)
\eeq
we have
\beq
h_{\mu \nu}(x) \rightarrow h_{\mu \nu}^{\prime}\left(x^{\prime}\right)=h_{\mu \nu}(x)-\left(\partial_{\mu} \xi_{\nu}+\partial_{\nu} \xi_{\mu}\right)
\eeq
and the smallness of $h_{\mu v}$ is preserved, so therefore these types of infinitesimal coordinate transformations are symmetries of the linearized theory.

This means that we can use this gauge redundancy to gauge away (gauge fix) 4 more d.o.f., leaving $(10-4)-4=6-4=2$ dynamical and physical d.o.f.

This allows us to go to the traceless and transverse gauge
\beq
ds^{2}=-d t^{2}+\left(\delta_{i j}+h_{i j}\right) d x^{i} d x^{j}
\eeq
where
\beq
h_{i}^{i}=0\,, ~\text { and } ~\partial^{j} h_{i j}=0\,.
\eeq

de Sitter spacetime is also a vacuum solution with no new d.o.f., as the cosmological constant cannot fluctuate. Therefore, the graviton in de Sitter again has two degrees of freedom, and we can write the linearized gravitational fluctuations of de Sitter spacetime as
\beq
ds^{2}=-d t^2+a^{2}(t)\left(\delta_{i j}+h_{i j}\right) d x^{i} d x^{j}
\eeq
again with $h_{i}^{i}=0, \partial^{j} h_{i j}=0$, and $a \propto e^{H t}$ with $H=$ cont.

In slow-roll inflation, we are no longer in vacuum, and while at the background level, $H$ is no longer constant, also at the level of linearized perturbations, the inflaton fluctuations carry an extra scalar degree of freedom
\beq
\phi(t, x)=\phi(t)+\delta \phi(t, {\bf x})~,
\eeq
where $\phi(t)$ is the homogeneous background inflaton field satisfying the slow-roll equations, and $\delta \phi(t, {\bf x})$ is the linear perturbation.

The gauge where $\delta \phi \neq 0$ and the linearized metric takes the form
\beq
d s^{2}=-d t^2+a^{2}(t)\left(\delta_{i j}+\gamma_{i j}\right) d x^{i} d x^{j}
\eeq
where $\gamma_{ij}$ is transverse and traceless,
\beq
\gamma^i{}i=0,\qquad \partial^i\gamma_{ij}=0\,,
\eeq
is called the flat gauge because there is no scalar curvature perturbation in this gauge.

However, with a time reparametrization
\beq
t \rightarrow \tilde{t}=t+\delta t
\eeq
then
\bea
\phi(\tilde{t})&=&\phi(t)+\dot{\phi}(t) \delta t
\nonumber\\
&\Rightarrow &\phi\left(\tilde t, x\right)=\phi(t)+\dot{\phi}(t) \delta t+\delta \phi(t, {\bf x})
\eea
So clearly, if we choose
\beq
\delta t=-\frac{\delta \phi}{\dot\phi}
\eeq
then at linear order, we have
\beq
\phi\left(\tilde{t}, {\bf x}\right)=\phi_c(t)
\eeq
and so the field is homogeneous. This is called the comoving gauge because in this gauge, the time slices are slices of constant $\phi$ and therefore comoving with $\phi$. In this gauge, it is clear that $\phi$ is acting as a clock.

In the comoving gauge, the scalar fluctuation appears in the metric as fluctuations of the scale factor
\beq
a(t) \rightarrow a(\tilde{t})=a(t)+\dot{a}(t) \delta t\,.
\eeq
Defining the scalar curvature perturbation

\beq
\zeta = \frac{\dot{a}}{a} \delta t=H \delta t \quad\left(=-\frac{H}{\dot\phi} \delta \phi\right)\,,
\eeq
the perturbed metric becomes
\beq
ds^{2}=-dt^{2}+a^{2}\left(\delta_{ij}(1+2\zeta)+\gamma_{ij}\right)dx^{i}dx^j
\eeq
to leading order in perturbations. At linear order, this is equivalent to writing
\beq
ds^{2}=-dt^{2}+a^{2}\left(e^{2\zeta}\delta_{ij}+\gamma_{ij}\right)dx^{i}dx^{j}
\eeq
which turns out to be a more convenient definition of $ \zeta$ at higher orders in perturbation theory\footnote{Similarly, at the non-linear level, it is also sometimes convenient to redefine the three-dimensional metric as $h_{ij} = [e^\gamma]_{ij}$ \cite{Seery:2008ax,Giddings:2010nc,Giddings:2011zd,Giddings:2011ze}.}.

During inflation $a(t)$ has the approximative form
\beq
a \propto e^{H t} \rightarrow a(\tilde{t})\propto e^{H(t+\delta t)}=e^{H t+\zeta}
\eeq

So, during inflation, if a long wavelength mode of $\zeta$ is constant, we can just view it as a shift in the normalization of the scale factor on small scales. This observation is going to be important later because $\zeta$, in fact, is conserved and constant on super-horizon scales in single-field attractor inflation.

The fact that $\zeta$ is conserved on super-horizon scales is the main motivation for working with this variable. Remember, the largest scales we observe today are the ones that exited the horizon earliest during inflation in an unfair "first out and last in." This means that the perturbations corresponding to the largest length scales observed today are insensitive to most of the universe's evolution between the early phase of inflation and today. This is good news for making predictions of inflation.

To see why they are conserved, let's derive their e.o.m. from their action. To find the action, we start by perturbing
\beq
S=S_{\text {grav}}+S_{G}=\frac{1}{2} \int d^{4}x \sqrt{-g}\left[R-\partial_{\mu} \phi\partial^{\mu} \phi-2 V(\phi)\right]
\eeq
 
The simplest way to proceed is to compare with the ADM formalism and write the metric on the ADM form
\beq
ds^{2}=-N^{2} d t^{2}+h_{i j}\left(d x^{i}+N^{i} d t\right)\left(d x^{j}+N^{j} d t\right)~.
\eeq
With the ADM ansatz for the metric, the action becomes
\beq
S=\frac{1}{2} \int d^{4}x \sqrt{h}\left[N R^{(3)}-2 N V+N^{-1}\left(E_{i j} E^{i j}-E^{2}\right)+N^{-1}\left(\dot{\phi}-N^{i} \partial_{i} \phi\right)^2-N h^{i j} \partial_{i} \phi \partial_j \phi\right] \\
\eeq
where
\bea
E_{i j}&=&\frac{1}{2}\left(\dot{h}_{i j}-\nabla_{i} N_{j}-\nabla_{j} N_{i}\right) \\
E&=&E_{i}^{i}
\eea
and related to the extrinsic curvature by $K_{i j}=N^{-1} E_{i j}$, is, up to boundary terms, defined as
\beq
\sqrt{-g} R=\sqrt{h} N\left(R^{(3)}+K_{i j} K^{ij}-K^{2}\right).
\eeq

When using the ADM formalism, one needs to be a bit careful, as it is not fully covariant, but only explicitly invariant under spatial coordinate transformations, and the invariance under the kind of time reparametrizations we did above to change gauge from flat to comoving gauge is enforced by thinking of $N$ and $N^{i}$ as Lagrange multipliers whose equations of motion become constraint equations enforcing the invariance. These equations are for $N^{i}$
\beq
\nabla_{i}\left[N^{-1}\left(E^{i}_j-\delta_{j}^{i} E\right)\right]=N^{-1}\left(\dot\phi-N^i\partial_i\phi\right)\partial_j\phi\,,
\eeq
and for $N$
\beq
R^{(3)}-2 V-N^{-2}\left(E_{i j} E^{i i}-E^{2}\right)-N^{-2} \left(\dot\phi-N^i\partial_i\phi\right)^2-h^{ij}\partial_i\phi\partial_j\phi=0
\eeq
which are also referred to as the \underline{momentum and Hamiltonian constraints}.

For the physical variables $h_{ij}$ and $\phi$, the two gauges discussed above are

\begin{enumerate}
  \item \underline{Comoving gauge}
\end{enumerate}

\beq
\begin{array}{ll}
\delta \phi=0, & h_{i j}=a^{2}\left[e^{2\zeta} \delta_{i j}+\gamma_{i j}\right], \\
\gamma_{i i}=0, & \partial_{i} \gamma_{i j}=0
\end{array}
\eeq

\begin{enumerate}
  \setcounter{enumi}{1}
  \item \underline{Flat gauge}
\end{enumerate}

\beq
\begin{array}{ll}
\delta \phi \neq 0 & h_{i j}=a^{2}\left[\delta_{i j}+\gamma_{i j}\right] \\
 \gamma_{i i}=0, & \partial_{i} \gamma_{i j}=0
\end{array}
\eeq

Now, solving the constraint equations using the comoving gauge yields to first order
\beq
N=1+\frac{1}{H} \dot{\zeta} \quad, \quad N_{i}=\partial_{i}\left(-\frac{\zeta}{H}+\frac{\dot{\phi}^{2}}{2 H^2} \partial^{-2} \dot{\zeta}\right)~.
\eeq
To get the quadratic action for $\zeta$ we only need $N$ and  $N_{i}$ to first order. Inserting it into the action gives after integration by parts and using the background equation of notion for $\phi$
\beq
S_{\zeta}=\frac{1}{2} \int dt d^{3}x \frac{\dot{\phi}^{2}}{H^{2}}\left[a^{3} \dot{\zeta}^{2}-a(\partial \zeta)^{2}\right]~.
\eeq
Note that no slow-roll approximation has yet been made\footnote{To compare with Mukhanov, Feldman, and Brandenberger use $v=-2\zeta$ \cite{Mukhanov:1990me}.}.

Also note that the action is proportional to $\epsilon=\dot{\phi}^{2}/(2H^{2})$, which means the at $\zeta$ becomes unphysical in pure de Sitter when $\epsilon \rightarrow 0$. The suppression by $\epsilon$ is only apparent after integration by parts. For a more systematic understanding of the slow-roll hierarchy, see \cite{Jarnhus:2007ia}.

The equation of motion for $\zeta$ is just the Euler-Lagrange equation obtained from
\bea
&&~~~ \delta S=0\\
&\Rightarrow & \quad \frac{\partial L}{\partial \zeta}-\partial_{t} \frac{\partial L}{\partial \dot{\zeta}}-\partial_{i} \frac{\partial L}{\partial \partial_{i} \zeta} =0 \\
&\Rightarrow & -\partial_{t}\left(a^{3} \frac{\dot{\phi}^{2}}{H^{2}} \dot{\zeta}\right)+a \frac{\dot{\phi}^{2}}{H^{2}} \partial^{2} \zeta  =0
\eea

Now let's analyze this equation in Fourier space
\beq
\zeta(t, \vect{x})=\int \frac{d^{3} x}{(2 \pi)^{3}} \zeta_{k}(t) e^{i {\bf k} \cdot {\bf x}}~.
\eeq
So in Fourier space $\partial^{2} \rightarrow-k^{2}$ and the equation of motion becomes 
\beq
-\partial_{t}\left(a^{3} \frac{\dot{\phi}^{2}}{H^{2}} \dot{\zeta}_{k}\right)-a \frac{\dot{\phi}^{2}}{H^{2}} k^{2} \zeta_{k}=0~.
\eeq
We see that on super-horizon scales $k<<aH$ obtained by the limit $k \rightarrow 0$ in the equation above
\beq
\p_t\left(\frac{1}{2}a^3\ep \dot\zeta_k\right)\to 0~,\qquad \textrm{for}\qquad  k\to 0
\eeq
thus, we either have $\dot\zeta_k \approx 0$ or $a^3 \ep\zeta \propto \textrm{constant}$ on super-horizon scales, which implies that the super-horizon solution must take the form
\beq
\zeta_k \approx \textrm{constant}+\textrm{fast~decaying~term.}
\eeq
The assumption that $\zeta_k$ behaves this way is also called the single-field attractor inflation assumption.

The most elegant way to analyze this is to redefine the field
\beq
\chi=a \frac{\dot{\phi}}{H} \zeta=a \frac{\phi^{\prime}}{\mathcal{H}} \zeta \equiv z \zeta
\eeq
where prime denotes derivative with respect to conformal time, $\tau$, and $\mathcal{H}=a^{\prime}/a$. We then have
\beq
\dot{\zeta}=\frac{1}{a} \zeta^{\prime}=\frac{1}{a}\left(\frac{1}{z} \chi\right)^{\prime}=\frac{1}{a}\left(\frac{\chi^{\prime}}{z}-\frac{z^{\prime}}{z} \frac{\chi}{z}\right)\, ,
\eeq
which we can rewrite as
\beq
a^{2} z^{2} \dot{\zeta}^{2}=\chi^{\prime 2}-\frac{z^{\prime}}{z}\left(\chi^{2}\right)^{\prime}+\left(\frac{z^{\prime}}{z}\right)^{2} \chi^{2}\, .
\eeq
After an integration by parts, we get
\bea
a^{2} z^{2} \dot{\zeta}^{2}&=& \chi^{\prime 2}+\left(\frac{z^{\prime \prime}}{z}-\frac{z^{\prime 2}}{z^{2}}\right) \chi^{2}+\left(\frac{z^{\prime}}{z}\right)^{2} \chi^{2}+\text { total derivative} \\
&=&\chi^{\prime 2}+\frac{z^{\prime \prime}}{z} \chi^{2}+\text { total derivative}
\eea

So in terms of $\chi$, after integration by parts and neglecting a total derivative, the action becomes that of a minimally coupled scalar field with time-dependent mass, $m^{2}(\tau)=-\frac{z^{\prime \prime}}{z}$
\beq
S_{\chi}=\frac{1}{2} \int d\tau d^{3} x\left[\chi^{\prime 2}-(\partial \chi)^{2}+\frac{z^{\prime \prime}}{z} \chi^{2}\right]
\eeq

Defining the Fourier modes as
\beq
\chi(t, \vect{x})=\int \frac{d^{3} x}{(2 \pi)^{3}} \chi_{\bf k}(t) e^{i {\bf k} \cdot {\bf x}}
\eeq
one finds
\beq
S_{\chi}=\frac{1}{2} \int d \tau \frac{d^{3} k}{(2\pi)^3}\left[\chi_{\bf k}^{\prime} \chi_{-{\bf k}}^{\prime}-k^{2} \chi_{\bf k} \chi_{-{\bf k}}+\frac{z^{\prime \prime}}{z} \chi_{\bf k} \chi_{-{\bf k}}\right]
\eeq
and the Euler-Lagrange equation
\beq
\frac{\partial L}{\partial \chi}-\partial_{\tau} \frac{\partial L}{\partial \chi^{\prime}}=0
\eeq
becomes
\beq
\chi^{\prime\prime}_{\bf k}+\left(k^{2}-\frac{z^{\prime \prime}}{z}\right) \chi_{\bf k}=0
\eeq

To leading order in slow-roll, we have\footnote{Strictly speaking, several different slow-roll conventions are common in the literature. Here $\epsilon\equiv-\dot H/H^2=\dot\phi^2/(2H^2)$, while $\eta$ in this expression denotes the potential slow-roll parameter $\eta_V\equiv V''/V$, evaluated to leading order in slow roll. Equivalently, if one uses the field-acceleration parameter $\eta_\phi\equiv-\ddot\phi/(H\dot\phi)$, then $\eta_\phi=\eta_V-\epsilon+\mathcal O(\epsilon^2)$, and the same result may be written as $
z''/z=a^2H^2\left(2+2\epsilon-3\eta_\phi\right)
=\tau^{-2}\left(2+6\epsilon-3\eta_\phi\right) $
to first order.
}
\beq
 \frac{z^{\prime \prime}}{z}  =2 a^{2} H^{2}\left(1+\frac{5}{2} \epsilon-\frac{3}{2} \eta\right)
\eeq
  \vspace{4pt}
    \hrule
  \vspace{4pt}
{\bf Exercise 3:} Show that this is true.
  \vspace{4pt}
    \hrule
  \vspace{4pt}

Rewriting the scale factor in terms of conformal time, using $a=-1/(H \tau(1-\epsilon)) $  now gives
\beq
 \frac{z^{\prime \prime}}{z}=\frac{1}{\tau^{2}}(2+9 \epsilon-3 \eta)
\eeq

Defining
\beq
\frac{z^{\prime \prime}}{z}= \frac{1}{\tau^{2}}\left(\nu^{2}-\frac{1}{4}\right)
\eeq
implies that
\beq
\nu=\frac{3}{2}+3 \epsilon-\eta
\eeq
which means that the e.o.m. becomes
\beq
\chi_{k}^{\prime \prime}+\left[k^{2}-\frac{1}{\tau^{2}}\left(\nu^{2}-\frac{1}{4}\right)\right] \chi_{k}=0
\eeq
This is the defining function for Hankel functions (linear combinations of  Bessel functions), which implies
\beq
\chi_{k}(\tau)=\sqrt{-\tau}\left[C_{1}(k) H_{\nu}^{(1)}(-k \tau)+C_{2}(k) H_{\nu}^{(2)}(-k \tau)\right]
\eeq
To fix the integration constants, we need a physical boundary condition. We are going to impose that the field was initially in a vacuum. In order to understand the initial vacuum state, we need to quantize the field.

\subsection{Canonical quantization}

In the time-dependent perturbation theory, it is often convenient to adopt the interaction picture. In the interaction picture, the free-field field theory is evolved in the Heisenberg picture. In general
\beq
H=H_{0}+H_{I}
\eeq
with $H_{0}$, the Hamiltonian of the free theory and $H_{I}$ that of the interactions, so
\beq
i\hbar \frac{d}{d t} A_{I}(t)=\left[A_{I}(t), H_{0}\right]
\eeq
for some operator in the interaction picture. While interactions are evolved in the Schrödinger picture
\beq
i \hbar \frac{d}{d t}\left|\psi_{I}(t)\right\rangle=H_{I}^{(I)}\left|\psi_{I}(t)\right\rangle
\eeq
for some interaction picture quantum state $\left|\psi_{I}(t)\right\rangle$ and $H_{I}^{(I)}$ the interaction Hamiltonian in the interaction picture. In the free theory, $H_{I}=0$ and states are time-independent. 

Now, in linear perturbation theory with our $H$ given by the quadratic Lagrangian of $\chi$ above, there are no interactions, so we just have $H_{0}$, and quantize $\chi$ in the Heisenberg picture. Later, when discussing non-Gaussianity and loop effects, we will have to go to higher orders and include $H_I$. 

Quantizing $\chi$ in the Heisenberg picture, using Canonical quantization, we start with promoting $\chi$ and its canonical conjugate field
\beq
\Pi_{\bf k}=\frac{\partial L}{\partial \chi_{\bf k}^{\prime}}=\chi_{-\bf k}^{\prime}
\eeq
to operators $\hat{\chi}, \hat{\Pi}$, and impose the equal time canonical commutation relations
\beq
\begin{aligned}
& {\left[\hat{\chi}(\tau, {\bf x}), \hat{\chi}\left(\tau, {\bf x}^{\prime}\right)\right]=\left[\hat{\Pi}(\tau, {\bf x}), \hat{\Pi}\left(\tau, {\bf x}^{\prime}\right)\right]=0} \\
& {\left[\hat{\chi}(\tau, {\bf x}), \hat{\Pi}\left(\tau, {\bf x}^{\prime}\right)\right]=i \delta\left({\bf x}-{\bf x}^{\prime}\right) \quad[\hbar \equiv 1]}
\end{aligned}
\eeq

The equation of $\chi_{k}$ is that of an harmonic oscillator
\beq
\chi_{k}^{\prime \prime}+\omega_{k}^{2}(\tau)=0
\eeq
with time-dependent frequency
\beq
\omega_{k}^{2}(\tau)=k^{2}-\frac{z^{\prime \prime}}{z}
\eeq
that only depends on norm of the Fourier mode $|{\bf k}|=k$.

Since $\chi$ is a real scalar field, it is hermitian $\chi^{\dagger}(\tau, {\bf x})=\chi(\tau, {\bf x})$, which imply $\chi_{\bf k}^{\dagger}=\chi_{-\bf k}$, so when quantizing, we can write $\hat{\chi}_{\bf k}$ in terms of raising and lowering operators
\beq
\hat{\chi}_{\bf k}(\tau)=\frac{1}{\sqrt{2k}}\left(\hat{c}_{\bf k}(\tau)+\hat{c}_{-\bf k}^{\dagger}(\tau)\right)
\eeq
with
\beq
\left[\hat{c}_{{\bf k}_{1}}, \hat{c}^{\dagger}_{{ \bf k}_{2}}\right]=(2\pi)^3\delta^{(3)}\left({\bf k}_{1}-{\bf k}_{2}\right)
\eeq
where the vacuum is the state
\beq
\hat{c}_{\vect{k}}(\tau)|0\rangle_{\tau}=0
\eeq
Now, clearly, the definition of the vacuum is time-dependent. The state, $|0\rangle_{\tau}$, is a part of a family of instantaneous vacuum states defined by the time-dependent operator $\hat{c}_{\vect{k}}(\tau)$. The state $|0\rangle_{\tau_0}$ is not annihilated by $\hat{c}_{\vect{k}}(\tau)$ at some later time $\tau>\tau_0$. This follows from the fact that the Hamiltonian
\beq
\hat{H}=\frac{1}{2} \int d^{3} x\left[\hat{\Pi}^{2}+(\partial \hat{\chi})^{2}-\frac{z^{\prime \prime}}{z} \hat{\chi}^{2}\right]
\eeq
has an explit time-dependence through $z(\tau)$, and so energy is not conserved. This is how a rich universe can be created out of the vacuum.

The standard way of dealing with this phenomenon is by means of a Bogolubov transformation
\beq
\begin{aligned}
& \hat{c}_{\bf{k}}(\tau)=\alpha_{k}(\tau) \hat{c}_{\bf{k}}\left(\tau_{0}\right)+\beta_{k}(\tau) \hat{c}_{-\bf{k}}^{\dagger}\left(\tau_{0}\right) \\
& \hat{c}_{\bf{k}}^{\dagger}(\tau)=\alpha_{k}^{*}(\tau) \hat{c}_{\bf{k}}^{\dagger}\left(\tau_{0}\right)+\beta_{k}^{*}(\tau) \hat{c}_{-\bf{k}}\left(\tau_{0}\right)
\end{aligned}
\eeq
where $\alpha_{k}, \beta_{k}$ are the Bogoliubov coefficients, which have to
satisfy
\beq
\left|\alpha_{k}\right|^{2}-\left.| \beta_{k}\right|^{2}=1
\eeq
for the commutation relation to be preserved in time.

Note that the number of particles at time $\tau$, if we are initially in the vacuum $|0\rangle_{\tau_{0}}$, is given by

\beq
{}_{\tau_{0}}\langle 0|\hat{N}_{k}| 0\rangle_{\tau_{0}}=(2\pi)^3\delta^{(3)}(0)\left|\beta_{k}\right|^{2}\,,
\eeq
where
\beq
\hat{N}_{k}=\hat{c}_{\bf{k}}^{\dagger}(\tau) \hat{c}_{\bf{k}}(\tau)\,.
\eeq
Dividing by the formal volume factor contained in $\delta^{(3)}(0)$, the occupation number is $n_k = \left|\beta_{k}\right|^{2}$.

The solution to the dynamical equation can be obtained through
\beq
f_{k}(\tau)=\frac{1}{\sqrt{2 k}}\left(\alpha_{k}(\tau)+\beta_{k}^{*}(\tau)\right)
\eeq
with
\beq
f_{k}(\tau) f_{k}^{* \prime}(\tau)-f_{k}^{\prime}(\tau) f_{k}^{*}(\tau)=i
\eeq
which is called the Wronskian condition and ensures that the canonical commutation relation is consistent with that of $\hat{c}_{k}, \hat{c}_{k}^{\dagger}$. We then have that
\beq
\begin{aligned}
& \hat{\chi}_{\bf k}(\tau)=f_{k}(\tau) \hat{c}_{\bf{k}}\left(\tau_{0}\right)+f_{k}^{*}(\tau) \hat c_{-\bf{k}}^{\dagger}\left(\tau_{0}\right) \\
\Rightarrow \quad& \hat{\chi}(\tau, {\bf x})=\int \frac{d^{3} \bf{k}}{(2 \pi)^{3}}\left[f_{k}(\tau) \hat{c}_{\bf k}(\tau_0) e^{i \bf{k} \cdot {\bf x}}+f_{k}^*(\tau) \hat c_{\bf{k}}^{\dagger}\left(\tau_{0}\right) e^{-i {\bf k} \cdot {\bf x}}\right]
\end{aligned}
\eeq

Inserting into the Heisenberg equation of motion
\beq
i \partial_{\tau} \hat{\Pi}=[\hat{\Pi}, \hat{H}]
\eeq
one can verify that $f_{k}(\tau)$ satisfy the classical equation ot motion with the solution
\beq
f_{k}(\tau)=\sqrt{-\tau}\left[C_{1} H_{\nu}^{(1)}(-k \tau)+C_{2} H_{\nu}^{(2)}(-k \tau)\right]~.
\eeq
Since we saw that
\beq
{}_{\tau_{0}} \langle 0 |\hat{N}_{k}| 0 \rangle_{\tau_{0}}=\left|\beta_{k}\right|^{2}
\eeq
and $\left|\alpha_{k}\right|^{2}-\left|\beta_{k}\right|^{2}=1$, as well as $f_{k}=\frac{1}{\sqrt{2k}}\left(\alpha_{k}+\beta_{k}^{*}\right)$, we see that in the instantaneous vacuum (instantaneous no particle state), $\left|\beta_{k}\right|^{2}=0 \Rightarrow\left|\alpha_{k}\right|^{2}=1 \Rightarrow f_{k}=\frac{1}{\sqrt{2 k}} \alpha_{k} \Rightarrow$ $\left|f_{k}\right|^{2}=\frac{1}{2 k} \Rightarrow f_{k}=\frac{1}{\sqrt{2 k}} e^{\mp i F(k, \tau)}$, where $F(k, \tau)$ is some real function of $k$ and $\tau$.

Now, as $\tau \rightarrow-\infty$, the physical wavelength $\lambda_{\textrm{phys}}=a/k \rightarrow 0$, so the wavelength $\lambda_{\textrm{phys}} \ll 1/H \Leftrightarrow k \gg aH$, so far inside the horizon at early
times, since the wavelength corresponding to the $k$-mode is tiny compared to the horizon, or the curvature of the spacetime. Hence, the modes are effectively in flat space, just like a tiny flat-earther doesn't realize the curvature of the Earth because Earth is much bigger -- too big for him/her to understand...

Now for $\tau \rightarrow-\infty$
\beq
\begin{aligned}
& H_{\nu}^{(1)}(-k \tau) \rightarrow \frac{\sqrt{2 / \pi}}{\sqrt{-\tau k}} e^{-i( k \tau+\frac{\nu\pi}{2}+\frac{\pi}{4})} ~,\qquad H_{\nu}^{(2)}(-k \tau) \rightarrow \frac{\sqrt{2 / \pi}}{\sqrt{-\tau k}} e^{i (k \tau+\frac{\nu\pi}{2}+\frac{\pi}{4})}
\end{aligned}
\eeq
Thus, we choose the constants of proportionality such that our definition of the vacuum agrees with the Minkowski vacuum at $\tau \rightarrow-\infty$
\beq
C_{1}=\frac{\sqrt{\pi}}{2} e^{i(\frac{\nu\pi}{2}+\frac{\pi}{4})}, \quad C_{2}=0\,.
\eeq

The annihilation operator multiplies the positive-frequency mode in the far-past Minkowski limit, while the creation operator multiplies its complex conjugate. If one tried to use negative-frequency modes as annihilation modes, one would encounter a sign pathology: either the one-particle states acquire negative norm, or, after forcing a positive-norm Fock space, the Hamiltonian becomes unbounded from below. The standard quantization avoids this by assigning positive-frequency modes to annihilation operators and negative-frequency modes to creation operators.\footnote{Since formally the norm is $(\chi, \chi)=i \int_{-\infty}^{\infty}dx\left(\chi^{*} \Pi-\chi \Pi^{*}\right)$ with $\Pi = \chi'$, then for positive frequency modes we have positive norm modes $(\chi_k,\chi_{k'})=\delta(k-k')$ while for negative frequency modes, we have negative norm modes $(\chi_k,\chi_{k'})=-\delta(k-k')$.}. We then have
\beq
\hat{\chi}_{k}(\tau) \rightarrow \frac{1}{\sqrt{2 k}} e^{-i k \tau} \hat{c}_{\bf{k}}\left(\tau_{0}\right)+\frac{1}{\sqrt{2 k}} e^{i k \tau} \hat{c}_{-\vect{k}}^{\dagger}(\tau_0)
\eeq
for $\tau_{0} \rightarrow -\infty$. 

The observables of a quantum field are, of course, things like expectation values and correlation functions. For $\hat{\chi}$, the two-point correlation function is
\beq
\begin{aligned}
&{}_{\tau_{0}}\langle 0|\hat{\chi}_{\bf{k}_{1}}^{\dagger} \hat{\chi}_{\bf{k}_{2}}| 0\rangle_{\tau_{0}} \equiv (2\pi)^3\delta^{(3)}\left(\bf{k}_{1}-\bf{k}_{2}\right) \frac{2 \pi^{2}}{k^{3}} \mathcal{P}_{\chi}(k) \\
& k=\left| \bf{k}_{1}\right|
\end{aligned}
\eeq

Inserting our normalized solution, we obtain
\beq
\mathcal{P}_{\chi}(k)=\frac{k^{3}}{2 \pi^{2}}\left|f_{k}\right|^{2}\,.
\eeq

Using that $\zeta_{k}=\chi_{k} / z$ and that on superhorizon scales, for $-k \tau \rightarrow 0$, we have
\beq
H_{\nu}^{(1)}(-k \tau) \sim \sqrt{\frac{2}{\pi}} e^{-i \frac{\pi}{2}} 2^{\nu-\frac{3}{2}} \frac{\Gamma(\nu)}{\Gamma(3 / 2)}(-k \tau)^{-\nu}\,,
\eeq
and we obtain on superhorizon scales
\beq
\begin{aligned}
\mathcal{P}_{\zeta}(k) & =\frac{k^{3}}{2 \pi^{2}} \frac{\left| f_{k}\right|^{2}}{z^{2}}  \\
& =\frac{2^{2 \nu-3}}{(2 \pi)^{2}}\left(\frac{\Gamma(\nu)}{\Gamma(3 / 2)}\right)^{2}\left(\frac{H}{a \dot{\phi}}\right)^2(-k \tau)^{3-2 \nu}(-\tau)^{-2} \\
& \propto k^{3-2 \nu}\,.
\end{aligned}
\eeq

Defining the spectral tilt to be
\beq
\begin{aligned}
n_{s}-1=\frac{d \ln \mathcal{P}_{\zeta}(k)}{d \ln k} & =3-2 \nu  =2 \eta-6 \epsilon\,.
\end{aligned}
\eeq
So
\beq
\boxed{
n_{s}-1=2 \eta-6 \epsilon}
\eeq
is one of the major predictions of inflation. Another important observable is the amplitude of perturbations. Since they are conserved on super-horizon scales, we can evaluate the amplitude of perturbations by evaluating the power spectrum at horizon exit
\beq
\mathcal{P}_{\zeta}(k)= 2^{2\nu-3}(1-\epsilon)^{2\nu-1}\left.\left(\frac{\Gamma(\nu)}{\Gamma(3/2)}\right)^2 \left(\frac{H}{\dot\phi}\right)^2\left(\frac{H}{2\pi}\right)^2\right|_{k=aH}~.
\eeq
Using $\nu\simeq3/2$ we then obtain
\beq
\boxed{ \left. \mathcal{P}_{\zeta}(k) \simeq\left(\frac{H}{2 \pi}\right)^{2}\left(\frac{H}{\dot \phi}\right)^{2}\right|_{k=a H}}
\eeq

\subsection{Tensor modes}

So far, we have calculated the spectrum of the scalar perturbations, $\zeta$. But we also have two tensor modes $\gamma_{ij}$. From the ADM action, we obtain at quadratic order
\beq
S_{\gamma}=\frac{1}{8} \int d t d^{3}x\left[a^{3} \dot{\gamma}_{i j} \dot{\gamma}_{i j}-a \partial_{l} \gamma_{i j} \partial_{l} \gamma_{i j}\right]
\eeq
which in conformal time yields
\beq
S_{\gamma}=\frac{1}{8} \int d \tau d^{3} x a^{2}\left[{\gamma}^\prime_{i j} {\gamma}^\prime_{i j}-\partial_{l} \gamma_{i j} \partial_{l} \gamma_{i j}\right]
\eeq

Now expanding in plane waves for the two polarization modes
\beq
\gamma_{i j}=\int \frac{d^{3} k}{(2 \pi)^{3}} \sum_{s= \pm} \epsilon_{i j}^{s}(k) \gamma_{\bf{k}}^{s}(\tau) e^{i \bf{k} \cdot {\bf x}}
\eeq
with the transverse condition $\epsilon_{i i}=k^{i} \epsilon_{i j}=0$  and  $\epsilon_{i j}^{s}(k) \epsilon_{i j}^{s^{\prime}}(k)=2 \delta_{s s^{\prime}} $, we have
\beq
S_{\gamma}=\frac{1}{4} \sum_{s= \pm} \int d \tau \frac{d^3 k}{(2\pi)^3} a^{2}\left[\gamma_{\bf{k}}^{\prime s} \gamma_{-\bf{k}}^{\prime s}-k^{2} \gamma_{\bf{k}}^{s} \gamma_{-\bf{k}}^{s}\right]~.
\eeq
Defining $h_{k}^{s}=\frac{1}{\sqrt{2}} a \gamma_{\bf{k}}^{s}$, it canbe written as
\beq
S_{\gamma}=\frac{1}{2} \sum_{s= \pm} \int d \tau \frac{d^3 k}{(2\pi)^3} \left[h_{\bf{k}}^{\prime^{s}} h_{-\bf{k}}^{\prime s}-\left(k^{2}-\frac{a^{\prime\prime}}{a}\right) h_{{\bf k}} h_{-\bf{k}}\right]~.
\eeq
This is the same action as for $\chi_{\bf k}$ except now $z \rightarrow a$ and so $z''/z \rightarrow a''/a \simeq 1/\tau^{2}(2+3 \epsilon)$. So, we can define
\beq
\nu_{T}=\frac{3}{2}+\epsilon~,
\eeq
so for tensor modes, we find
\beq
n_{T}=\frac{d \ln P_{\gamma}}{d \ln k}=3-2 \nu_{T} \,,
\eeq
and we find
\beq
\boxed{
n_{T}=-2 \epsilon}
\eeq
and\footnote{The factor of $8$ comes from  $8=2 \times 2 \times 2$ where $
\sum_{s s^{\prime}}\left\langle\gamma^{s} \gamma^{s^{\prime}}\right\rangle  =\sum_{s s^{\prime}} 2/a^2\left\langle h^{s} h^{s^{\prime}}\right\rangle=4/a^2\left(\left\langle h^{+} h^{+}+h^- h^{-}\right\rangle\right. \\
 =8/a^2\left\langle h^{s} h^{s}\right\rangle .
$}
\beq
\boxed{
\left.\mathcal{P}_{\gamma} \simeq 8\left(\frac{H}{2 \pi}\right)^{2}\right|_{k=a H}}
\eeq

\subsection{Observational tests of inflation}

We have not measured the tensor modes from inflation yet. But from the non-observation, we get important constraints. Let us consider the ratio, called the tensor-to-scalar ratio
$$
r=\frac{\mathcal{P}_{\gamma}}{\mathcal{P}_{\zeta}}=\frac{8}{H^{2} / \dot\phi^{2}}=16 \epsilon=-8 n_{T}
$$
This is also called the single-field slow-roll consistency relation and is an important prediction. In $R^{2}$-model (Starobinsky model) \cite{Starobinsky:1980te}
$$
n_{s}-1 \simeq -\frac{2}{N}\,,\qquad r \simeq \frac{12}{N^{2}}
$$
so for $N \sim 60$ we obtain
$$
n_{s} \simeq 0.967\,,\qquad r \simeq 0.0033
$$
in this model.

\subsubsection{Curvaton}

In the curvaton model \cite{Enqvist:2001zp,Lyth:2001nq,Moroi:2001ct}, there is an inflaton field $\phi$ and a curvaton field $\sigma$, which, for simplicity, we can assume to have the simple, basic potential
\beq\label{curvpot1}
V(\phi, \sigma)=\frac{1}{2} M^{2} \phi^{2}+\frac{1}{2} m^{2} \sigma^{2}\,.
\eeq
The curvaton is very light and subdominant during inflation, and some curvature perturbation is created by the inflaton, like before. But imagine that the amplitude is too small to fit the data, for instance, if $H$ during inflation is small.

Then all the observed curvature perturbations can instead come from the curvaton field, which is just a spectator during inflation, if after the inflaton has decayed into radiation with energy density $\rho_{r} \propto 1 / a^{4}$, and $H$ start decreasing as $H \propto 1 / a^{2}$, the curvaton mass will be come heavier compared to $H$, such that $m \sim H$ at which point the curvaton will start to oscillate in its potential with an energy density $\rho_{\sigma} \propto 1/a^{3}$ and soon dominate the energy density. At that point, the density perturbations can easily be computed by computing $\delta\rho_\sigma/\rho_\sigma$ during inflation and noting that it will stay frozen for superhorizon modes
\beq
\frac{\delta \rho_{\sigma}}{\rho_{\sigma}}=\frac{m^{2} \sigma \delta \sigma}{\frac{1}{2} m^{2} \sigma^{2}}=2 \frac{\delta \sigma}{\sigma}\,,
\eeq
which, by a gauge transformation from this flat gauge into the comoving gauge, yields
\beq
\begin{aligned}
& \delta \rho_\sigma
\left(\tau^{\prime}\right)=\delta \rho_{\sigma}(\tau)+\dot{\rho}_{\sigma} \delta t=0 \\
& \Rightarrow \quad \delta t=-\frac{\delta \rho_{\sigma}}{\dot\rho_{\sigma}}\,,
\end{aligned}
\eeq
from which we obtain the comoving curvature perturbation induced by the curvaton
\beq
\zeta_{\sigma}=\frac{\dot{a}}{a} \delta t=H \delta t=-\frac{H}{\dot{\rho}_{\sigma}} \delta \rho_{\sigma}
\eeq
by an argument very similar to how we found $\zeta_{\phi}$ for the inflation.

For the curvaton, the action is just
\beq
S=\frac{1}{2} \int d\tau d^{3} x a^{2}\left[\sigma^{\prime 2}-(\partial \sigma)^{2}-m^{2} a^{2} \sigma^{2}\right]\,.
\eeq
So, doing the field redefinition
\beq
\chi_{\sigma}=a \sigma
\eeq
we find
\beq
\begin{aligned}
& S=\frac{1}{2}\int d\tau d^{3} x \left[\chi_{\sigma}^{\prime 2}-\left(\partial \chi_{\sigma}\right)^{2}-\left(m^{2} a^{2}-\frac{a^{\prime \prime}}{a}\right) \chi^{2}\right] \,,
\end{aligned}
\eeq
and so for the curvaton, we can take
\beq
\frac{z^{\prime \prime}}{z} \rightarrow \frac{a^{\prime \prime}}{a}-a^2m^{2}=\frac{1}{\tau^{2}}(2+3 \epsilon-3 \eta_{\sigma})
\eeq
with $\eta_{\sigma} \equiv {m}^{2}/(3 H^{2})$. Thus, to leading order in slow-roll, we have 
\beq
\nu_{\sigma}= \frac{3}{2}+\epsilon-\eta_{\sigma}\,.
\eeq

So, for the simplest curvaton model, we find
\beq
n_{s}-1=\frac{d \ln \mathcal{P}_{\zeta_\sigma}(k)}{d \ln k}=3-2 \nu_{\sigma}\,,
\eeq
or
\beq
\boxed{n_{s}-1=-2 \epsilon+2 \eta_{\sigma} \geq-2 \epsilon}
\eeq

If we consider the case where the inflaton potential is a simple mass term,  $\phi^{2}$ inflation, then $\epsilon=\frac{1}{2 N}$ and since we need $N\approx 60$ number of $e$-folds, we obtain
\beq
n_{s} \approx 0.98\,.
\eeq

In a mixed model, $n_{s}$ could have a contribution from the inflation, and defining the ratio
\beq
R=\frac{\mathcal{P}_{\zeta_\sigma}}{\mathcal{P}_{\zeta_{\phi}}}\,,
\eeq
we obtain
\beq
\begin{aligned}
& n_{s}=1-\frac{1}{1+R} \frac{8}{4 N+2}+\frac{R}{1+R}\left[-2 \epsilon+2 \eta_{\sigma}\right] \,,\\
& r=\frac{16 \epsilon}{1+R}\,,
\end{aligned}
\eeq
for the potential in (\ref{curvpot1}).

Also, as mentioned in the first lecture, pre-big bang and ekpyrotic scenarios require a curvaton, as the contraction phases do not lead to a scale-invariant spectrum.

\subsubsection{Observations}
Most important constraints on $n_s$ and $r$ come from Planck. Assuming $\Lambda$CDM, they found clearly that Starobinsky-type inflation is favored (see Figure (\ref{planck}) below).
\begin{figure}
\begin{center}
\includegraphics[max width=\textwidth]{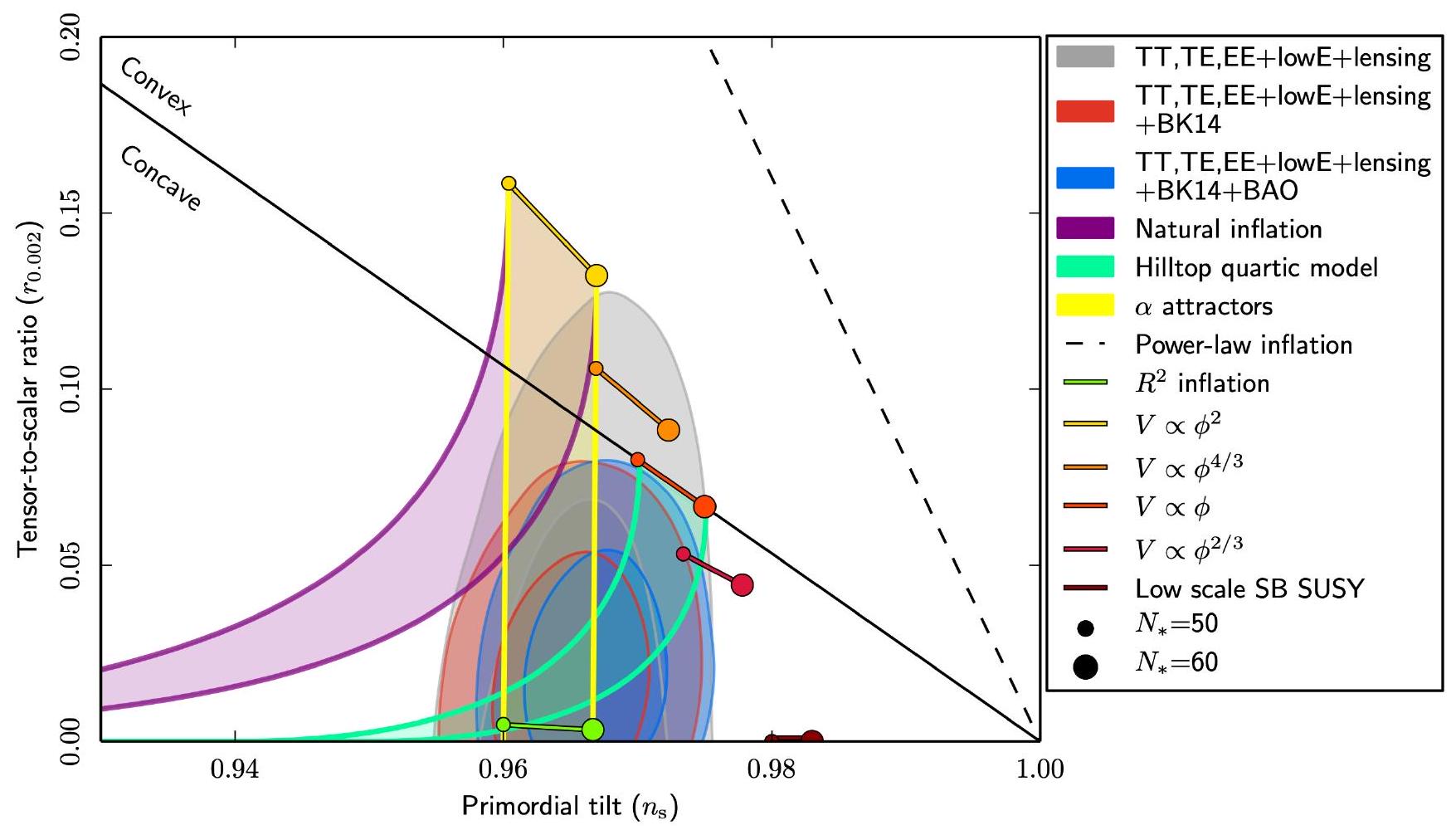}
\end{center}
\caption{Marginalized joint $68 \%$ and $95 \%$ CL regions for $n_{\mathrm{s}}$ and $r$ at $k=0.002 \mathrm{Mpc}^{-1}$ from Planck alone and in combination with BK14 or BK14 plus BAO data, compared to the theoretical predictions of selected inflationary models. Note that the marginalized joint $68 \%$ and $95 \%$ CL regions assume $d n_{\mathrm{s}} / d \ln k=0$. Figure from \cite{Planck:2018jri}.}
\label{planck}
\end{figure}

Most people thought this result is very robust, since perturbations are conserved on super-horizon scales and are therefore insensitive to early-universe physics before the creation of the CMB, and the late-time evolution of the universe is very constrained.

However, the Hubble tension is sending another message. The Hubble tension is a significant disagreement in the measurement of the Hubble constant today when assuming $\Lambda$CDM and using CMB data, and when measuring it directly using supernovae.

\begin{figure}
\begin{center}
\includegraphics[max width=\textwidth]{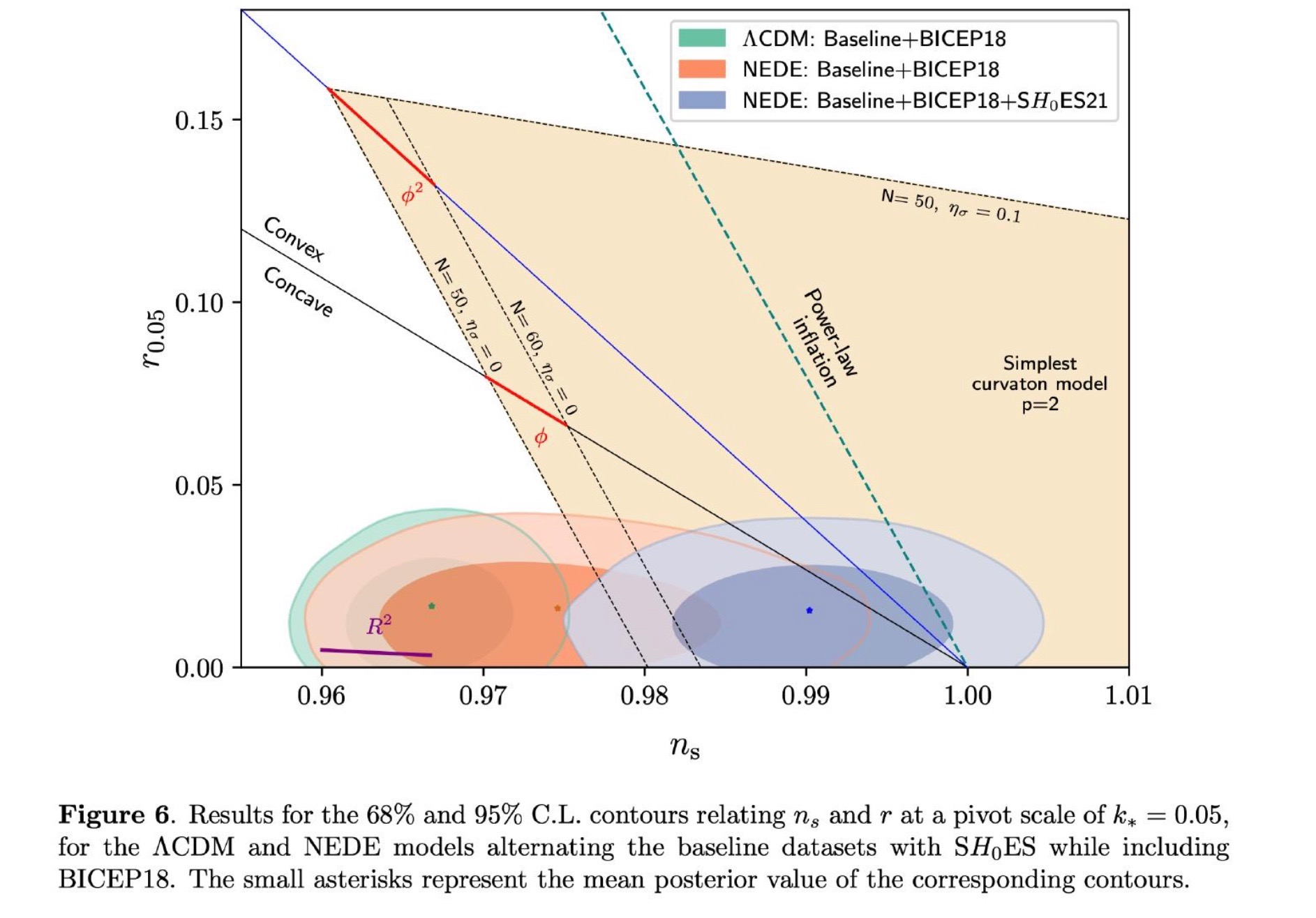}
\end{center}
\caption{NEDE plot from \cite{Cruz:2022oqk}.}
\label{nede}
\end{figure}

New Early Dark Energy (NEDE) is a promising framework for addressing the tension by adding new physics to $\Lambda$CDM \cite{Niedermann:2019olb,Niedermann:2020dwg}. It involves a 1st order phase transition in Dark Energy just before recombination. NEDE, however, implies that $n_{s} \gtrsim 0.98$, which rules out Starobinsky inflation, but favors the simplest curvaton model. See Fig.(\ref{nede})

\newpage

\section{Lecture 3: Beyond linear perturbation theory}

So far, we have only considered the equation of motion of the perturbations, like $\zeta$, to linear order, which we derived from their quadratic action. This is equivalent to saying that we have only treated the free-field theory of the perturbations and ignored interactions. Going to higher orders in perturbation theory means including interactions. When quantizing the free field theory, we obtain Gaussian quantum fluctuations, which, when stretched to large scales with high occupation numbers, become Gaussian random variables. Gaussian fluctuations are completely characterized by their two-point function, which we already calculated. But going to higher orders in perturbation theory and including interactions, we will find deviations from Gaussianity, characterized by a non-vanishing 3-point correlation function and a non-vanishing connected 4-point function. We can also have loop corrections to the two-point function. We will now discuss these issues in turn, starting with non-Gaussianity.

\subsection{Non-Gaussianity}

Let's start by considering the 3-point function of the curvature perturbation
\beq
\left\langle\zeta_{\vect{k}_{1}}  \zeta_{\vect{k}_{2}} \zeta_{\vect{k}_{3}} \right\rangle \equiv(2 \pi)^{3} \delta\left(\Sigma_{a} \vect{k}_{a}\right) B_{\zeta}\left(\vect{k}_{1}, \vect{k}_{2}, \vect{k}_{3}\right)\,,
\eeq
where we have introduced the bispectrum $B_{\zeta}$, which is a function of the triangle formed by the three momenta, $\vect{k}_{1}, \vect{k}_{2}, \vect{k}_{3}$, due to momentum conservation.

There are three extreme shapes, which are typically used as templates and which embody different limits of the underlying physics:

\noindent {\underline{\bf Squeezed (local)}: $\left|\vect{k}_{1}\right| \ll\left|\vect{k}_{2}\right|,\left|\vect{k}_{3}\right|$

\begin{center}
\includegraphics[max width=5cm]{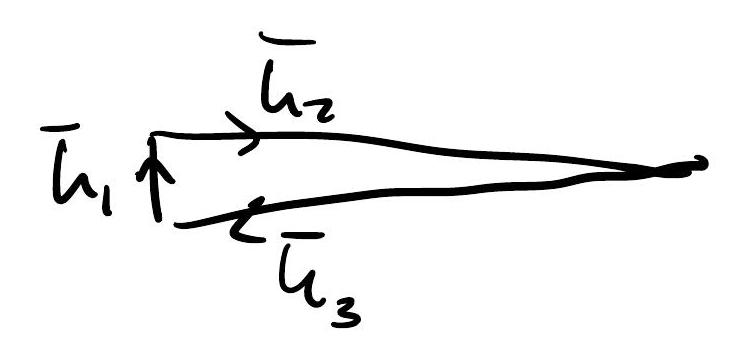}
\end{center}

\noindent {\underline{\bf Equilateral}}: $\quad\left| \vect{k}_{1}\right| \simeq\left|\vect{k}_{2}\right| \simeq\left|\vect{k}_{3}\right|$

\begin{center}
\includegraphics[max width=5cm]{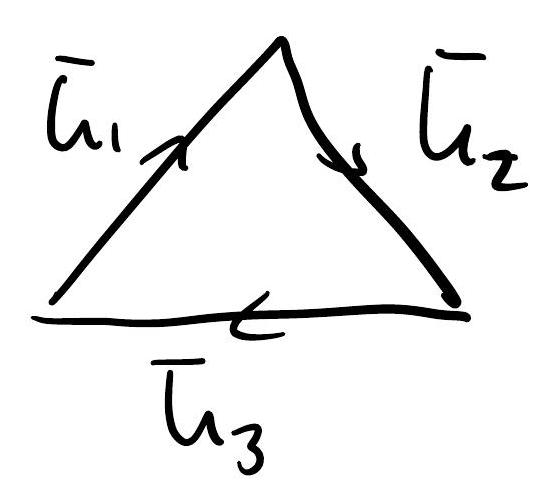}
\end{center}

\noindent {\underline{\bf  Folded}}: $\quad\left| \vect{k}_{3}\right| \simeq 2\left|\vect{k}_{1}\right| \simeq 2\left|\vect{k}_{2}\right|$

\begin{center}
\includegraphics[max width=5cm]{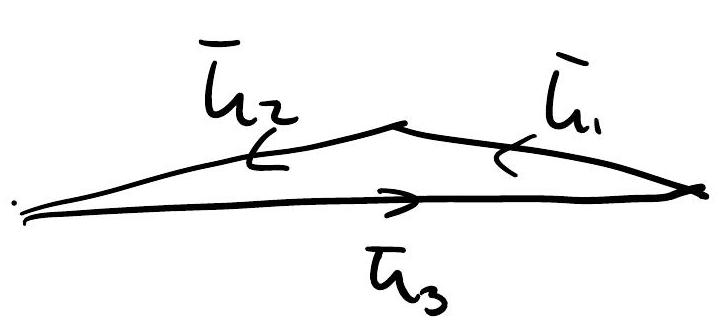}
\end{center}

The single-field models of inflation with a standard kinetic term and the curvaton model have a bispectrum that is peaked around the local squeezed limit, whereas some higher-derivative theories, such as DBI inflation, are peaked in the equilateral shape, and, for example, modifications to the initial vacuum could lead to a maximum of the bispectrum in the folded shape.

Since the simpler "standard" single-field models of slow-roll inflation and the curvaton model have a bispectrum that is maximal in the local shape, we are going to focus on local non-Gaussianity here.

\subsubsection{Local non-Gaussianity}

As a parametrization of the strength of non-Gaussianity, one usually introduces the dimensionless now-linarity parameter, which in general can depend on momenta
\beq\label{bispectrum}
B_{\zeta} \equiv \frac{6}{5} f_{N L} \sum_{a<b} P_{\zeta}\left(k_{a}\right) P_{\zeta}\left(k_{b}\right)
\eeq
where the power spectrum
\beq
P_{\zeta}(k) \equiv  \frac{2\pi^2}{k^{3}} \mathcal{P}_{\zeta}(k)
\eeq
is defined so 
\beq
\left\langle \zeta_{\vect{k}_{1}}, \zeta_{\vect{k}_{2}}\right\rangle=(2 \pi)^{3} \delta\left(\vect{k}_{1}+\vect{k}_{2}\right) P_{\zeta}\left(k_{1}\right)
\eeq

The local shape of non-Gaussianity corresponds to the case where $f_{N L}$ is independent of momenta, in which case $B_{\zeta}$ can be obtained from the simple ansatz
\beq
\zeta=\zeta_{g}+\frac{3}{5} f_{N L}^{\text {local }}\left(\zeta_{g}^{2}-\left\langle\zeta_{g}^{2}\right\rangle\right)\,,
\eeq
where $\zeta_{g}$ is the Gaussian linear perturbation.

In single-field inflation, $f_{N L}^{\text {local }}$ is given by the Maldacena consistency relation \cite{Maldacena:2002vr}. If we consider $\left\langle \zeta_{\vect{k}_{1}}  \zeta_{\vect{k}_{2}} \zeta_{\vect{k}_{3}}\right\rangle$ in the squeezed limit $k_{1} \ll k_{2}$, $k_{3}$, we can think of the long wavelength mode as locally rescaling the background for the short wavelength modes
\begin{center}
\includegraphics[max width=8cm]{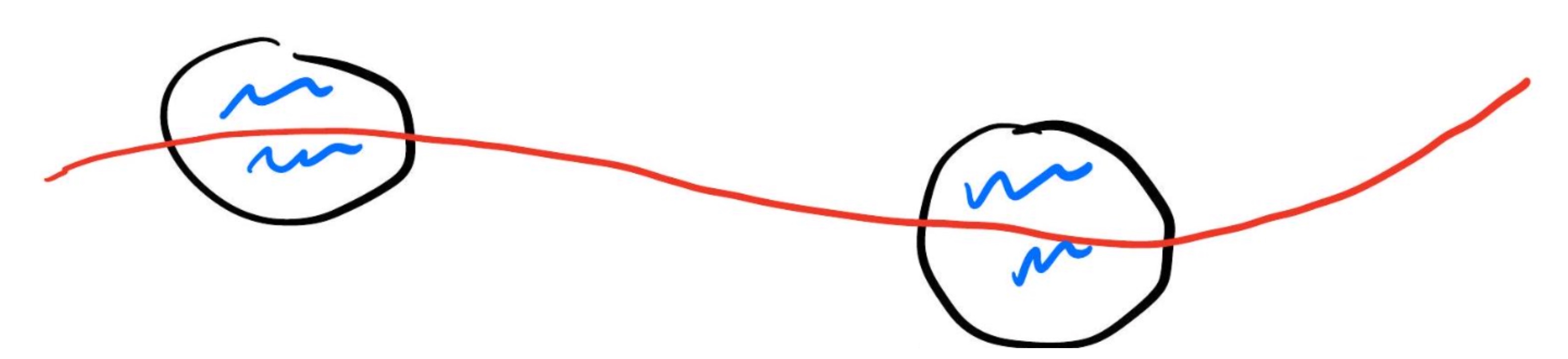}
\end{center}
Since, if on the scale of $\vect{k}_{2}$ and $\vect{k}_{3}$,  $\zeta(\vect{k}_1)$ looks as a constant $\zeta_{B}$, we see from
\beq
d s^{2}=-d t^{2}+a^{2} e^{2 \zeta_{B}} d \vect{x}^{2}\,,
\eeq
that it can locally be absorbed into the background
\beq
d x \rightarrow d \tilde x=e^{\zeta_{B}} d x\,,
\eeq
which corresponds to
\beq
k \rightarrow \tilde{k}=e^{-\zeta_{B}} k\,.
\eeq

We can then Taylor expand the local two-point function to leading order in the long wavelength mode
\beq
\left\langle\zeta\left(\vect{x}_{2}\right) \zeta\left(\vect{x}_{3}\right)\right\rangle_{\zeta_{B}}=\left\langle\zeta\left(\vect{x}_{2}\right) \zeta\left(\vect{x}_{3}\right)\right\rangle_0+\left.\zeta_{B}\left(\vect{x}_{1}\right) \frac{\partial}{\partial \zeta}\left\langle\zeta\left(\tilde{\vect{x}}_{2}\right) \zeta\left(\tilde{\vect{x}}_{3}\right)\right\rangle\right|_{\zeta_B=0}+\ldots
\eeq
Since $\zeta_{B}$ is almost constant on the length scale $\left|\vect{x}_{2}-\vect{x}_{3}\right|$, we can choose $x_{1}$ freely between $x_{2}$ and $x_{3}$, but take for simplicity $\vect{x}_{1}=\left(\vect{x}_{2}+\vect{x}_{3}\right) / 2$ Now, going to Fourrier space, we have
\bea
 \left\langle\zeta_{\vect{k}_{2}} \zeta_{\vect{k}_{3}}\right\rangle_{\zeta_B}&=&\int d^{3} x_{2} \int d^{3} x_{3} e^{-i \vect{x}_{2} \cdot \vect{k}_{2}} e^{-i \vect{x}_{3} \cdot \vect{k}_{3}} \left\langle \zeta\left(\vect{x}_{2}\right) \zeta\left(\vect{x}_{3}\right)\right\rangle_{\zeta_B}\nonumber \\
& =&\left\langle \zeta_{\vect{k}_{2}} \zeta_{\vect{k}_{3}}\right\rangle_{0} 
 +\int d^{3} x_{2} \int d^{3} x_{3} e^{-i \vect{x}_{2} \cdot\vect{k}_{2}} e^{-i \vect{x}_{3} \cdot \vect{k}_{3}}   \zeta_{B}\left(\vect{x}_{1}\right) \frac{\partial}{\partial \zeta_{B}}\left[\iint \frac{d^{3} \tilde{q}_{2}}{(2 \pi)^{3}} \frac{d^{3} \tilde{q}_{3}}{(2 \pi)^{3}} e^{i {\tilde{\vect x}}_{i} \cdot{\tilde{\vect q}}_{2}} e^{i {\tilde{\vect x}}_{3} \cdot{\tilde{\vect q}}_{3}}\left\langle\zeta_{\tilde{\vect q_{2}}} \zeta_{\tilde{\vect{q}}_{3}}\right\rangle\right]_{\zeta_{{B}}=0} +\ldots \nonumber\\ 
& =&\left\langle \zeta_{\vect k_2} \zeta_{\vect k_3}\right\rangle_{0} 
 +\int d^{3} x_{2} \int d^{3} x_{3} e^{-i \vect{x}_{2} \cdot \vect{k}_{2}} e^{-i \vect{x}_{3} \cdot \vect{k}_{3}} \int \frac{d^{3} q_{B}}{(2 \pi)^{3}} e^{i \vect{x}_{1} \cdot {\vect{q}}_{B}}\zeta_{B}\left(\vect{q}_{B}\right)\nonumber\\
  &&\times \frac{\partial}{\partial \zeta_{B}}\left.\left[\iint \frac{d^{3} q_{2}}{(2 \pi)^{3}} e^{-3 \zeta_{B}} \frac{d^{3} q_{3}}{(2 \pi)^{3}} e^{-3 \zeta_{B}}e^{ i \vect{x}_{2} \cdot \vect{q}_{2}} e^{i \vect{x}_{3} \cdot \vect{q}_{3}}\left\langle \zeta\left(\vect{q}_{2} e^{-\zeta_{B}}\right) \zeta\left(\vect{q}_{3} e^{-\zeta_{B}}\right)\right\rangle\right]\right|_{\zeta_B=0}\nonumber\\
& =&\left\langle \zeta_{\vect{k}_2} \zeta_{\vect{k}_3}\right\rangle_{0} 
 +\int d^{3} x_{2} \int d^{3} x_{3} \int \frac{d^{3} q_{B}}{(2 \pi)^{3}} e^{-i \vect{x}_{2}\cdot(\vect{k}_{2}-\frac{1}{2} \vect{q}_{B})} e^{-i \vect{x}_{3}\cdot(\vect{k}_{3}-\frac{1}{2} \vect{q}_{B})}  \zeta_{B}\left(\vect{q}_{B}\right)\nonumber\\
& &\times \frac{\partial}{\partial \zeta_{B}}\left[\iint \frac{d^{3} q_{2}}{(2 \pi)^{3}} e^{-3 \zeta_{B}} \frac{d^{3} q_{3}}{(2 \pi)^{3}} e^{-3 \zeta_{B}} e^{i \vect{x}_{2} \cdot\vect{q}_{2}} e^{i \vect{x}_{3}\cdot \vect{q}_{3}}\left\langle \zeta\left(\vect{q}_{2} e^{-\zeta_{B}}\right) \zeta\left(\vect{q}_{3} e^{-\zeta_{B}}\right)\right\rangle\right|_{\zeta_{B}=0} \nonumber\\
& =&\left\langle \zeta_{\vect{k}_{2}} \zeta_{\vect{k}_{3}}\right\rangle_{0} +\int \frac{d^{3} q_{B}}{(2 \pi)^{3}} \int \frac{d^{3} q_{2}}{(2 \pi)^{3}} \frac{d^{3} q_{3}}{(2 \pi)^{3}}(2 \pi)^{3} \delta^{3}\left(\vect{k}_{2}-\frac{1}{2} \vect{q}_{B}-\vect{q}_{2}\right)(2 \pi)^{3} \delta^{3}\left(\vect{k}_{3}-\frac{1}{2} \vect{q}_{B}-\vect{q}_{3}\right)\nonumber \\
& &\times \zeta_{B}\left(\vect{q}_{B}\right) \frac{\partial}{\partial \zeta_{B}}\left[\left.e^{-6\zeta_{B}}\left\langle\zeta\left(\vect{q}_{2} e^{-\zeta_{B}}\right) \zeta\left(\vect{q}_{3} e^{-\zeta_{B}}\right)\right\rangle\right]\right|_{\zeta_B= 0} \nonumber\\
& =&\left\langle\zeta_{\vect{k}_{2}} \zeta_{\vect{k}_{3}}\right\rangle_{0} +\int \frac{d^{3} q_{B}}{(2 \pi)^{3}} \zeta_{B}\left(\vect{q}_{B}\right) \frac{\partial}{\partial \zeta_{B}}\left[e^{-6 \zeta_{B}}\left\langle \zeta\left(\vect{q}_{2} e^{-\zeta_{B}}\right) \zeta\left(\vect{q}_{3} e^{-\zeta_{B}}\right)\right\rangle\right]\left.\right|_{\zeta_{B}=0} \nonumber\\
& =&\left\langle \zeta_{\vect{k}_{2}} \zeta_{\vect{k}_{3}}\right\rangle_{0} -(n_s-1)\int \frac{d^3 q_B}{(2\pi)^3} \zeta_B\left(\vect{q}_B\right)  \left\langle \zeta_{\vect{k}_2-\frac{1}{2}\vect{q}_B} \zeta_{\vect{k}_3-\frac{1}{2}\vect{q}_B} \right\rangle\,.\nonumber
\eea
with $q_{2}=\vect{k}_{2}-\frac{1}{2} \vect{q}_{B}$, $q_{3}=\vect{k}_{3}-\frac{1}{2} \vect{q}_{B}$ after integrating out the delta-functions after the fifth equality.

Now taking the squeezed limit 
\bea
 \lim _{k_{1} \rightarrow 0}\left\langle\zeta_{\vect{k}_{1}} \zeta_{\vect{k}_{2}} \zeta_{\vect{k}_{3}}\right\rangle & = & \lim _{k_{1} \rightarrow 0}\left\langle\zeta_{\vect{k}_{1}}\langle \zeta_{\vect{k}_{2}} \zeta_{\vect{k}_{3}}\rangle_{\zeta_B}\right\rangle\nonumber \\
 &=&-\left(n_{s}-1\right) \int \frac{d^{3} q_{B}}{(2 \pi)^{3}}\left\langle\zeta_{\vect{k}_{1}} \zeta_{\vect{q}_{B}}\right\rangle \left\langle \zeta_{\vect{k}_2-\frac{1}{2}\vect{q}_B} \zeta_{\vect{k}_3-\frac{1}{2}\vect{q}_B} \right\rangle \nonumber \\
& =&-\left(n_{s}-1\right)\left\langle\zeta_{\vect{k}_{1}} \zeta_{-\vect{k}_{1}}\right\rangle\left\langle\zeta_{\vect{k}_{2}+\frac{1}{2}\vect{k}_1} \zeta_{\vect{k}_{3}+\frac{1}{2}\vect{k}_1}\right\rangle\nonumber \\
& =&-\left(n_{s}-1\right)(2 \pi)^{3} \delta\left(\vect{k}_1+\vect{k}_{2}+\vect{k}_{3}\right) P_{\zeta_{k_{1}}} P_{\zeta_{k_{S}}} \nonumber\\
& \Rightarrow& \boxed{f_{N L}^{\text {local }}=-\frac{5}{12}\left(n_{s}-1\right)}\,.
\eea
where in the last line, we used that the first two terms of the sum in (\ref{bispectrum}) involving $P_\zeta(k_1)$ dominate in the squeezed limit, and in the line before that, we introduce $\vect{k}_S = (\vect{k}_2-\vect{k}_3)/2$. So, in single-field inflation, $f_{N L}$ is pretty small, of order the slow-roll parameters. 

Note that Maldacena verified this by calculating the full 3-point function $\langle \zeta_{\vect{k}_1}\zeta_{\vect{k}_2}\zeta_{\vect{k}_3}\rangle$ using the in-in formalism in QFT\footnote{For an introduction to in-in QFT computations in cosmology and diagrammatic rules, see \cite{Giddings:2010ui}.}, and then taking the squeezed limit of the full result \cite{Maldacena:2002vr}.

It is straightforward to generalize the Maldacena consistency relation to an $n$-point function with
\bea
&& k_{1}, \ldots, k_{l} \gg k_{l+1}, \ldots, k_{n}\nonumber \\
& \Rightarrow& P^{n}\left(k_{1}, \ldots, k_{n}\right)=-\left(n_{l}-1\right) P^{l}\left(k_{1}, \ldots, k_{l}\right)  P^{n-l+1}\left(k_{1}+ \ldots +k_{l}, k_{l+1}, \ldots\right)
\eea
(see Chen, Huang, and Shiu \cite{Chen:2006dfn}),
or double-squeezed limit with two external legs being long wavelength (soft modes). 

Now let's compare with the curvaton.

\subsection{Curvaton}

In the case of the curvaton, we previously found
\beq\label{zetalin}
\zeta_{\sigma}=-\frac{H}{\dot{\rho}_{\sigma}} \delta \rho_{\sigma}\,.
\eeq
During matter domination, we have $\dot\rho_{\sigma}=-3 H\left(\rho_{\sigma}+p_{\sigma}\right)=-3 H \rho_{\sigma}$, which implies
\beq\label{zetalin2}
\zeta_{\sigma} \simeq \frac{\delta \rho_{\sigma}}{3 \rho_{\sigma}}\,.
\eeq
Earlier, we showed that to linear order
\beq
\frac{\delta \rho_{\sigma}}{\rho_{\sigma}}=\frac{m^{2} \sigma \delta \sigma}{\frac{1}{2} m^{2} \sigma^{2}}=2 \frac{\delta \sigma}{\sigma}
\eeq
where $\delta \sigma$ is Gaussian. However, going to second order, we have
\beq\label{deltarhosigma}
\frac{\delta \rho_{\sigma}}{\rho_{\sigma}}=\frac{m^{2} \sigma \delta \sigma+\frac{1}{2} m^2\delta \sigma^2}{\frac{1}{2} m^{2} \sigma^{2}}=2\left(\frac{\delta \sigma}{\sigma}+\frac{1}{2} \frac{\delta \sigma^{2}}{\sigma^{2}}\right)\,.
\eeq
Since the second term in the bracket is the square of a Gaussian fluctuation, it is non-Gaussian. This is the intrinsic non-Gaussianity of the curvaton energy density perturbations in the spatially flat gauge where the comoving curvature perturbation vanishes ($\zeta =0$). 

However, there is an extra contribution to the non-Gaussianity which comes from extending the gauge transformation going from spatially flat gauge to uniform-density gauge $\delta \rho=0$ to second order. More precisely, we need to extend (\ref{zetalin}) to second order, and as we will see, the non-linearity of the relation extended to second order introduces additional non-Gaussianity.

Making the gauge transformation $t \to t +\delta t$ takes $\ln a \to \ln a +H\delta t$, or when identifying $\zeta = H\delta t$, the gauge transformation takes $a\to ae^{\zeta}$. When the curvaton decays into radiation, the total energy density of the universe consists of radiation and the non-relativistic curvaton, $\rho = \rho_r +\rho_\sigma$. Since radiation scales as $1/a^4$, the gauge transformation takes $\rho_r \to \rho_r e^{-4\zeta}$ and similarly we have $\rho_\sigma \to \rho_\sigma e^{-3\zeta}$.

In the uniform-density gauge, we have $\delta \rho=0$, which implies, after the gauge transformation from the spatially flat gauge to the uniform-density gauge, we must have
\beq
\bar\rho= (\bar\rho_r +\delta\rho_r) e^{-4\zeta}+ (\bar\rho_\sigma +\delta\rho_\sigma) e^{-3\zeta}\,,
\eeq
where, to avoid confusion, we have temporarily introduced the notation $\bar\rho$ for the homogeneous background value, i.e $\rho(t,\vect{x}) = \bar\rho(t) +\delta\rho(t,\vect{x}) $, and in the above expression in the uniform-density gauge we have demanded $\delta\rho=0$, which when expanding to second order, implies
\beq
\left(1-\frac{\bar\rho_\sigma}{\bar\rho}\right)(1-4\zeta+8\zeta^2)+\frac{\bar\rho_\sigma}{\bar\rho}\left(1+\frac{\delta\rho_\sigma}{\bar\rho_\sigma}\right)(1-3\zeta+\frac{9}{2}\zeta^2) =1\,.
\eeq

Solving for $\zeta$ order by order, we obtain the equivalent of the relation in (\ref{zetalin2}) to second order in perturbation theory
\beq
\zeta = \frac{r_{\mathrm{dec}}}{3}\frac{\delta\rho_\sigma}{\bar \rho_\sigma} - \frac{r_{\mathrm{dec}}^2(r_{\mathrm{dec}}+2)}{18}\frac{\delta\rho_\sigma^2}{\bar \rho_\sigma^2} +\ldots
\eeq
where the curvaton contributes a fraction $r_{\mathrm{dec}}=3\rho_{\sigma}/(4\rho_r+3\rho_\sigma)$ when it decays. If the curvaton dominates,$r_{\mathrm{dec}} =1$, we recover the expression above in (\ref{zetalin2}) at linear order in perturbation theory.

Now, accounting also for the intrinsic non-Gaussianity in the curvaton energy density and inserting the expression from (\ref{deltarhosigma}) in the equation above, we obtain
\beq
\zeta = \frac{2r_{\mathrm{dec}}}{3}\frac{\delta\sigma}{\sigma} +\left(\frac{r_{\mathrm{dec}}}{3}  -      \frac{2 r_{\mathrm{dec}}^2(r_{\mathrm{dec}}+2)}{9}\right)\frac{\delta\sigma^2}{\sigma^2} +\ldots
\eeq
which then implies that
\beq
\zeta=\zeta_{g}+\left(\frac{3}{4r_{\mathrm{dec}}}-1- \frac{r_{\mathrm{dec}}}{2} \right)\zeta_{g}^{2}\,,
\eeq
with the Gaussian linear perturbation defined as $\zeta_g \equiv (2r_{\mathrm{dec}}/{3})(\delta\sigma/{\sigma})$.

When comparing with the definition
\beq
\zeta=\zeta_{g}+\frac{3}{5} f_{N L}^{\text {local }}\left(\zeta_{g}^{2}-\left\langle\zeta_{g}^{2}\right\rangle\right)\,,
\eeq
we then obtain
\beq
\boxed{f_{N L}^{\text {local }}=\frac{5}{4} \frac{1}{r_{\mathrm{dec}}}-\frac{5}{3}-\frac{5}{6}r_{\mathrm{dec}}}
\eeq
We notice that when the curvaton dominates, when it decays for $r_{\mathrm{dec}}=1$, one finds $f_{N L}^{\text {local }}=-\frac{5}{4}$, so generally we expect
\beq
\left|f_{N L}^{\text {local }}\right|\sim \mathcal{O}(1)\,,
\eeq
which is much larger than what we found in single-field slow-roll inflation. It can potentially be measured within the next 10 years.
\begin{center}
\begin{figure}
\includegraphics[max width=\textwidth]{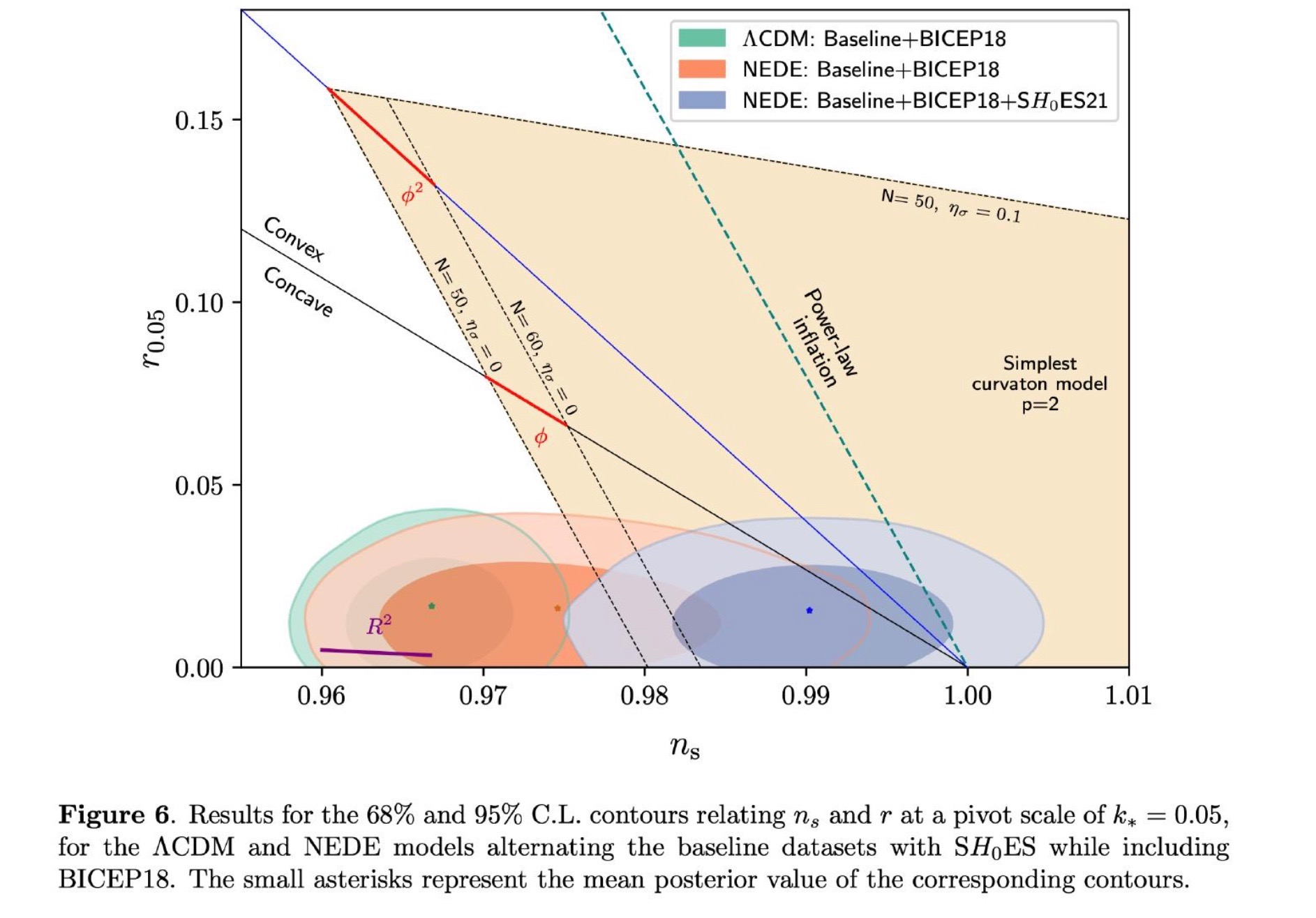}
\caption{Results for the $68 \%$ and $95 \%$ C.L. contours relating $n_{s}$ and $r$ at a pivot scale of $k_{*}=0.05$, for the $\Lambda \mathrm{CDM}$ and NEDE models alternating the baseline datasets with $\mathrm{S} H_{0} \mathrm{ES}$ while including BICEP18. The small asterisks represent the mean posterior value of the corresponding contours. Ficure taken from \cite{Cruz:2022oqk}}\label{curvfig}
\end{figure}
\end{center}
So, coming back to our plot. In the plot in Figure \ref{curvfig}, we used the relations for the simplest curvaton model
\beq
V(\phi, \sigma)=\frac{1}{2} M^{2} \phi^{2}+\frac{1}{2} m^{2} \sigma^{2}
\eeq
which implies \cite{Enqvist:2013paa,Bartolo:2003jx}
\beq
 n_{s}=1-\frac{1}{1+R} \frac{8}{4 N+2}+\frac{R}{1+R}\left[-2 \epsilon+2 \eta_{\sigma}\right] \,,\quad f_{\mathrm{NL}}=-\left(\frac{R}{1+R}\right)^{2}\left[\frac{5}{3}-\frac{5}{4 r_{\mathrm{dec}}}+\frac{5}{6} r_{\mathrm{dec}}\right] 
\eeq
where $r=(16 \epsilon)/(1+R)$ and $R=\mathcal{P}_{\zeta_{\text {curvaton }}}/\mathcal{P}_{\zeta_{\text {inflaton }}}$, so when \\
\noindent{\underline{the inflation dominates}}
\bea
 R &\rightarrow& 0 \\
 r &\rightarrow& 16 \epsilon \\
 f_{\mathrm{NL}} &\rightarrow& 0\,,\qquad\qquad\qquad\qquad\qquad{}
\eea
where we ignored the slow-roll suppressed contribution to $f_{\mathrm{NL}}$. On the other hand, when \\
\noindent{\underline{the curvaton dominates}}
\bea
 R& \rightarrow& \infty\\
r &\rightarrow& 0 \\
 f_{\mathrm{NL}} &\rightarrow& -5/4 \qquad\qquad(r_{\mathrm{dec}}=1)\,.
\eea
Thus, tensor modes, parametrized by $r$, and non-Gaussianity, parametrized by $f_{\mathrm{NL}}$ are complementary probes of the big yellow region in the plot.

\subsection{Exchange consistency relation}

One can think of the Maldacena consistency relation for the 3-point function to be pictorially of the form
\begin{center}
\includegraphics[max width=5cm]{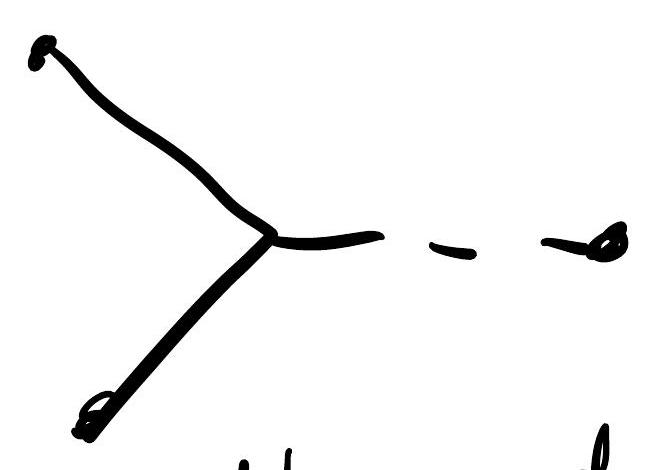}
\end{center}
where the dashed line is a long mode, and the solid lines are short wavelength modes. Now it was first understood by Seery, Sloth, and Vernizzi (SSV) \cite{Seery:2008ax} that the four-point function in the counter-collinear limit, also called collapsed shape, satisfies a new consistency relation for the exchange diagram contribution

\includegraphics[max width=7cm, center]{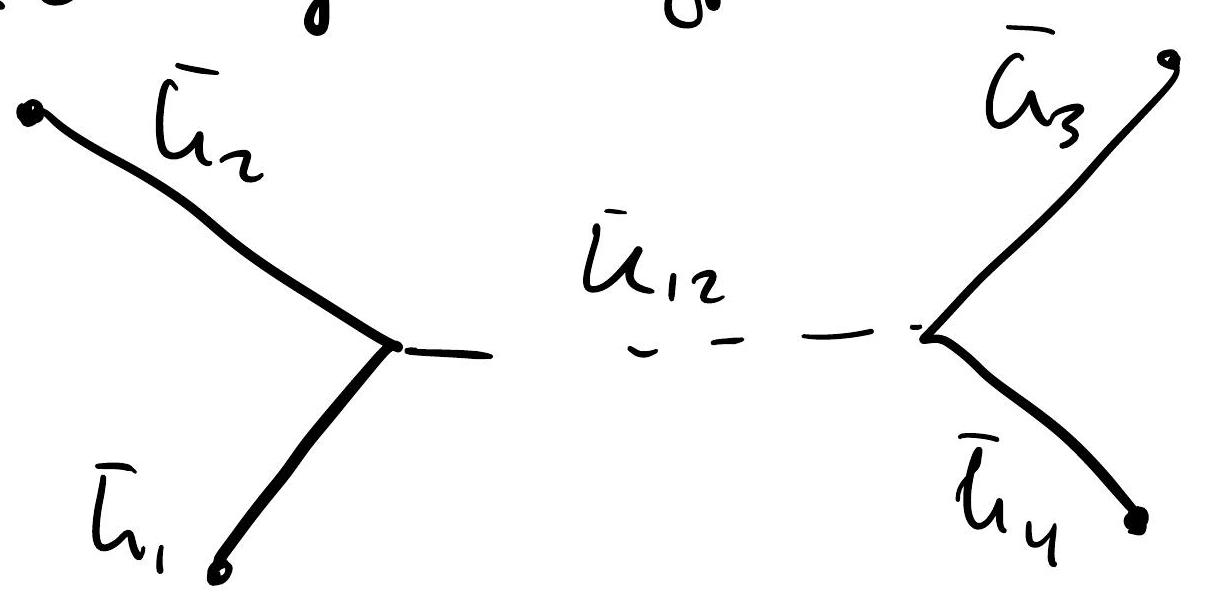}
with $\vect{k}_{12}=\vect{k}_{1}+\vect{k}_{2}$ for momentum conservation reasons and we assume $k_{12} \ll k_{1} \approx k_{2}$, $k_{3} \approx k_{4}$ like a folded kite or a parallelogram.
\begin{center}
\includegraphics[max width=7cm]{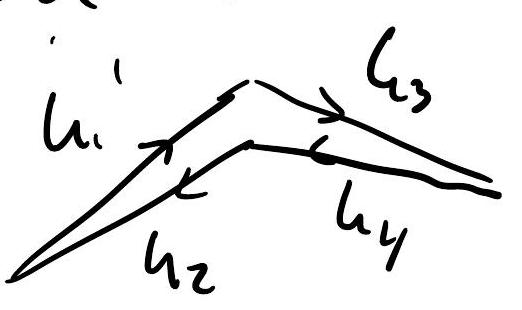}
\end{center}
\begin{center}
\includegraphics[max width=7cm]{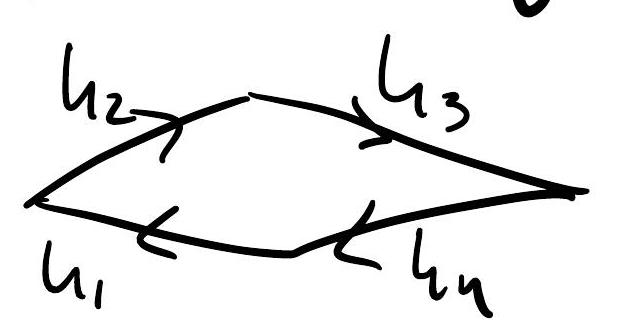}
\end{center}

The SSV consistency relation \cite{Seery:2008ax} for the 4-point exchange diagram in the counter-collinear limit tells us that the diagram can be cut
\begin{center}
\includegraphics[max width=7cm]{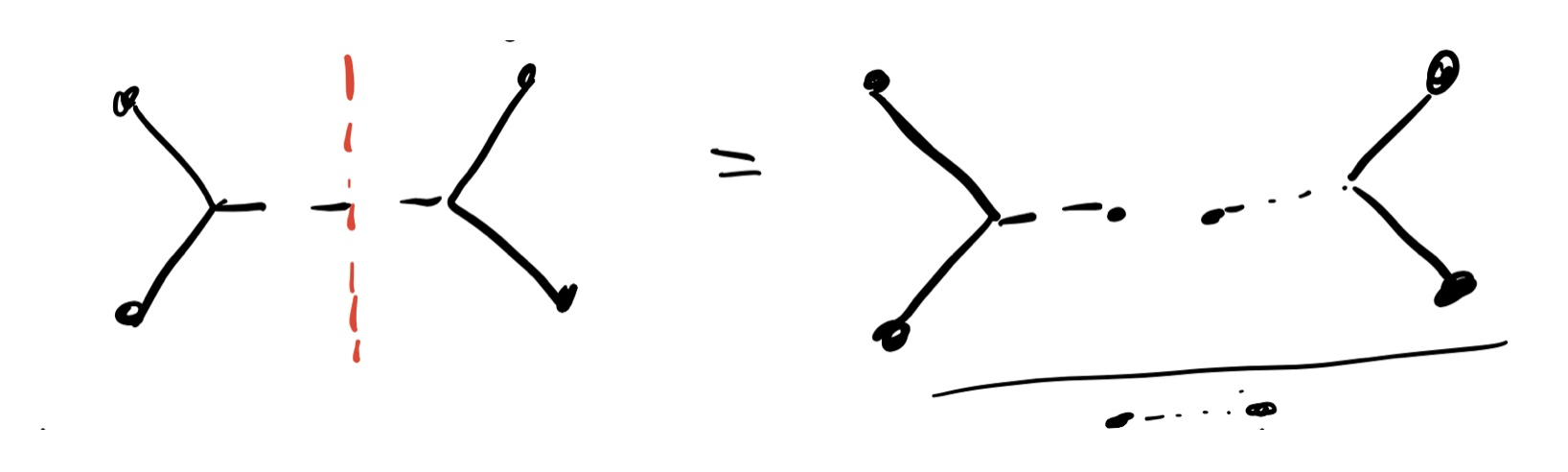}
\end{center}
so that it satisfies a relation
\begin{center}
\includegraphics[max width=7cm]{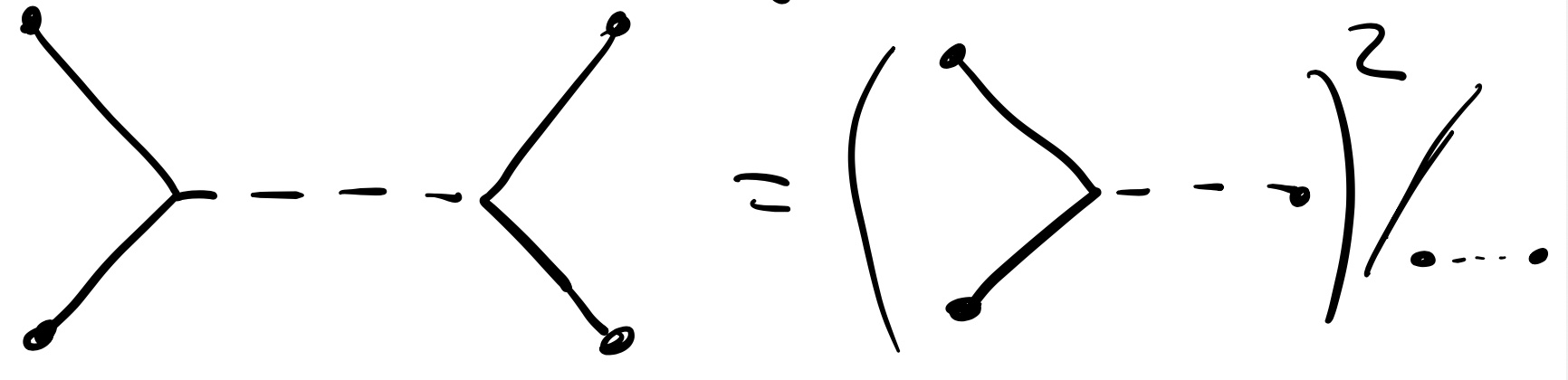}
\end{center}
in the following sense: remember, we had in the presence of a long/soft mode $\zeta_{B}$
\bea
\left\langle\zeta_{\vect{k}_{1}}, \zeta_{\vect{k}_{2}}\right\rangle_{\zeta_{B}}&=&\left\langle\zeta_{\vect{k}_{1}} \zeta_{\vect{k}_{2}}\right\rangle_{0}+\left.\zeta_{B} \frac{\partial}{\partial \zeta_{B}}\left\langle\zeta_{\vect{k}_{1}} \zeta_{\vect{k}_{2}}\right\rangle\right|_{\zeta_B=0}+\ldots \nonumber\\
& =&\left\langle\zeta_{\vect{k}_{1}} \zeta_{\vect{k}_{2}}\right\rangle_{0}
-\left(n_{s}-1\right) \int \frac{d^{3} q_{B}}{(2 \pi)^{3}} \zeta_{B}\left(\vect{q}_{B}\right)\left\langle\zeta_{\vect{k}_{1}-\frac{1}{2} \vect{q}_{B}} \zeta_{\vect{k}_{2}-\frac{1}{2} \vect{q}_{B}}\right\rangle
\eea

So now we can obtain the semiclassical contribution to the four-point function from the correlation of a pair of two-point functions along the long mode
\bea
\lim _{k_{12}\rightarrow 0}\left\langle \zeta_{\vect{k}_{1}} \zeta_{\vect{k}_{2}} \zeta_{\vect{k}_{3}} \zeta_{\vect{k}_{4}}\right\rangle&=&\left\langle\left\langle \zeta_{\vect{k}_{1}} \zeta_{\vect{k}_{2}} \right\rangle_{\zeta_B}\left\langle \zeta_{\vect{k}_{3}} \zeta_{\vect{k}_{4}}\right\rangle_{\zeta_B}\right\rangle\nonumber \\
&=& 
 \left(n_{s}-1\right)^{2}\left\langle\zeta_{\vect{k}_{1}} \zeta_{\vect{k}_{2}}\right\rangle\left\langle\zeta_{\vect{k}_{12}} \zeta_{-\vect{k}_{12}}\right\rangle\left\langle\zeta_{\vect{k}_{3}}, \zeta_{\vect{k}_{4}}\right\rangle\,.
\eea
This is the scalar version of the SSV consistency relation, which can also be written in terms of the trispectrum
\bea
\lim _{k_{12} \rightarrow 0} T\left(\vect{k}_{1}, \vect{k}_{2}, \vect{k}_{3}, \vect{k}_{4}\right) & =&\left(n_{s}-1\right)^{2} P_{\zeta}\left(k_{1}\right) P_{\zeta}\left(k_{3}\right) P_{\zeta}\left(k_{12}\right) \\
& =&4 \tau_{N L}^{\text {local }} P_{\zeta}\left(k_{1}\right) P_{\zeta}\left(k_{3}\right) P_{\zeta}\left(k_{12}\right)
\eea
where $\tau_{NL}^{\text {local}}=\left(\frac{6}{5} f_{NL}^{\text {local }}\right)^{2}$, with the trispectrum, $T$, defined in terms of the connected four-point function as
\beq
\left\langle \zeta_{\vect{k}_{1}} \zeta_{\vect{k}_{2}} \zeta_{\vect{k}_{3}} \zeta_{\vect{k}_{4}}\right\rangle_c=(2 \pi)^{3} \delta\left(\sum_{a} \vect{k}_{a}\right) T_{\zeta}\left(\vect{k}_{1}, \vect{k}_{2}, \vect{k}_{3}, \vect{k}_{4}\right)
\eeq
The SSV consistency relation was originally derived to include the exchange of gravitons, which is less slow-roll suppressed, and can be generalized also in other ways \cite{Seery:2008ax,Giddings:2010nc} (in particular, see the appendix of \cite{Giddings:2010nc} for a small review of consistency relations). Note that the full trispectrum was calculated in \cite{Seery:2008ax,Seery:2006vu}, and the SSV exchange consistency relation was verified by a full in-in QFT calculation (for an introduction to in-in diagrammatic QFT techniques, see \cite{Giddings:2010ui} ).

\subsection{Semiclassical consistency relations: loops and IR effects}

At higher orders in perturbation theory, we can also have loop corrections to, for example, the two-point correlation function $\left\langle\zeta_{\vect{k}_{1}}, \zeta_{\vect{k}_{2}}\right\rangle$. This has been a hot topic of discussion because of apparent IR divergences that need to be dealt with correctly, but may also teach us about the global nature of inflationary spacetimes.

It was first shown by Giddings and Sloth (GS) \cite{Giddings:2010nc} that one can similarly use semiclassical relations (soft theorems) to extract the IR contribution from loops.

The basic insight of GS is that if one goes to one order higher in the background expansion along the long mode
\bea
\left\langle \zeta_{\vect{k}_{1}} \zeta_{\vect{k}_{2}}\right\rangle_{\zeta_{B}}= & \left\langle \zeta_{\vect{k}_{1}} \zeta_{\vect{k}_{2}}\right\rangle_{0}+\zeta_{B} \frac{\partial}{\partial \zeta_{B}}\langle\left. \zeta_{\vect{k}_{1}} \zeta_{\vect{k}_{2}}\rangle\right|_{\zeta_B=0} +\frac{1}{2} \zeta_{B}^{2} \frac{\partial^{2}}{\partial \zeta_{B}^{2}}\langle\left. \zeta_{\vect{k}_{1}} \zeta_{\vect{k}_{2}} \rangle\right|_{\zeta_B=0}+\ldots
\eea
and then take the average over the long modes
\beq
\left\langle\left\langle \zeta_{\vect{k}_{1}} \zeta_{\vect{k}_{2}}\right\rangle\right\rangle_{\zeta_B}=\left\langle \zeta_{\vect{k}_{1}} \zeta_{\vect{k}_{2}}\right\rangle_{0}+\frac{1}{2}\left\langle\zeta_{B}^{2}\right\rangle \frac{\partial^{2}}{\partial \zeta_{B}^{2}}\left.\left\langle\zeta_{\vect{k}_{1}}\zeta_{\vect{k}_{2}} \right\rangle_{\zeta_B}\right|_{\zeta_B=0}+\ldots
\eeq
Using from earlier
\beq
\left\langle \zeta_{\vect{k}_{1}} \zeta_{\vect{k}_{2}}\right\rangle_{\zeta_{B}}=e^{-6 \zeta_{B}}\left\langle\zeta\left( e^{-\zeta_{B}} \vect{k}_{1}\right) \zeta\left(e^{-\zeta_{B}} \vect{k}_{2}\right)\right\rangle
\eeq
we obtain the GS consistency relation for IR 1-loop contributions
\beq
\left\langle\left\langle \zeta_{\vect{k}_{1}} \zeta_{\vect{k}_{2}}\right\rangle_{\zeta_{B}}\right\rangle=\left\langle \zeta_{\vect{k}_{1}} \zeta_{\vect{k}_{2}}\right\rangle_{0}+\left(\frac{1}{2}\left(n_{s}-1\right)^{2}+\alpha_{s}\right)\left\langle \zeta_{\vect{k}_{1}} \zeta_{\vect{k}_{2}}\right\rangle_0 \left\langle\zeta_{B}^{2}(x)\right\rangle_{*}
\eeq
where
\beq
\left\langle\zeta_{B}^{2}(x)\right\rangle_{*} \approx \int_{a_{i} H_i}^{a_{*} H_*} \frac{d q}{q} \frac{1}{2 \epsilon} \frac{H^{2}}{(2 \pi)^{2}}
\eeq
and $\alpha_{s}=d n_{s} / d \ln (k)$ is the running of the spectral index.

In principle, the integral is IR divergent, but the IR cutoff should be the largest relevant length scale/box size, and the upper cutoff $\left(a_{*} H_*\right)$ is when the short modes cross the horizon $a_{*} H_* \approx k_{1} \sim k_{2}$. For a more detailed discussion of how to deal with the IR divergence and the derivation of an IR renormalization group equation showing how the correlation function changes with the box size/observable region of interest, see \cite{Giddings:2011zd}.

This can be generalized to higher orders by going to higher orders in the Taylor expansion. 

\subsection{The infrared triangle}

The infrared triangle was put forward to illustrate the infrared structure of gauge theory and gravity mainly in the context of black hole physics by Strominger \cite{Strominger:2017zoo} as an outcome of attempts to understand the black hole information paradox in his work with Perry and Hawking \cite{Hawking:2016msc}. However, in the work of Ferreira, Sandora, and Sloth (FSS) \cite{Ferreira:2016hee} (see also \cite{Sloth:2025nan}), it was argued that a similar relation exists for inflationary spacetimes.\footnote{
The SSV and GS consistency relations for soft exchange diagrams and soft loop
effects \cite{Seery:2008ax,Giddings:2010nc}, together with the de Sitter
Infrared-triangle perspective of FSS
\cite{Ferreira:2016hee,Ferreira:2017ogo,Ferreira:2017erz}, predate and are
conceptually closely related to more recent work on cosmological soft theorems
and the cosmological bootstrap. In particular, the use of soft limits and
consistency relations to constrain inflationary correlators, including exchange
diagrams, is part of the same broader set of ideas; see e.g.
\cite{Pajer:2020wxk,Baumann:2022jpr,Arkani-Hamed:2018kmz}. For related
flat-space developments involving soft theorems, asymptotic symmetries, and
celestial amplitudes, see \cite{McLoughlin:2022ljp}.}

The infrared triangle relates the semiclassical consistency relations, also sometimes called soft theorems, to gravitational memory effects and asymptotic symmetries.

\begin{center}
\underline{\Large{\bf Infrared triangle of de Sitter spacetime}}
\end{center}

\begin{center}
\includegraphics[max width=\textwidth]{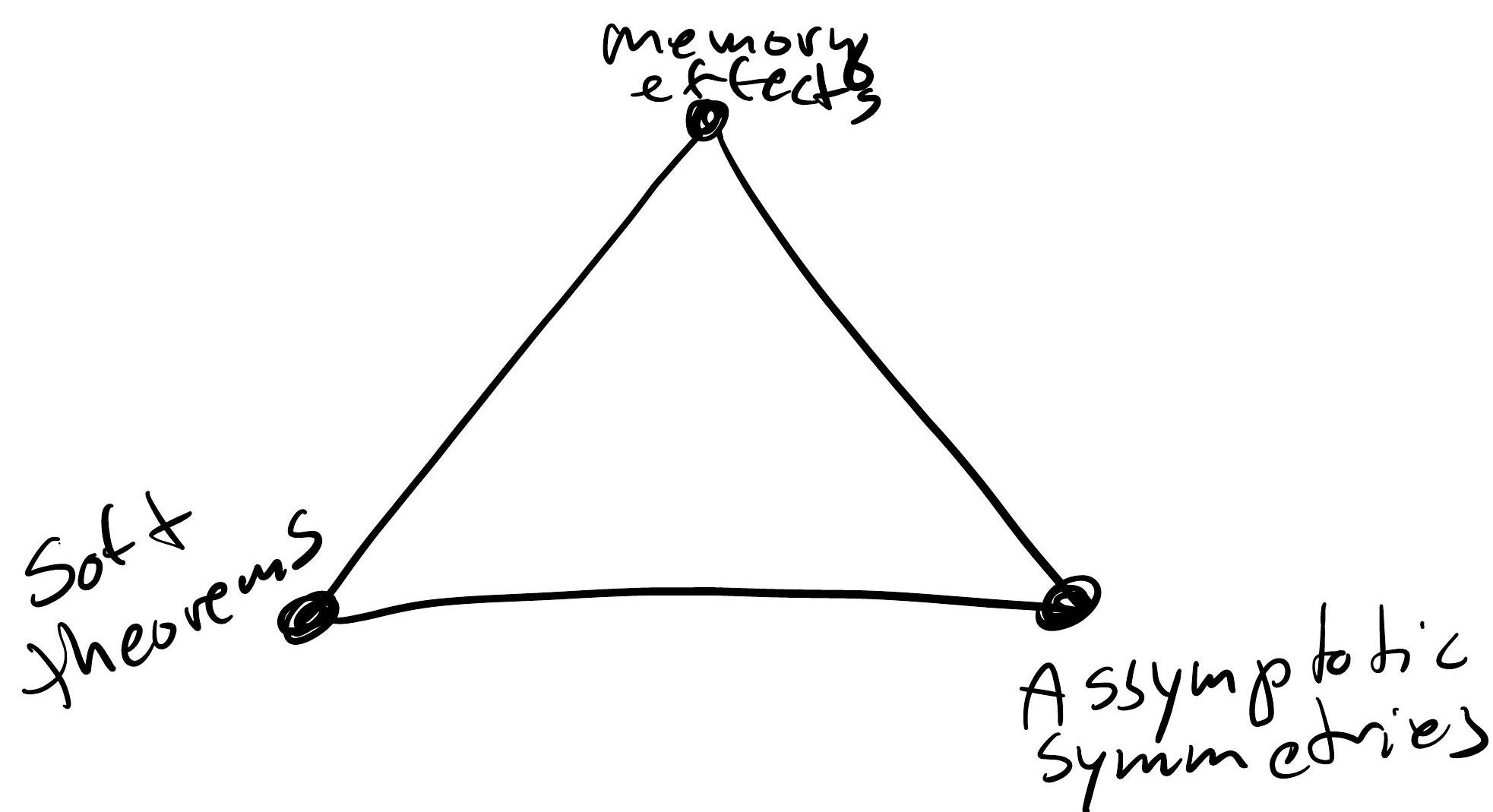}
\end{center}

(see Ferreira, Sandora, Sloth 2016 \& 2017 \cite{Ferreira:2016hee,Ferreira:2017ogo,Ferreira:2017erz}, Sloth 2025 \cite{Sloth:2025nan},  and Anninos, Ng, Strominger 2010 \cite{Anninos:2010zf} )

\subsection{Asymptotic symmetries}

Since gravitons freeze and become constant on superhorizon scales, they can locally be viewed as a rescaling of the local coordinate, equivalent to a large gauge transformation.

However, this is not true when comparing different patches across variations in the long mode. The presence of the long mode at super-horizon scales (approaching asymptotic infinity where the asymptotic symmetries act) spontaneously breaks the asymptotic symmetry group, and the long (soft) mode can be viewed as the Goldstone mode of the spontaneously broken asymptotic symmetry. More precisely, at the linearized level, a constant soft graviton corresponds to broken symmetric traceless anisotropic rescalings (shears) inside the spatial diffeomorphism group, which is the asymptotic symmetry group of de Sitter \cite{Sloth:2025nan}.

Focussing on tensor modes (a similar story holds for $\zeta$) in the transverse and traceless gauge
\beq
d s^{2}=-d t^{2}+a^{2}\left[e^{\gamma}\right]_{i j} d x^{i} d x^{j}
\eeq
where we choose the exponentiated parametrization, with $\gamma_{ij}$ transverse and traceless. Since modes freeze and become constant on super-horizon scales, the addition of a super-horizon soft mode is a large gauge transformation (gauge transformation not falling off at infinity) corresponding to a spatial diffeomorphism on $\mathbb{R}^{3}$ (the asymptotic symmetry group of de Sitter spacetime).
\begin{center}
\includegraphics[max width=\textwidth]{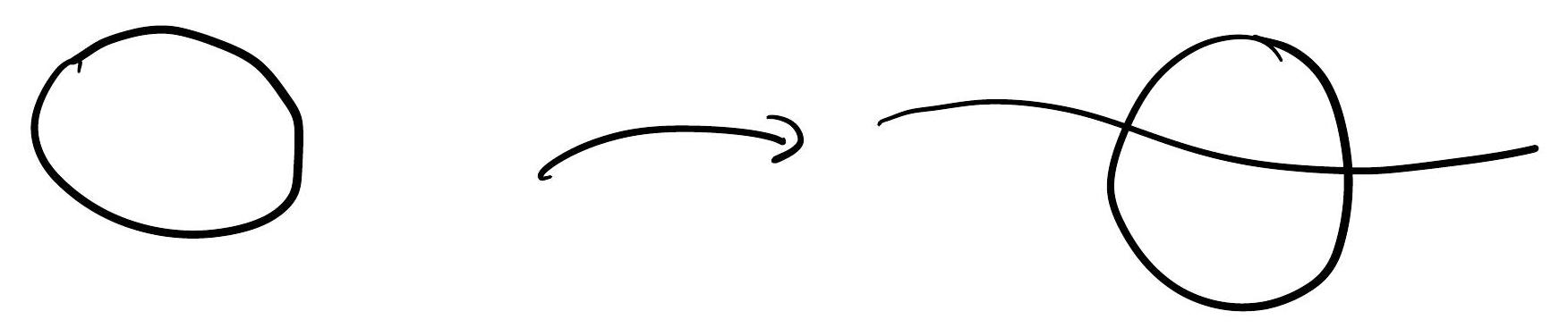}
\end{center}
If the charge, $Q$, generating the asymptotic symmetry transformation corresponding to adding a long mode, acts non-trivially on the vacuum, the asymptotic symmetry is spontaneously broken
\bea
 x^{i} \qquad&\rightarrow&\qquad{\left[e^{\gamma_{L} / 2}\right]^{i}}_j x^{j} \\
 |0\rangle \qquad&\rightarrow& \qquad\left|0^{\prime}\right\rangle=e^{i Q}|0\rangle\,.
\eea

From the definition of the Noether charge related to a variation, $\delta \gamma_{ij}$, of the canonical variable $\gamma_{ij}$, if the variation is a symmetry transformation, we have ($\pi$ = canonical conjugate momentum)
\beq
Q=\frac{1}{2} \int d^{3} x \pi^{i j} \delta \gamma_{i j}\quad+{ h.c. }
\eeq
Demanding that the field variation corresponds to a large gauge transformation, in the form of a spatial diffeomorphism
\beq
\mathcal{L}_{\xi}\left[e^{\gamma}\right]_{i j}=\delta\left[e^{\gamma}\right]_{i j}
\eeq
gives to first order
\beq
\delta \gamma_{i j}=\partial_i \xi_j +\partial_j \xi_i-\frac{2}{3}\delta_{ij}\partial^a\xi_a+\xi_{a} \partial^{a} \gamma_{i j}\,.
\eeq

If the large gauge transformation equals adding a long wavelength soft graviton mode, then we must have
\beq
\xi_{i}=-\frac{1}{2} \gamma^L_{i j} x^{j}
\eeq
where the transverse and traceless condition of $\gamma^L_{i j}$ ensures $\delta\gamma_{ij}=-\gamma^L_{ij}$, and we obtain $\left(\pi_{i j}=\frac{1}{4} a^{3} \dot{\gamma}_{i j}\right)$
\beq
Q=-\frac{a^{3}}{16} \int d^{3} x \dot{\gamma}_{i j} \gamma_{a b}^{L} x^{b} \partial^{a} \gamma_{i j}\quad +\mathrm{h.c.} \,,
\eeq
$\left(M_{p} \equiv 1\right)$.

If the asymptotic symmetry is spontaneously broken, then the charge associated with it will act non-trivially on the vacuum
\beq
e^{i Q}|0\rangle=\left|0^{\prime}\right\rangle \neq|0\rangle
\eeq
To see that this really changes the vacuum $|0\rangle$ into $|0'\rangle$ which now includes a soft mode $\gamma_{L}$, consider the 3-point graviton amplitude in the squeezed limit
\beq
\left\langle\gamma_{\vect{q}_{1}}^{s_{1}} \gamma_{\vect{q}_{2}}^{s_{2}} \gamma_{\vect{q}_{3}}^{s_{3}}\right\rangle
\eeq
where $q_{1} \ll q_{2}, q_{3}$.

The free-Gaussian three-point correlation function is zero, so
\beq
\left\langle 0\left|\gamma_{\vect{q}_{1}}^{s_{1}} \gamma_{\vect{q}_{2}}^{s_{2}} \gamma_{\vect{q}_{3}}^{s_{3}}\right| 0\right\rangle=0\,.
\eeq
Of course, the actual physical tree-level three-point function from the cubic graviton interaction is not zero. In the standard in-in picture, one must insert one interaction Hamiltonian, whereas in the charge picture, the transformed vacuum already contains the soft mode \cite{Ferreira:2016hee}.

Now, let's consider the same correlation function when a soft mode is added in the transformed vacuum
\bea
 \left\langle 0^{\prime}\left|\gamma_{\vect{q}_{1}}^{s_{1}} \gamma_{\vect{q}_{2}}^{s_{2}} \gamma_{\vect{q}_{3}}^{s_{3}}\right| 0^{\prime}\right\rangle &=& \left\langle 0\left|e^{-i Q} \gamma_{\vect{q}_{1}}^{s_{1}} \gamma_{\vect{q}_{2}}^{s_{2}} \gamma_{\vect{q}_{3}}^{s_{3}} e^{i Q}\right| 0\right\rangle \\
 &=& \left\langle 0\left|\gamma_{\vect{q}_{1}}^{s_{1}} \gamma_{\vect{q}_{2}}^{s_{2}} \gamma_{\vect{q}_{3}}^{s_{3}}\right| 0\right\rangle-i\left\langle 0\left|\left[Q, \gamma_{\vect{q}_{1}}^{s_{1}} \gamma_{\vect{q}_{2}}^{s_{2}} \gamma_{\vect{q}_{3}}^{s_{3}}\right]\right| 0\right\rangle  +\mathcal{O}\left(Q^{2}\right)\label{dot1}
\eea
:\\
:
\bea\label{dot2}
\qquad\qquad\qquad\qquad~~&=&\frac{3-n_{t}}{2} \frac{\epsilon^{s_1}_{i j} q_2^{i} q_2^{j}}{q_2^{2}}\left\langle\gamma^{s_1}_{\vect{q}_{1}} \gamma^{s_1}_{-\vect{q}_{1}}\right\rangle\left\langle\gamma^{s_2}_{\vect{q}_{2}} \gamma^{s_3}_{\vect{q}_{3}}\right\rangle\,.\qquad\qquad\qquad
\eea
This reproduces Maldacena 2002 \cite{Maldacena:2002vr}, where it was computed with the Maldacena consistency relation (soft theorem). We can therefore conclude
\beq
\boxed{\textrm{Soft theorems} \longleftrightarrow \textrm{asymptotic symmetries!}}
\eeq
  \vspace{4pt}
    \hrule
   \vspace{4pt}
\noindent {\bf Exercise 4:} Connect the dots above (between eq.(\ref{dot1}) and eq.(\ref{dot2})). [hint: have a look at \cite{Ferreira:2016hee} sect.4.1]
  \vspace{4pt}
    \hrule
  \vspace{4pt}
  \vspace{4pt}
  \vspace{4pt}
\subsection{Gravitational memory}

To complete the infrared triangle of de Sitter, let's briefly consider gravitational memory. As a thought experiment, we will introduce the concept of a patient observer as one who carries gravitational memory. One example of a patient observer, connecting to the other two corners of the triangle, is an observer who records the initial state before the soft mode is created, and is around long enough to compare with the final state after the soft mode has left the horizon.
\begin{center}
\includegraphics[max width=5cm]{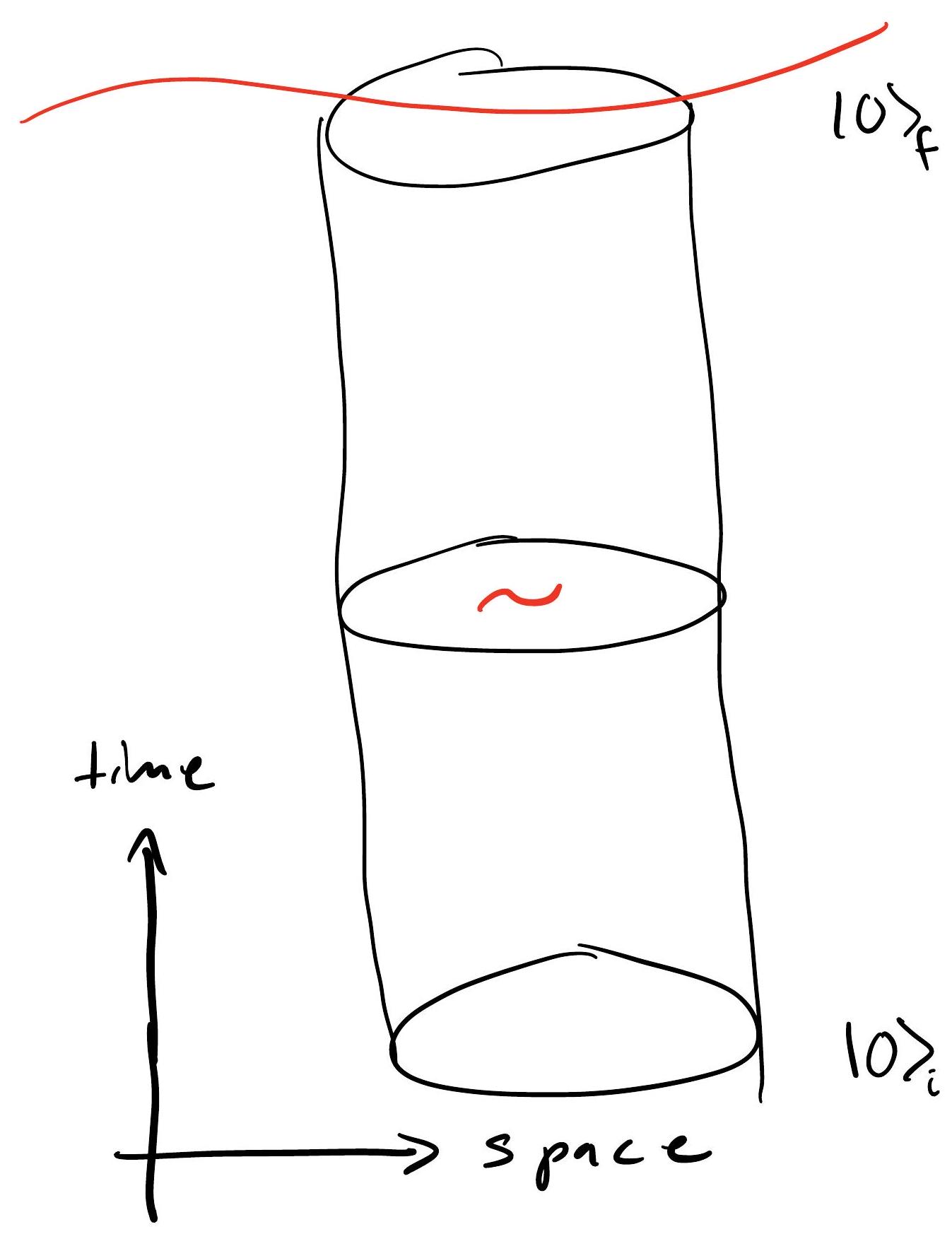}
\end{center}

Such an observer will see
\beq
{ }_{i}\langle 0 \mid 0\rangle_{f} \neq 1
\eeq
since
\beq
|0\rangle_{f}=e^{i Q}|0\rangle_{i}
\eeq

A patient observer could be a circular array of satellites very carefully bound together in the radial direction, but not preventing them from feeling shear effects.
When a long mode is added, the spatial distance between the satellites changes by $d s^{2}=a^{2} \delta_{i j} d x^{i} d x^{j} \rightarrow d s^{\prime 2}=a^{2}\left(e^{\gamma_{L}}\right)_{i j} d x^{i} d x^{j}$
\begin{center}
\includegraphics[max width=8cm]{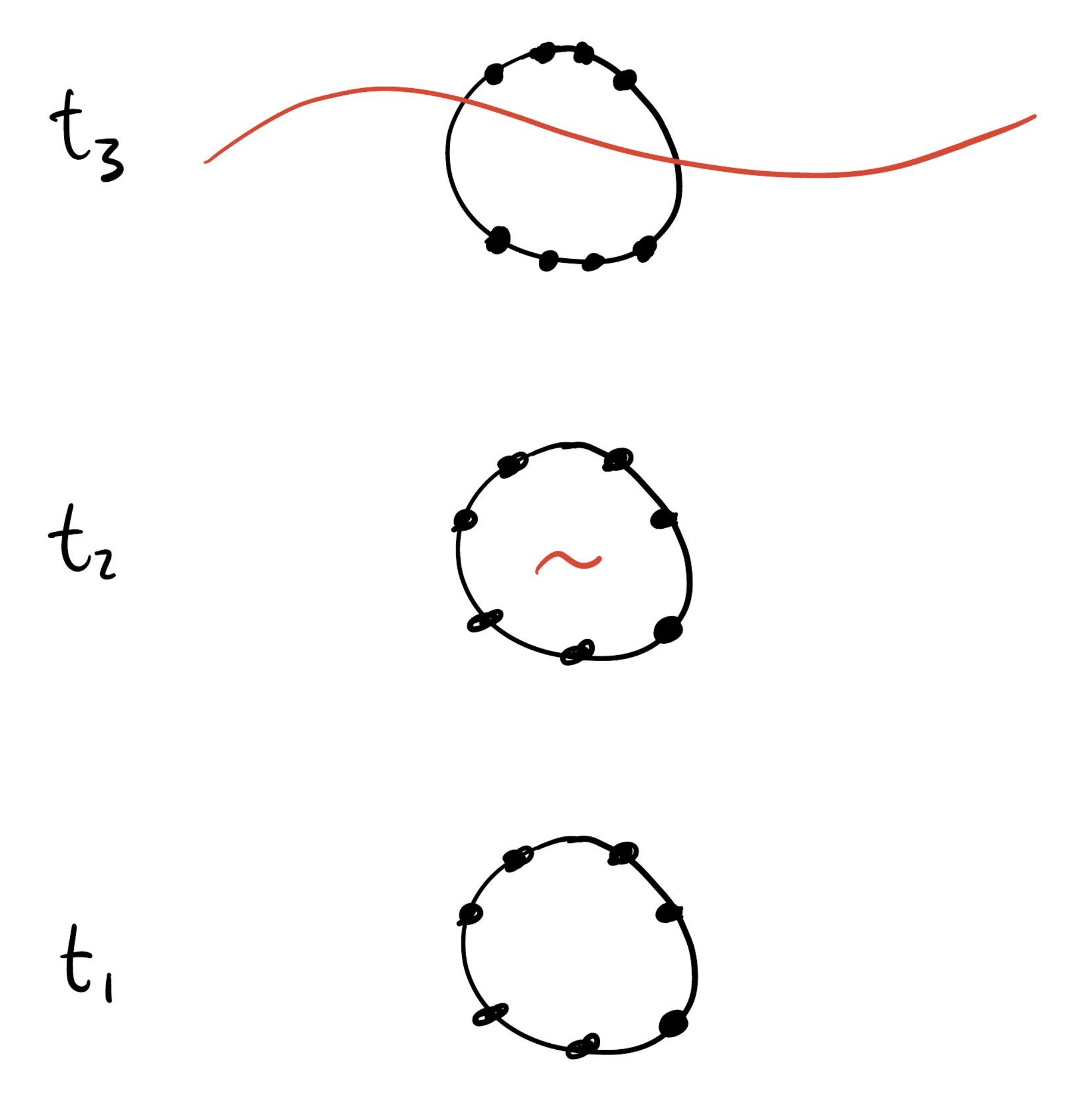}
\end{center}
This thought experiment connects gravitational memory effects to soft theorems and asymptotic symmetries. For details and challenges, see Ferreira, Sandora, Sloth, $2016~ \&~ 2017$ \cite{Ferreira:2016hee,Ferreira:2017ogo,Ferreira:2017erz}.

\subsubsection*{Acknowledgements}

I thank the organizers of the {\it Nordita Winter School 2024 - Particle Physics and Cosmology}, as well as the students who pointed out typos, which should (hopefully) be mostly fixed in this version.

\newpage
\bibliographystyle{unsrturl}
\bibliography{inflation-refs} 

\end{document}